\documentclass[12pt]{article}
\usepackage{jheppub}

\pdfoutput=1

\usepackage{amsmath,bbm,array,amsfonts,graphicx,wrapfig,lscape,float,mathtools,multirow,longtable}
\usepackage[dvipsnames]{xcolor}

\newcommand{\be}{\begin{equation}}
\newcommand{\ee}{\end{equation}}
\newcommand{\beq}{\begin{equation}}
\newcommand{\beql}[1]{\begin{equation}\label{#1}}
\newcommand{\eeq}{\end{equation}}
\newcommand{\ba}{\begin{array}}
\newcommand{\ea}{\end{array}}
\newcommand{\bea}{\begin{eqnarray}}
\newcommand{\beal}[1]{\begin{eqnarray}\label{#1}}
\newcommand{\eea}{\end{eqnarray}}
\newcommand{\ben}{\begin{enumerate}}
\newcommand{\een}{\end{enumerate}}
\newcommand{\bean}{\begin{eqnarray*}}
\newcommand{\eean}{\end{eqnarray*}}
\newcommand{\eref}[1]{(\ref{#1})}
\newcommand{\sref}[1]{\S\ref{#1}}
\newcommand{\tref}[1]{Table~\ref{#1}}

\newcommand{\fref}[1]{Figure \ref{#1}}
\newcommand{\btab}[1]{\begin{tabular}{#1}}
\newcommand{\etab}{\end{tabular}}

\def \Black {\pdfsetcolor{0 0 0 1}}

\newcommand{\comment}[1]{}

\newcommand{\qed}{\nobreak \ifvmode \relax \else
      \ifdim\lastskip<1.5em \hskip-\lastskip
      \hskip1.5em plus0em minus0.5em \fi \nobreak
      \vrule height0.75em width0.5em depth0.25em\fi}

\definecolor{darkspringgreen}{rgb}{0.09, 0.45, 0.27}
\definecolor{forestgreen}{rgb}{0.13, 0.55, 0.13}

\usepackage{array}
\usepackage{physics}
\usepackage{subcaption}
\usepackage{tikz,tikz-3dplot}
\usepackage[colorlinks=true]{hyperref}
\newcolumntype{C}[1]{>{\centering\let\newline\\\arraybackslash\hspace{0pt}}m{#1}}

\definecolor{yellow2}{rgb}{0.98, 0.80, 0.20}

\title{$3d$ Printing of $2d$ $\mathcal{N}=(0,2)$ Gauge Theories} 

\author[a,b]{Sebasti\'an Franco,} 
\author[a,b]{Azeem Hasan}

\affiliation[a]{
Physics Department, The City College of the CUNY \\
160 Convent Avenue, New York, NY 10031, USA}

\affiliation[b]{The Graduate School and University Center, The City University of New York  \\
365 Fifth Avenue, New York NY 10016, USA}

\emailAdd{sfranco@ccny.cuny.edu}
\emailAdd{ahasan@gradcenter.cuny.edu}

\preprint{
\begin{flushright}
CCNY-HEP-18-01
\end{flushright}
}

\abstract{We introduce {\it $3d$ printing}, a new algorithm for generating $2d$ $\mathcal{N}=(0, 2)$ gauge theories on D1-branes probing singular toric Calabi-Yau 4-folds using $4d$ $\mathcal{N}=1$ gauge theories on D3-branes probing toric Calabi-Yau 3-folds as starting points. Equivalently, this method produces brane brick models starting from brane tilings. $3d$ printing represents a significant improvement with respect to previously available tools, allowing a straightforward determination of gauge theories for geometries that until now could only be tackled using partial resolution. We investigate the interplay between triality, an IR equivalence between different $2d$ $\mathcal{N}=(0, 2)$ gauge theories, and the freedom in $3d$ printing given an underlying Calabi-Yau 4-fold. Finally, we present the first discussion of the consistency and reduction of brane brick models. 
}

\begin{document}

\maketitle

\section{Introduction}

In recent years we have witnessed considerable progress in the understanding of $2d$ $(0,2)$ gauge theories. These developments motivated a program aimed at realizing $2d$ $(0,2)$ theories in terms of branes and, in turn, exploiting such branes configuration to study the field theory dynamics. The engineering of $2d$ $(0,2)$ theories on the worldvolume of D1-branes probing singular toric Calabi-Yau (CY) 4-folds was developed in \cite{Franco:2015tna}, following the seminal work in \cite{GarciaCompean:1998kh}. {\it Brane brick models}, a new class of brane configurations in Type IIA string theory that are connected to D1-branes probing toric CY 4-folds by T-duality, were introduced and developed in \cite{Franco:2015tya,Franco:2016nwv,Franco:2016qxh}. Brane brick models streamline the map between gauge theory and geometry. They are analogous to brane tilings for $4d$ $\mathcal{N}=1$ theories on D3-branes probing toric CY 3-folds \cite{Franco:2005rj,Franco:2005sm,Franco:2017jeo}. 

This program produced numerous additional results, including: the realization of triality \cite{Gadde:2013lxa} in terms of brane brick models \cite{Franco:2016nwv} and geometric transitions in the mirror CY \cite{Franco:2016qxh}, a detailed understanding of these theories in terms of mirror symmetry \cite{Franco:2016qxh}, a field theoretic and geometric computation of the elliptic genus \cite{Franco:2017cjj}, the proposal of a new duality for $0d$ $\mathcal{N}=1$ theories \cite{Franco:2016tcm} and the development of an algebraic framework that underlies minimally supersymmetric theories in different dimensions and unifies their dualities \cite{Franco:2017lpa}.

There are several clear directions for further progress. On the practical front, it is desirable to develop more efficient methods for generating brane brick models associated to general toric singularities and, conversely, for rapidly finding the geometry corresponding to a brane brick model. In addition, there are various formal questions regarding brane brick models, which include: developing their mathematical and combinatorial understanding, clarifying the notions of consistency and reducibility and connecting dualities of gauge theories in different dimensions. The purpose of this paper is to shed light on all these issues. To do so, we will introduce {\it $3d$ printing}, a method for generating $2d$ $(0, 2)$ gauge theories on D1-branes probing singular toric CY 4-folds starting from $4d$ $\mathcal{N} = 1$ gauge theories on D3-branes probing toric CY 3-folds. In other words, this procedure generates brane brick models starting from brane tilings. $3d$ printing significantly generalizes {\it orbifold reduction} \cite{Franco:2016fxm}, which was an earlier step in this direction.

Various additional advances deserve to be mentioned. Constructions of $2d$ $(0,2)$ theories in other corners of string theory and F-theory have been presented in \cite{Tatar:2015sga,Benini:2015bwz, Schafer-Nameki:2016cfr, Apruzzi:2016iac,Apruzzi:2016nfr,Lawrie:2016axq,Lawrie:2016rqe}. The corresponding Green-Schwarz mechanism for anomaly cancellation was studied in \cite{Weigand:2017gwb}. AdS$_3$/CFT$_2$ pairs were constructed in \cite{Couzens:2017way,Couzens:2017nnr}. Finally, $2d$ theories with exotic SUSY were constructed in \cite{Florakis:2017zep}.

This paper is organized as follows. \sref{section_BBMs} reviews the basics of brane brick models. \sref{section_3d_printing} introduces $3d$ printing. \sref{section_3d_printing_and_geometry} explains how the CY$_4$ geometry emerges from $3d$ printing. The new ideas are illustrated with several explicit examples in \sref{section_examples}. \sref{section_phases_Q111/Z2} investigates the relation between $3d$ printing and triality, presenting a full classification of the toric phases of $\mathbb{Q}^{1,1,1}/\mathbb{Z}_2$. \sref{section_consistency_and_reduction} contains the first discussion of consistency and reduction of brane brick models. We present our conclusions in \sref{section_conclusions}. The periodic quivers for all toric phases of $\mathbb{Q}^{1,1,1}/\mathbb{Z}_2$ are presented in an appendix.

\section{Brane Brick Models}

\label{section_BBMs}

For completeness, we present here a brief review of brane brick models. We refer the reader to \cite{Franco:2015tna,Franco:2015tya,Franco:2016nwv,Franco:2016qxh} for detailed presentations. 

Brane brick models are obtained from D1-branes at $\text{CY}_4$ singularities by T-duality. A brane brick model is a Type IIA brane configuration consisting of D4-branes wrapping a 3-torus $\mathbb{T}^3$ and suspended from an NS5-brane that wraps a holomorphic surface $\Sigma$ intersecting with $\mathbb{T}^3$ as summarized in \tref{Brane brick-config}. The holomorphic surface $\Sigma$ is the zero locus of the Newton polynomial of the $\text{CY}_4$. 

\begin{table}[ht!!]
\centering
\begin{tabular}{l|cccccccccc}
\; & 0 & 1 & 2 & 3 & 4 & 5 & 6 & 7 & 8 & 9 \\
\hline
$\text{D4}$ & $\times$ & $\times$ & $\times$ & $\cdot$ & $\times$ & $\cdot$ & $\times$ & $\cdot$ & $\cdot$ & $\cdot$  \\
$\text{NS5}$ & $\times$ & $\times$ & \multicolumn{6}{c}{----------- \ $\Sigma$ \ ------------} & $\cdot$ & $\cdot$ \\
\end{tabular}
\caption{Brane brick model configuration.}
\label{Brane brick-config}
\end{table}

\begin{table}[H]
\centering
\resizebox{\hsize}{!}{
\begin{tabular}{|l|l|l|}
\hline
{\bf Brane Brick Model} \ \ &  {\bf Gauge Theory} \ \ \ \ \ \ \  & {\bf Periodic Quiver} \ \ \ 
\\
\hline\hline
Brick  & Gauge group & Node \\
\hline
Oriented face  & Bifundamental chiral field & Oriented (black) arrow 
\\
between bricks $i$ and $j$ & from node $i$ to node $j$  & from node $i$ to node $j$ \\
\hline
Unoriented square face  & Bifundamental Fermi field & Unoriented (red) line \\
between bricks $i$ and $j$ & between nodes $i$ and $j$ & between nodes $i$ and $j$  \\
\hline
Edge  & Interaction by $J$- or $E$-term & Plaquette encoding \\ 
& & a $J$- or an $E$-term \\
\hline
\end{tabular}
}
\caption{
Dictionary between brane brick models and $2d$ gauge theories.
\label{tbrick}
}
\end{table}

Brane brick models, or equivalently their dual periodic quivers, fully encode the $2d$ $(0,2)$ quiver gauge theories on the worldvolume of D1-branes probing toric CY 4-folds. The dictionary between the brane brick models and the gauge theories is summarized in \tref{tbrick}.

\section{$3d$ Printing}

\label{section_3d_printing}

Brane brick models considerably simplify the connection between the geometry of toric CY 4-folds and the $2d$ $(0,2)$ gauge theories living on the worldvolume of D1-branes probing them. Every toric CY$_4$ is in general associated to a class of brane brick models, which are related by triality. 

Given a toric CY$_4$, there are various systematic ways of constructing a brane brick model, i.e. a $2d$ $(0,2)$ gauge theory, associated to it. {\it Partial resolution} produces the unknown gauge theory by embedding the desired geometry into a larger one, for which the gauge theory is known. A standard class of starting points for partial resolution is given by orbifolds of $\mathbb{C}^4$. Multiple examples of this method can be found in \cite{Franco:2015tna}. An alternative approach is the {\it fast inverse algorithm} for brane brane brick models \cite{Franco:2015tya}. In this case, brane brick models are constructed from phase boundaries, which are the analogues of zig-zag paths for brane tilings. Finally, brane brick models can also be constructed using mirror symmetry, as explained in \cite{Franco:2016qxh}.

In this section we introduce {\it $3d$ printing}, another algorithmic procedure, which generates brane brick models starting from brane tilings describing the $4d$ $\mathcal{N}=1$ gauge theories associated to toric CY 3-folds. It significantly generalizes dimensional reduction, orbifolding and the recently introduced {\it orbifold reduction} \cite{Franco:2016fxm}. This method is attractive due to its simplicity and because it provides a novel conceptual perspective on the physics and combinatorics of brane brick models, by relating them to brane tilings.

\subsection{Dimensional Reduction}
 
 \label{section_dimensional_reduction}
 
$3d$ printing is a natural generalization of dimensional reduction for toric theories. We thus begin with a brief review of dimensional reduction. $4d$ $\mathcal{N}=1$ vector and chiral multiplets reduce to $2d$ $(0,2)$ multiplets as follows:

\begin{itemize}
\item \underline{$4d$ $\mathcal{N}=1$ vector $\mathcal{V}_i$} $\rightarrow$ $2d$ $(0,2)$ vector $V_i$ + $2d$ $(0,2)$ adjoint chiral $\Phi_{ii}$
\item \underline{$4d$ $\mathcal{N}=1$ chiral $\mathcal{X}_{ij}$} $\rightarrow$ $2d$ $(0,2)$ chiral $X_{ij}$ + $2d$ $(0,2)$ Fermi $\Lambda_{ij}$
\end{itemize}  
 
 The $J$-terms of the $2d$ theory descend from the $4d$ F-terms and are given by
 \bea
J_{\Lambda_{ij}} = \frac{\partial W}{\partial X_{ij}} \, ,
\label{J_dim_red}
\eea
with $W$ the $4d$ superpotential. In this expression, we understand the $J$-terms and $W$ as functions of the $2d$ $(0,2)$ chiral multiplets coming from the $4d$ chiral multiplets.
 
The $E$-terms follow from the $4d$ gauge interactions, and are given by
\beq
E_{ij} = \Phi_{ii} X_{ij} - X_{ij} \Phi_{jj} \, .
\label{E_dim_red}
\eeq

\subsection{$3d$ Printing CY 4-Folds}

Before introducing $3d$ printing, it is useful to explain in general lines the relation between the toric CY$_3$ and CY$_4$ connected by it. The relation between the two geometries will be discussed in detail in \sref{section_3d_printing_and_geometry}.

\begin{figure}[ht]
	\centering
	\includegraphics[width=11cm]{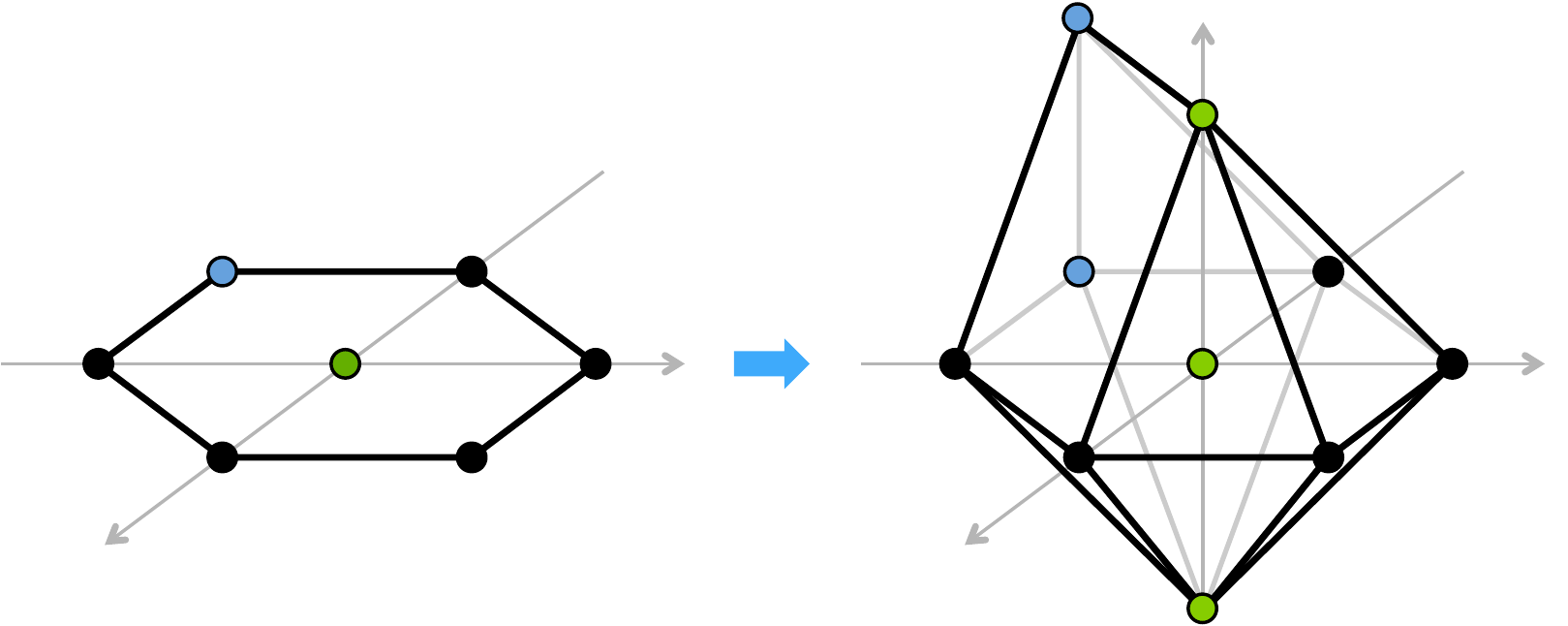}
\caption{Example of lift of two points in the toric diagram of the complex cone over $dP_3$.}
	\label{example_3d_printing_toric}
\end{figure}

Recall that the toric diagram of a CY$_3$ $T_{\rm{CY}_3}$ is 2-dimensional and every point in it corresponds to a (collection of) perfect matching(s) in an associated brane tiling. $3d$ printing turns $T_{\rm{CY}_3}$ into the $3d$ toric diagram of a CY$_4$, $T_{\rm{CY}_4}$, by growing a third dimension, e.g. the $z$-direction, as follows. We can simultaneously take several points in $T_{\rm{CY}_3}$ and expand each of them into an arbitrary number of points along $z$ with an also arbitrary shift with respect to the $x-y$ plane. It is important to note that any points inside the resulting convex hull are automatically lifted, properly generating a convex $T_{\rm{CY}_4}$. \fref{example_3d_printing_toric} shows an example that starts from the toric diagram of the complex cone over $dP_3$ and lifts two points. The blue and green points give rise to two and three points, respectively.

The toric diagram of of a wide class of CY$_4$'s can be reached by this procedure.\footnote{It would be interesting to characterize the class of geometries that cannot be generated by $3d$ printing in its current form and, if possible, to generalize it to produce arbitrary toric CY$_4$'s. An interesting class of CY$_4$'s that seems to be out of reach of $3d$ printing is given by those without crepant resolutions. These are geometries for which the normalized volume of the toric diagram cannot be obtained by a triangulation in terms of minimal tetrahedra {\it with vertices on points in the toric diagram}. A simple example is the $\mathbb{C}^4/\mathbb{Z}_2$ orbifold with action $(1,1,1,1)$ \cite{Franco:2015tna}.} This makes it a powerful tool for studying the corresponding brane configurations and gauge theories. It represents a significant improvement over orbifold reduction, in which a single point in $T_{\rm{CY}_3}$ could be expanded into the third dimension. We will further elaborate on the connection between the two procedures in \sref{section_connection_to_orbred}.

\begin{figure}[ht]
	\centering
	\includegraphics[width=10cm]{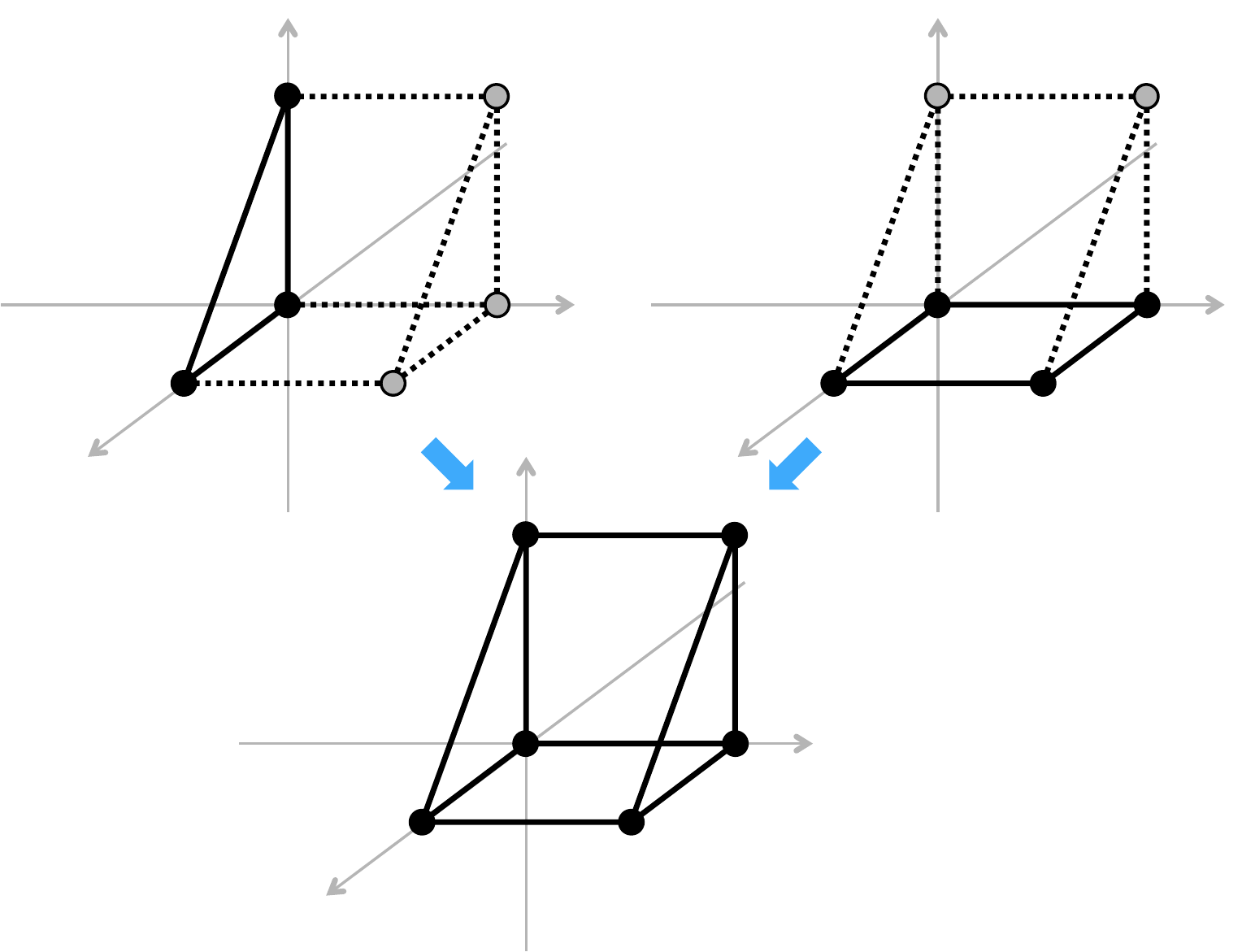}
\caption{The $D_3$ toric diagram can be reached by lifting three points in $\mathbb{C}^3$ or two points in the conifold.}
	\label{C3_and_conifold_to_D3}
\end{figure}

Even after this preliminary introduction, it becomes clear that a given CY$_4$ can be reached by starting from different CY$_3$'s. \fref{C3_and_conifold_to_D3} shows an example, in which the so-called $D_3$ geometry \cite{Franco:2015tna} is obtained from $\mathbb{C}^3$ and the conifold. In order to simplify the comparison, we placed the toric diagram of $\mathbb{C}^3$ on the $x-z$ plane and lifted it along the $y$-axis. This example will be studied in detail below. We will also discuss additional sources of freedom in the construction, e.g. the choice of perfect matchings, which reflects the richness of the resulting brane brick models.

\subsection{$3d$ Printing of Periodic Quivers}

In this section we discuss the action of $3d$ printing on the gauge theory. $3d$ printing can be regarded as a procedure for generating brane brick models starting from brane tilings. In practice, it is more convenient to formulate it in terms of the dual periodic quivers, going from a quiver on $\mathbb{T}^2$ for the $4d$ $\mathcal{N}=1$ theory\footnote{More generally, as we mention below, it might be possible to use multiple quivers related by Seiberg duality, and hence associated to the same underlying CY$_3$, as building blocks.} to one on $\mathbb{T}^3$ for the $2d$ $(0,2)$ theory.

\subsubsection{Quiver Blocks}

Let us first recall some basics regarding brane tilings. We refer the reader to \cite{Franco:2005rj} for further details. The periodic quiver dual to a brane tiling is such that plaquettes are in one-to-one correspondence with terms in the superpotential. A perfect matching $p$ is a collection of edges in a brane tiling such that every node is the endpoint of exactly one edge in $p$. Equivalently, we can regard perfect matchings as collections of chiral fields containing exactly one chiral field per plaquette in the periodic quiver, i.e. per term in the superpotential.

The elementary building block for $3d$ printing is a quiver $\mathcal{S}(\mathcal{Q},p)$ on $\mathbb{T}^2 \times I$, with $I$ a line interval along the vertical direction. Here $\mathcal{Q}$ is the periodic quiver on $\mathbb{T}^2$ corresponding to a $4d$ $\mathcal{N} = 1$ gauge theory associated to a toric CY$_3$ and $p$ is a perfect matching of $\mathcal{Q}$. We refer to these objects as {\it quiver blocks}. $\mathcal{S}(\mathcal{Q},p)$ is constructed from $\mathcal{Q}$ and $p$ by a process that closely resembles dimensional reduction and orbifold reduction \cite{Franco:2016fxm}, the details of which are as follows:        
\begin{itemize}
\item On each of the two boundary $\mathbb{T}^2$'s, place a copy of $\mathcal{Q}$ but replace the chiral fields that belong to $p$ by Fermi fields.\footnote{In the explicit examples that follow, identifying the perfect matching using for constructing each quiver block is hence straightforward. It simply corresponds to the Fermi fields on the $\mathbb{T}^2$ boundaries.} The chiral fields that remain unaffected should now be understood as $2d$ chiral fields.

\item Label the two copies of every $4d$ gauge group $i$ in $\mathcal{Q}$ as $\underline{i}$ and $\overline{i}$, depending on which boundary of the quiver block they live on.
            
\item For every $4d$ chiral field $\mathcal{X}_{ij}$ in $\mathcal{Q}$ that is in $p$, add a $2d$ chiral field $X_{\overline{i},\underline{j}}$.
            
\item For every $4d$ chiral field $\mathcal{X}_{ij}$ in $\mathcal{Q}$ that is not in $p$, add a $2d$ Fermi field $\Lambda_{\underline{i},\overline{j}}$.   
            
\item For every $4d$ gauge group $i$ of $\mathcal{Q}$ add a $2d$ chiral field $X_{\underline{i},\overline{i}}$. 
\end{itemize}
For clarity, here and in what follows, we use $\mathcal{X}$ to refer to $4d$ chiral fields and $X$ for $2d$ chiral fields. 

It is useful to note the similarities between this construction and dimensional reduction. The $X_{\overline{i},\underline{j}}$ and $\Lambda_{\underline{i},\overline{j}}$ are analogous to, and in the appropriate cases correspond to, the $2d$ chiral and Fermi fields that a $4d$ chiral field reduces to. Similarly, the $X_{\underline{i},\overline{i}}$ are related to the $2d$ chiral fields in the dimensional reduction of a $4d$ vector multiplet.

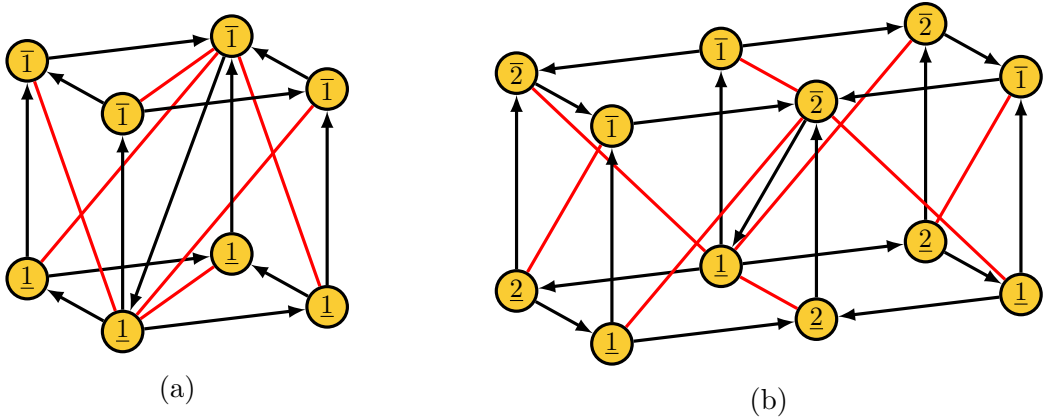
\begin{figure}
             \begin{subfigure}[c]{0.4\textwidth}
             \centering 
                \tdplotsetmaincoords{75}{110}
                \tdplotsetrotatedcoords{0}{0}{135}
                \begin{tikzpicture}[scale=3,tdplot_rotated_coords]
                    \tikzstyle{every node}=[circle,very thick,fill=yellow2,draw,inner sep=2pt,font=\footnotesize]
                    \draw (0,0,0) node(o){$\underline{1}$};
                    \draw (0,1,0) node(y){$\underline{1}$};
                    \draw (1,0,0) node(x){$\underline{1}$};
                    \draw (1,1,0) node(xy){$\underline{1}$};
                    \draw (0,0,1) node(z){$\overline{1}$};
                    \draw (0,1,1) node(yz){$\overline{1}$};
                    \draw (1,0,1) node(xz){$\overline{1}$};
                    \draw (1,1,1) node(xyz){$\overline{1}$};
                    \draw[very thick,-latex] (o) -- (y);
                    \draw[very thick,-latex] (y) -- (xy);
                    \draw[very thick,-latex] (x) -- (xy);
                    \draw[very thick,-latex] (o) -- (x);
                    \draw[very thick,-latex] (z) -- (yz);
                    \draw[very thick,-latex] (yz) -- (xyz);
                    \draw[very thick,-latex] (xz) -- (xyz);
                    \draw[very thick,-latex] (xyz) -- (o);
                    \draw[very thick,red] (o) -- (xz);
                    \draw[very thick,red] (o) -- (xy);
                    \draw[very thick,red] (z) -- (xyz);
                    \draw[very thick,red] (x) -- (xyz);
                    \draw[very thick,red] (y) -- (xyz);
                    \draw[very thick,red] (o) -- (yz);
                    \draw[very thick,-latex] (o) -- (z);
                    \draw[very thick,-latex] (x) -- (xz);
                    \draw[very thick,-latex] (y) -- (yz);
                    \draw[very thick,-latex] (xy) -- (xyz);
                    \draw[very thick,-latex] (z) -- (xz);
                \end{tikzpicture}
                 \caption{} 
             \end{subfigure}
           \begin{subfigure}[c]{0.6\textwidth}
           \centering 
              \tdplotsetmaincoords{75}{110} 
              \tdplotsetrotatedcoords{0}{0}{135} 
              \begin{tikzpicture}[scale=3,tdplot_rotated_coords] 
                \tikzstyle{every node}=[circle,very thick,fill=yellow2,draw,inner sep=2pt,font=\footnotesize]
                \draw (0,0,0) node(a1){$\underline{1}$};
                \draw (0,1,0) node(a2){$\underline{2}$};
                \draw (1,0,0) node(a3){$\underline{2}$};
                \draw (1,1,0) node(a4){$\underline{1}$};
                \draw (2,0,0) node(a5){$\underline{1}$};
                \draw (2,1,0) node(a6){$\underline{2}$};
                \draw (0,0,1) node(a7){$\overline{1}$};
                \draw (0,1,1) node(a8){$\overline{2}$};
                \draw (1,0,1) node(a9){$\overline{2}$};
                \draw (1,1,1) node(a10){$\overline{1}$};
                \draw (2,0,1) node(a11){$\overline{1}$};
                \draw (2,1,1) node(a12){$\overline{2}$};
                \draw[very thick,-latex](a6)--(a12);
                \draw[very thick,-latex](a10)--(a12);
                \draw[very thick,red](a4)--(a12);
                \draw[very thick,-latex](a4)--(a6);
                \draw[very thick,-latex](a4)--(a10);
                \draw[very thick,-latex](a10)--(a8);
                \draw[very thick,red](a4)--(a8);
                \draw[very thick,-latex](a4)--(a2);
                \draw[very thick,-latex](a2)--(a8);
                \draw[very thick,-latex](a12)--(a11);
                \draw[very thick,red](a6)--(a11);
                \draw[very thick,-latex](a6)--(a5);
                \draw[very thick,red](a9)--(a10);
                \draw[very thick,-latex](a9)--(a4);
                \draw[very thick,red](a3)--(a4);
                \draw[very thick,-latex](a8)--(a7);
                \draw[very thick,red](a2)--(a7);
                \draw[very thick,-latex](a2)--(a1);
                \draw[very thick,-latex](a5)--(a11);
                \draw[very thick,-latex](a11)--(a9);
                \draw[very thick,red](a5)--(a9);
                \draw[very thick,-latex](a5)--(a3);
                \draw[very thick,-latex](a3)--(a9);
                \draw[very thick,-latex](a7)--(a9);
                \draw[very thick,red](a1)--(a9);
                \draw[very thick,-latex](a1)--(a3);
                \draw[very thick,-latex](a1)--(a7);
              \end{tikzpicture} 
              \caption{}
            \end{subfigure}
            \caption{Examples of quiver blocks for: a) $\mathbb{C}^3$ and b) the conifold.}
            \label{exls}      
 \end{figure}

\fref{exls} shows examples of quiver blocks for $\mathbb{C}^3$ and the conifold. When representing $2d$ $(0,2)$ quivers, every node correspond to a $U(N_i)$ gauge group, black arrows correspond to chiral fields and red lines correspond to Fermi fields. Fermi lines are unoriented due to the $\Lambda_a \leftrightarrow \overline{\Lambda}_a$ symmetry of $2d$ $(0,2)$ theories.

\subsubsection{Building the Third Dimension of the Periodic Quiver}

We include one quiver block $\mathcal{S}(\mathcal{Q},p)$ for every image of the point $T_{\rm{CY}_3}$ associated to $p$ that we want to generate. In order to give rise to an image along the positive $z$ direction, the quiver block must be oriented as in the examples in \fref{exls}, namely with the $\overline{i}$ nodes at the top and the $\underline{i}$ nodes at the bottom. We refer to such a configuration as a $(+)$ quiver block. Conversely, to generate an image in the negative $z$ direction, we flip the vertical orientation of the quiver block, putting the $\underline{i}$ layer on top of the $\overline{i}$ one. We call this a $(-)$ quiver block.

It is well known that, generically, multiple perfect matching can correspond to the same point in a $T_{\rm{CY}_3}$. Thus, quiver blocks for different perfect matchings can be simultaneously used to generate various images along $z$ of the same point in the original toric diagram.

In what follows, we will restrict to quiver blocks coming from a single periodic quiver $\mathcal{Q}$. It would be interesting to determine if, and if so under what conditions, it is possible to combine quiver blocks associated to different quivers related by Seiberg duality. We leave this question for future investigation.

The quiver blocks are stacked along the $z$ direction and glued along their boundaries. The first and last boundaries are also identified. This process generates the periodic quiver on $\mathbb{T}^3$ associated to the desired CY$_4$ as follows:
        \begin{itemize}
            \item
                Identify overlapping nodes.
            \item
                Identify overlapping pairs of chiral or Fermi fields.
            \item
                Delete overlapping chiral-Fermi pairs, since they correspond to massive pairs of fields.    
       \end{itemize}
            
The order in which the quiver blocks are stacked along the $z$ direction is arbitrary. Generically, each ordering gives rise to a different brane brick model, i.e. a different $2d$ $(0,2)$ gauge theory, associated to the same CY$_4$. All such theories are related by sequences of trialities. An extremely rich combinatorics arises as a result of both the ordering and the freedom in choosing different sets of perfect matchings for lifting the same points in the toric diagram. This freedom was studied in \cite{Franco:2016fxm} in the far more restricted context of orbifold reduction, which uses a single perfect matching and where the only freedom is the relative ordering of the $(+)$ and $(-)$ quiver blocks.       
       
\begin{figure}[ht]
	\centering
	\includegraphics[width=12cm]{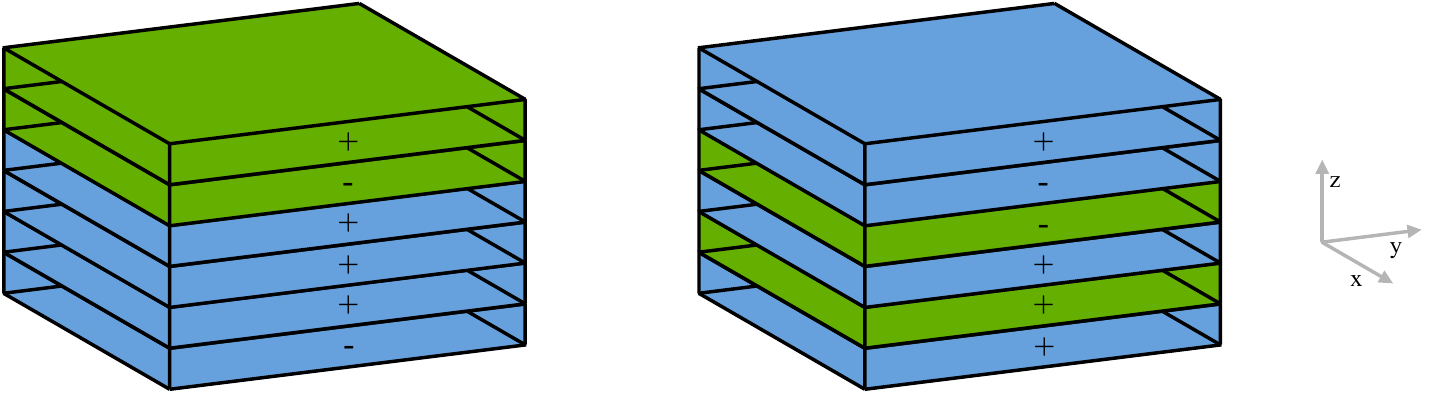}
\caption{Two possible arrangements along the $z$ direction of quiver blocks associated to the a pair of perfect matchings leading to the same CY$_4$.}
	\label{2_stackings}
\end{figure}

\fref{2_stackings} is a schematic illustration of two possible orderings in an example involving two perfect matchings $p$ and $q$. Let us denote $k_{p,\pm}$ and $k_{q,\pm}$ the numbers of quiver blocks with a given sign for each of these perfect matchings. In this case, $k_{p,+}=3$ and $k_{p,-}=1$ and the corresponding quiver blocks are represented by blue boxes. Similarly, $k_{q,+}=1$ and $k_{q,-}=1$ and its quiver blocks are shown in green. If $p$ and $q$ come from different points in $T_{\rm{CY}_3}$, $p$ would generate three new points in the positive $z$ direction and one point in the negative $z$ direction. Similarly, $q$ would give rise to one point above the plane and one point below the plane. If, instead, $p$ and $q$ correspond to the same point in the original toric diagram, this configuration would generate $k_{p,+}+k_{q,+}=4$ points over the original point and $k_{p,-}+k_{q,-}=2$ below it. In \sref{section_3d_printing_and_geometry} and \sref{section_consistency_and_reduction} we will explain that, in order to generate reduced, i.e. consistent, brane brick models we can at most use two different perfect matchings for a single point and the corresponding quiver blocks must have different signs.      
         
The alert reader may notice that $3d$ printing, as just explained, can sometimes give rise to a number of gauge groups that is larger than the expected one, which is equal to the normalized volume of the toric diagram. This is never the case for orbifold reduction \cite{Franco:2016fxm}. We will revisit this issue in \sref{section_consistency_and_reduction}, where we will discuss its relation to reducibility and how to deal with it. Until then, we will consider examples in which this phenomenon does not arise.

    \subsection{Anomaly Cancellation}

In this section we show that theories constructed via $3d$ printing are automatically free of non-abelian gauge anomalies.\footnote{Theories on D1-branes at singularities might have non-vanishing abelian gauge anomalies. We expect they are cancelled via interactions with bulk RR fields, as shown in \cite{Mohri:1997ef} for orbifolds of $\mathbb{C}^4$.}
 
Let us first consider general $2d$ $(0,2)$ quivers. We denote $n_{ij}^\chi $ the number of chiral arrows from node $i$ to node $j$, $n_{ij}^F$ the number of Fermi lines between $i$ and $j$, and $a_i^{\chi/F} $ the number of adjoint chiral/Fermi lines attached to node $i$. The cancellation of $SU(N)_i^2$ anomalies takes the form
\beq
\sum_{j\neq i} \left( n_{ji}^\chi N_j  +  n_{ij}^\chi N_j -  n_{ij}^F N_j \right) + 2 (a^{\chi}_i -  a^F_i)N_i = 2N_i \, .
\label{anomaly_cancellation}
\eeq
Let us focus on the case in which the ranks of all gauge groups are equal, namely $N_i=N$, which arises when a CY$_4$ is probed with a stack of $N$ regular D1-branes. In this case, \eref{anomaly_cancellation} reduces to        
\beq
n^{\chi}_i - n^{F}_i = 2 \, ,
\label{anomaly_cancellation_quiver}
\eeq       
where $n^{\chi}_i$ and $n^{F}_i$ are the total number of incoming plus outgoing chiral and the number of Fermi fields at node $i$, respectively
        
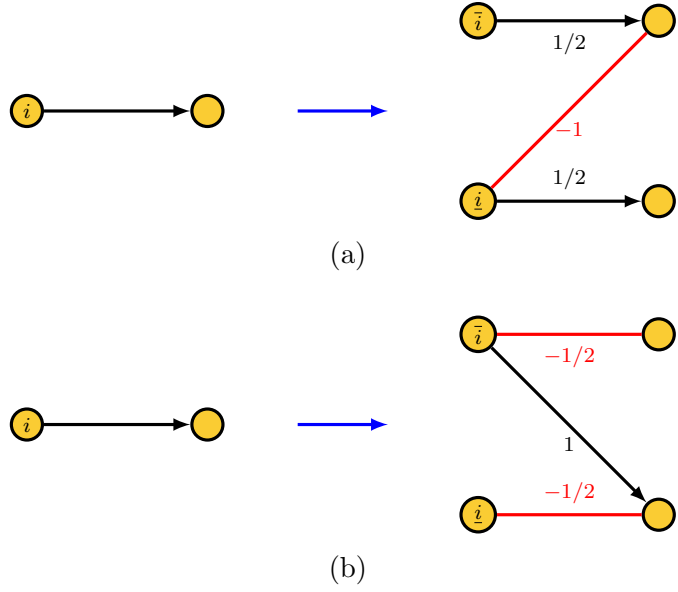
\begin{figure}
             \begin{subfigure}[c]{1\textwidth}
             \centering
                 \begin{tikzpicture}[scale=2.4]
                    \tikzstyle{every node}=[circle,very thick,fill=yellow2,draw,minimum size = 1em , inner sep=2pt,font=\scriptsize]
                    \draw (0,0) node (o){$i$};
                    \draw (1,0) node (n){};
                    \draw [very thick , -latex](o) -- (n);
                    \draw [very thick , blue , -latex] (1.5,0) -- (2,0);
                    \draw (2.5,0.5) node (oa){${\overline{i}}$};
                    \draw (2.5,-0.5) node (ob) {$\underline{i}$};
                    \draw (3.5,0.5) node (na){};
                    \draw (3.5,-0.5) node (nb) {};
                    \tikzstyle{every node}=[font=\scriptsize]
                    \draw [very thick , -latex](oa) -- node[below]{$1/2$} (na);
                    \draw [very thick , -latex](ob) -- node[above]{$1/2$} (nb);
                    \draw [very thick , red] (ob) -- node [below]{$-1$}(na);
                 \end{tikzpicture}
                 \label{inchiral}
                 \caption{}
             \end{subfigure}
         
         \vspace{.5cm}
         
             \begin{subfigure}[c]{1\textwidth}
             \centering
                \begin{tikzpicture}[scale=2.4]
                    \tikzstyle{every node}=[circle,very thick,fill=yellow2,draw,minimum size = 1em , inner sep=2pt,font=\scriptsize]
                    \draw (0,0) node (o){$i$};
                    \draw (1,0) node (n){};
                    \draw [very thick , -latex](o) -- (n);
                    \draw [very thick , blue , -latex] (1.5,0) -- (2,0);
                    \draw (2.5,0.5) node (oa){${\overline{i}}$};
                    \draw (2.5,-0.5) node (ob) {$\underline{i}$};
                    \draw (3.5,0.5) node (na){};
                    \draw (3.5,-0.5) node (nb) {};
                    \tikzstyle{every node}=[font=\scriptsize]
                    \draw [very thick , red](oa) -- node[below]{$-1/2$} (na);
                    \draw [very thick , red](ob) -- node[above]{$-1/2$} (nb);
                    \draw [very thick , -latex] (oa) -- node [below]{$1$}(nb);
                 \end{tikzpicture}
                \label{inFermi}
                \caption{} 
            \end{subfigure}
            \caption{$2d$ descendants of an outgoing $4d$ chiral field at node $i$ that: a) is not in $p$ and b) is in $p$. In both cases the net anomaly weight is $-\frac{1}{2}$ at node $\underline{i}$ and $\frac{1}{2}$ at node $\overline{i}$.}
            \label{anomalyout}
        \end{figure}

\begin{figure}
             \begin{subfigure}[c]{1\textwidth}
             \centering
                \begin{tikzpicture}[scale=2.4]
                    \tikzstyle{every node}=[circle,very thick,fill=yellow2,draw,minimum size = 1em , inner sep=2pt,font=\scriptsize]
                    \draw (0,0) node (o){$i$};
                    \draw (1,0) node (n){};
                    \draw [very thick , latex-](o) -- (n);
                    \draw [very thick , blue , -latex] (1.5,0) -- (2,0);
                    \draw (2.5,0.5) node (oa){${\overline{i}}$};
                    \draw (2.5,-0.5) node (ob) {$\underline{i}$};
                    \draw (3.5,0.5) node (na){};
                    \draw (3.5,-0.5) node (nb) {};
                    \tikzstyle{every node}=[font=\scriptsize]
                    \draw [very thick , latex-](oa) -- node[below]{$1/2$} (na);
                    \draw [very thick , latex-](ob) -- node[above]{$1/2$} (nb);
                    \draw [very thick , red] (oa) -- node [above]{$-1$}(nb);
                 \end{tikzpicture}
                 \label{inchiral}
                 \caption{}
             \end{subfigure}
         
         \vspace{.5cm}
         
             \begin{subfigure}[c]{1\textwidth}
             \centering
                 \begin{tikzpicture}[scale=2.4]
                    \tikzstyle{every node}=[circle,very thick,fill=yellow2,draw,minimum size = 1em , inner sep=2pt,font=\scriptsize]
                    \draw (0,0) node (o){$i$};
                    \draw (1,0) node (n){};
                    \draw [very thick , latex-](o) -- (n);
                    \draw [very thick , blue , -latex] (1.5,0) -- (2,0);
                    \draw (2.5,0.5) node (oa){${\overline{i}}$};
                    \draw (2.5,-0.5) node (ob) {$\underline{i}$};
                    \draw (3.5,0.5) node (na){};
                    \draw (3.5,-0.5) node (nb) {};
                    \tikzstyle{every node}=[font=\scriptsize]
                    \draw [very thick , red](oa) -- node[below]{$-1/2$} (na);
                    \draw [very thick , red](ob) -- node[above]{$-1/2$} (nb);
                    \draw [very thick , -latex] (na) -- node [above]{$1$}(ob);
                 \end{tikzpicture}
                \label{inFermi}
                \caption{} 
            \end{subfigure}
            \caption{$2d$ descendants of an incoming $4d$ chiral field at node $i$ that: a) is not in $p$ and b) is in $p$. In both cases the net anomaly weight is $\frac{1}{2}$ at node $\underline{i}$ and $-\frac{1}{2}$ at node $\overline{i}$.}
            \label{anomalyin}
        \end{figure}
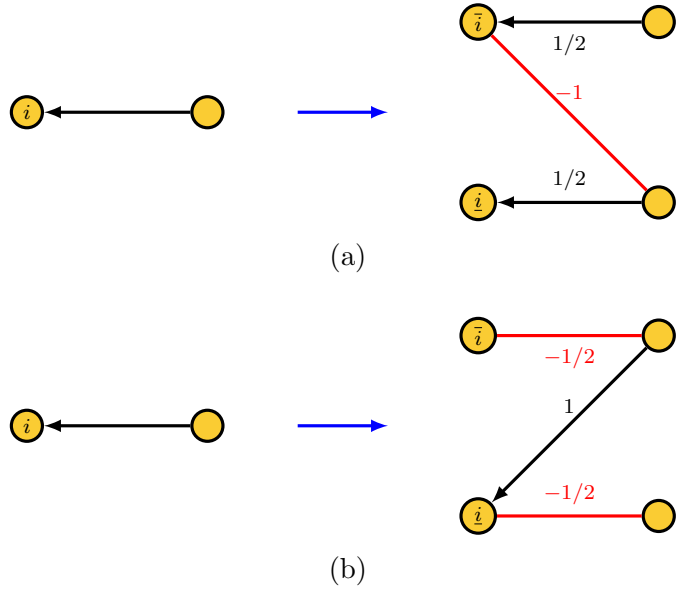

We are now ready to prove the cancellation of non-abelian anomalies in $3d$ printed theories. Consider an arbitrary quiver block $\mathcal{S}(\mathcal{Q},p)$. In order to keep track of anomalies, it is convenient to assign chiral (Fermi) fields that lie on a boundary $\mathbb{T}^2$ a weight $w$ equal to $\frac{1}{2}$ $\left(-\frac{1}{2}\right)$. We introduce semi-integer weights in order to split the contributions of such fields to the anomaly between adjacent quiver blocks. Similarly, we assign weights $1$ ($-1$) to chiral (Fermi) fields in the bulk of the quiver block. \fref{anomalyout} shows that the $2d$ descendants of an outgoing $4d$ chiral field $\mathcal{X}_{ij} \in \mathcal{Q}$ contribute a net weight $-\frac{1}{2}$ at a node $\underline{i}$. Analogously, \fref{anomalyin} shows that the $2d$ descendants of an incoming $4d$ chiral field $\mathcal{X}_{ki}$ contribute a net weight $\frac{1}{2}$. Since anomaly cancellation in $4d$ quivers requires an equal number of incoming and outgoing arrows at every node, the total contribution of all descendants of $4d$ chiral fields vanishes. Therefore, the only non-vanishing contribution to the net weight comes from $2d$ chiral fields $X_{\underline{i},\overline{i}}$. Denoting by $\mathcal{F}_{\underline{i}}$ the set of all $2d$ fields charged under the gauge group $\underline{i}$, we have
        \begin{align}
            \sum_{s\in\mathcal{F}_{\underline{i}}}w_{s} = 1 \, , \label{blockanomaly}       
        \end{align} 
for every node $\underline{i}$. A similar argument shows that the same equation is satisfied by all $\overline{i}$ nodes.

Next, let us consider what happens when gluing two quiver blocks $\mathcal{S}(\mathcal{Q},p)$ and $\mathcal{S}(\mathcal{Q},q)$ along their boundaries. Then, a node $a$ on the boundary of $\mathcal{S}(\mathcal{Q},p)$ becomes identified with a node $b$ on the boundary of $\mathcal{S}(\mathcal{Q},q)$. Summing the contributions of the form \eref{blockanomaly} for $a$ and $b$, we obtain
        \begin{align}
           \sum_{r\in\mathcal{F}_{a}}w_{r}  + \sum_{s\in \mathcal{F}_{b}}w_{s} = 2 \, .
\label{anomaly_weights}
        \end{align}
 This equation is independent of the relative orientation in the $z$ direction of the two quiver blocks, i.e. its valid for $(+,+)$, $(-,-)$ and $(+,-)$ pairs.
 
 The weights in the previous discussion precisely give rise to the contribution to anomalies of all fields. In particular, whenever two copies of $2d$ chiral or Fermi fields overlap on a glued boundary, we obtain the corresponding integer contributions. Similarly, the contribution of an overlapping chiral-Fermi pair is zero, as expected. Equation \eref{anomaly_weights} is thus equivalent to the anomaly cancellation condition \eref{anomaly_cancellation_quiver} for all nodes in the quiver.

\subsection{$J$- and $E$-Terms}
 
 The $J$- and $E$-terms of the $3d$ printed theory are encoded in terms of minimal plaquettes of the periodic quiver. Below we provide a prescription for directly constructing them. It closely follows dimensional reduction, which we reviewed in \sref{section_dimensional_reduction}. 
 
As we explained earlier, for a quiver block $\mathcal{S}(\mathcal{Q},p)$ every $4d$ chiral field $\mathcal{X}_{ij}$ that is not in $p$ gives rise to two $2d$ chiral fields $X_{\underline{i}, \underline{j}}$ and $X_{\overline{i},\overline{j}}$, and one $2d$ Fermi field $\Lambda_{\underline{i},\overline{j}}$. If, instead, $\mathcal{X}_{ij}$ is in $p$, it gives rise to two $2d$ Fermi fields $\Lambda_{\underline{i},\underline{j}}$ and $\Lambda_{\overline{i},\overline{j}}$, and one $2d$ chiral field $X_{\overline{i}\underline{j}}$. 

The $J$-terms for the Fermi fields in $\mathcal{S}(\mathcal{Q},p)$ are given by generalizing \eref{J_dim_red} as follows:
        \begin{itemize}
            \item
                 $J$-term of $\Lambda_{\underline{i},\underline{j}}$:
\bea
J_{\Lambda_{\underline{i},\underline{j}}} = \frac{\partial W}{\partial X_{\underline{i}, \underline{j}}} \, .
\label{J_underi_underi}
\eea

                \item
                $J$-term of $\Lambda_{\overline{i},\overline{j}}$: 
\bea
J_{\Lambda_{\overline{i},\overline{j}}} = \frac{\partial W}{\partial X_{\overline{i},\overline{j}}} \, .
\label{J_ibar_ibar}
\eea                            
                
            \item 
                 For the $J$-terms of $\Lambda_{\underline{i},\overline{j}}$, every monomial in the superpotential $W$ containing the $4d$ chiral field $\mathcal{X}_{ij}$ gives rise to a plaquette that has the general form
                 \begin{align}
                    \Lambda_{\underline{i} \overline{j}} A_{\overline{j}\overline{k}}P_{\overline{k}\underline{l}}B_{\underline{l} \underline{i}} \, ,      
\label{plaquette_J_term}
                 \end{align}    
                 where $A_{\overline{j}\overline{k}}$ and $B_{\underline{l} \underline{i}}$ are monomials in $2d$ chiral fields, each of them living on a different boundary of the quiver block. $P_{kl}$ is a $2d$ chiral field connecting the two boundaries, which descends from the single $4d$ chiral field in this superpotential term that is contained in $p$. The toric condition on $W$ implies that there are two plaquettes of the form \eref{plaquette_J_term}, leading to
\beq
J_{\Lambda_{\underline{i} \overline{j}}} = A_{\overline{j}\overline{k}}P_{\overline{k}\underline{l}}B_{\underline{l} \underline{i}} - \tilde{A}_{\overline{j}\overline{k}}\tilde{P}_{\overline{k}\underline{l}}\tilde{B}_{\underline{l} \underline{i}} \, .      
\eeq
        \end{itemize}
       
        The $E$-terms are similarly given by a generalization of \eref{E_dim_red}. We obtain:
 \begin{itemize}   
\item $E$-term for $\Lambda_{\underline{i},\overline{j}}$: 
\begin{align}
            E_{\Lambda_{\underline{i},\overline{j}}} = s(X_{\underline{i},\overline{i}}X_{\overline{i},\overline{j}} - X_{\underline{i},\underline{j}}X_{\underline{j},\overline{j}}) \, ,
\end{align}
where the sign $s$ is $+1$ for $(+)$ blocks and $-1$ for $(-)$ blocks. 
        
\item For the Fermi fields on the boundaries of the quiver blocks, $\Lambda_{\underline{i},\underline{j}}$ and $\Lambda_{\overline{i},\overline{j}}$, we get a single contribution to the $E$-terms of the form
        \begin{align}
            E_{\Lambda_{\underline{i},\underline{j}}} &=  s\,  X_{\underline{i},\overline{i}}X_{\overline{i},\underline{j}} \\
            E_{\Lambda_{\overline{i},\overline{j}}} &= -s \, X_{\overline{i},\underline{j}}X_{\underline{j},\overline{j}} 
            \label{E_Fermi_boundary}
        \end{align}
The second monomials in these $E$-terms come from the adjacent quiver blocks.
\end{itemize}        
   
The previous contributions to $J$- and $E$-terms are identified or combined when two quiver blocks $\mathcal{S}(\mathcal{Q},p)$ and $\mathcal{S}(\mathcal{Q},q)$ are glued. If a $4d$ chiral field is in precisely one of $p$ and $q$, then a Fermi and a chiral descendants of the same field overlap on the boundary. This results in a new minimal plaquette and we assign to it a sign that is opposite to the existing one. This new plaquette makes this chiral-Fermi pair massive.  
        
It is straight forward but lengthy, to show that with these prescriptions for $J$- and $E$-terms the condition
        \begin{align}
            \sum_{a} J_{a}E_{a}= 0 \, ,
            \label{vatr}
        \end{align}
where $a$ runs over all Fermis, is satisfied. This condition is required by supersymmetry.        

To illustrate the previous discussion, we now present the contributions to $J$- and $E$-terms for the quiver block examples in \fref{exls}. For the $\mathbbm{C}^{3}$ quiver block, we get
        \begin{align}
            \renewcommand{\arraystretch}{1.2}
            \begin{array}{rccccrcl} & & J & &\phantom{abcde}& & E & \\\Lambda_{\underline{1},\underline{1}}: & X_{\underline{1},\underline{1}}Y_{\underline{1},\underline{1}} & - & Y_{\underline{1},\underline{1}}X_{\underline{1},\underline{1}}  &&  Z_{\underline{1},\overline{1}}D_{\overline{1},\underline{1}} && \\\Lambda_{\underline{1},\overline{1}}^{(1)}: & Y_{\overline{1},\overline{1}}D_{\overline{1},\underline{1}} & - & D_{\overline{1},\underline{1}}Y_{\underline{1},\underline{1}}  &&  Z_{\underline{1},\overline{1}}X_{\overline{1},\overline{1}} & - & X_{\underline{1},\underline{1}}Z_{\underline{1},\overline{1}}\\\Lambda_{\underline{1},\overline{1}}^{(2)}: & D_{\overline{1},\underline{1}}X_{\underline{1},\underline{1}} & - & X_{\overline{1},\overline{1}}D_{\overline{1},\underline{1}}  &&  Z_{\underline{1},\overline{1}}Y_{\overline{1},\overline{1}} & - & Y_{\underline{1},\underline{1}}Z_{\underline{1},\overline{1}}\\\Lambda_{\overline{1},\overline{1}}: & X_{\overline{1},\overline{1}}Y_{\overline{1},\overline{1}} & - & Y_{\overline{1},\overline{1}}X_{\overline{1},\overline{1}}  &&   & - & D_{\overline{1},\underline{1}}Z_{\underline{1},\overline{1}}\end{array}
        \end{align}
 Similarly, for the conifold quiver block, we have
        \begin{align}
            \renewcommand{\arraystretch}{1.2}
            \begin{array}{rccccrcl} & & J & &\phantom{abcde}& & E & \\\Lambda_{\underline{1},\overline{2}}^{(1)}: & X_{\overline{2},\overline{1}}Y_{\overline{1},\overline{2}}Y_{\overline{2},\underline{1}} & - & Y_{\overline{2},\underline{1}}Y_{\underline{1},\underline{2}}X_{\underline{2},\underline{1}}  &&  Z_{\underline{1},\overline{1}}X_{\overline{1},\overline{2}} & - & X_{\underline{1},\underline{2}}Z_{\underline{2},\overline{2}}\\\Lambda_{\underline{1},\overline{2}}^{(2)}: & Y_{\overline{2},\underline{1}}X_{\underline{1},\underline{2}}X_{\underline{2},\underline{1}} & - & X_{\overline{2},\overline{1}}X_{\overline{1},\overline{2}}Y_{\overline{2},\underline{1}}  &&  Z_{\underline{1},\overline{1}}Y_{\overline{1},\overline{2}} & - & Y_{\underline{1},\underline{2}}Z_{\underline{2},\overline{2}}\\\Lambda_{\underline{2},\underline{1}}: & X_{\underline{1},\underline{2}}X_{\underline{2},\underline{1}}Y_{\underline{1},\underline{2}} & - & Y_{\underline{1},\underline{2}}X_{\underline{2},\underline{1}}X_{\underline{1},\underline{2}}  &&  Z_{\underline{2},\overline{2}}Y_{\overline{2},\underline{1}} && \\\Lambda_{\underline{2},\overline{1}}: & Y_{\overline{1},\overline{2}}Y_{\overline{2},\underline{1}}X_{\underline{1},\underline{2}} & - & X_{\overline{1},\overline{2}}Y_{\overline{2},\underline{1}}Y_{\underline{1},\underline{2}}  &&  Z_{\underline{2},\overline{2}}X_{\overline{2},\overline{1}} & - & X_{\underline{2},\underline{1}}Z_{\underline{1},\overline{1}}\\\Lambda_{\overline{2},\overline{1}}: & X_{\overline{1},\overline{2}}X_{\overline{2},\overline{1}}Y_{\overline{1},\overline{2}} & - & Y_{\overline{1},\overline{2}}X_{\overline{2},\overline{1}}X_{\overline{1},\overline{2}}  &&   & - & Y_{\overline{2},\underline{1}}Z_{\underline{1},\overline{1}}\end{array}
        \end{align}

\subsection{Relation to Orbifold Reduction}
    
\label{section_connection_to_orbred}

\begin{figure}[ht]
	\centering
	\includegraphics[width=\textwidth]{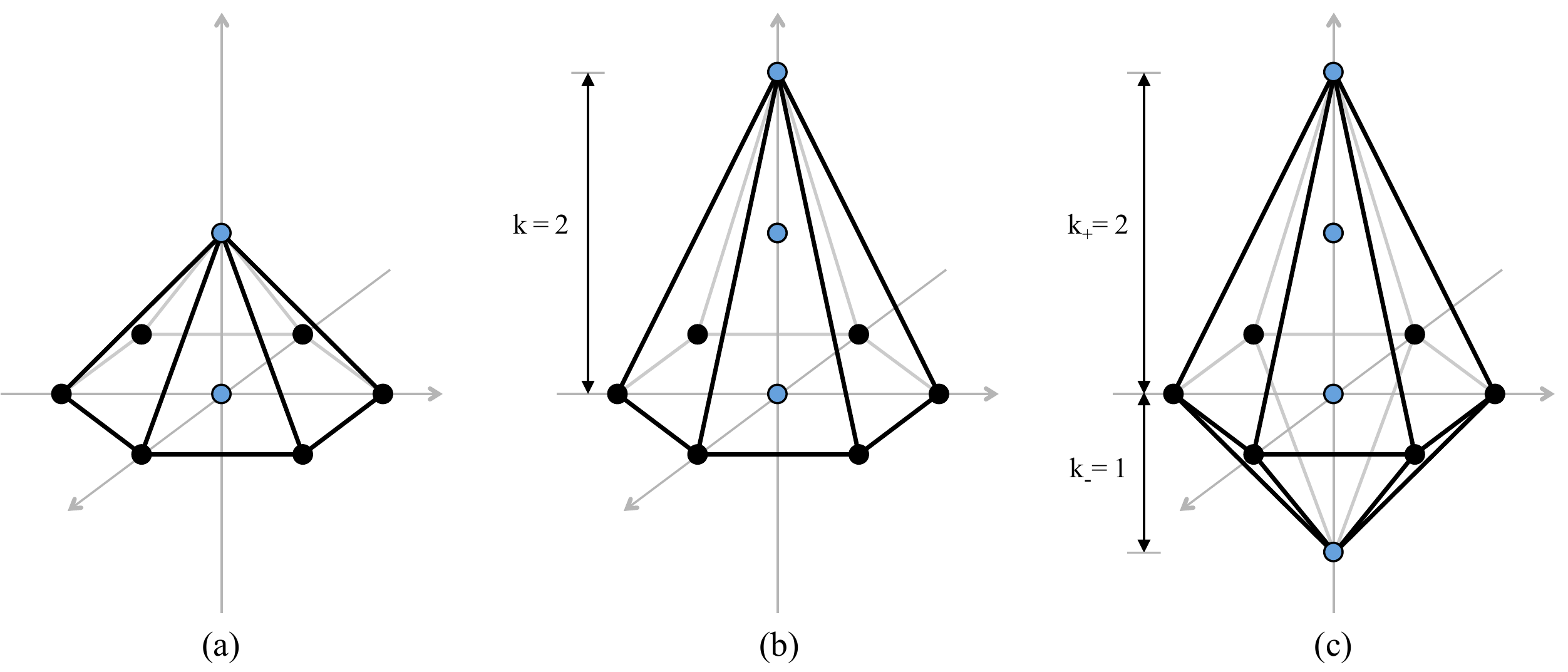}
\caption{Toric diagrams for: a) the dimensional reduction of $dP_3$ to $dP_3\times \mathbb{C}$, b) a $(dP_3\times \mathbb{C})/\mathbb{Z}_k$ orbifold with $k=2$ and c) an orbifold reduction of $dP_3$ with $k_+=2$ and $k_-=1$.}
	\label{red_orb_orbred}
\end{figure}

$3d$ printing considerably supersedes orbifold reduction, which was introduced in \cite{Franco:2016fxm} as a generalization of both dimensional reduction and orbifolding. \fref{red_orb_orbred} illustrates the effect of these three operations on a toric diagram.

Below we briefly explain how all these operations are special cases within the framework of $3d$ printing.
\begin{itemize}
\item {\bf Dimensional reduction.} Using a single quiver block $\mathcal{S}(\mathcal{Q},p)$ we generate a periodic quiver associated to the $2d$ $(2,2)$ gauge theory that corresponds to $\rm{CY}_4=\rm{CY}_3\times \mathbb{C}$. The result is independent of the perfect matching $p$ used for the lift. 
\item {\bf Orbifolding.} Using $k$ copies of the same quiver block $\mathcal{S}(\mathcal{Q},p)$, we generate the periodic quiver for an orbifold of the form $\rm{CY}_4=(\rm{CY}_3\times \mathbb{C})/\mathbb{Z}_k$.
\item {\bf Orbifold reduction.} This process uses $k_+$ copies of a quiver block $\mathcal{S}(\mathcal{Q},p)$ with $(+)$ orientation and $k_-$ copies of it with $(-)$ orientation.
\end{itemize}
It is important to emphasize that both orbifolding and orbifold reduction employ multiple copies of a single quiver block $\mathcal{S}(\mathcal{Q},p)$. $3d$ printing is far more general. In particular, it can lead to the same lifts of a point in $T_{\rm{CY}_3}$, but using multiple perfect matchings associated to it.\footnote{In general, brane brick model obtained by lifting multiple perfect matchings for the same point need to be reduced at a final stage. This can be done systematically as explained in \sref{section_consistency_and_reduction}.} In addition, $3d$ printing makes it possible to lift several points in $T_{\rm{CY}_3}$.

\section{$3d$ Printing and Geometry}

\label{section_3d_printing_and_geometry}

In this section we explain how $3d$ printing gives rise to the desired toric diagram. We do so in terms of {\it brick matchings}, which are the generalizations of perfect matchings to brane brick models. We refer the reader to \cite{Franco:2015tya} for a detailed presentation of these objects and their properties.

In order to provide a combinatorial definition of brick matchings, it is useful to complete $J_a$- and $E_a$-terms into pairs of gauge invariant plaquettes by including the corresponding Fermi fields, i.e. by appropriately multiplying them by either $\Lambda_a$ or $\bar{\Lambda}_a$. Brick matchings are defined such that for every Fermi field $\Lambda_a$, the chiral fields in the brick matching cover {\it either} each of the two $J_a$-term plaquettes {\it or} each of the two $E_a$-term plaquettes exactly once. Brick matchings also contain Fermi fields, but they are not important for our discussion.

Brick matchings map to points in the CY$_4$ toric diagram (as usual, this map can be many to one), turning the determination of the geometry associated to a gauge theory into a combinatorial problem. The position of a brick matching in $T_{\rm{CY}_4}$ is given by the net intersection number between the corresponding faces in the brane brick model and the edges of the unit cell, counted with orientation \cite{Franco:2015tya}. Equivalently, it is given by the chiral arrows in the dual periodic quiver that sit along the edges of the unit cell, again considered with orientations.

We will now discuss how the toric diagram is lifted via $3d$ printing. More concretely, we will explain how certain brick matchings originate from the perfect matchings used for the lift. For simplicity, let us assume that a single point in $T_{\rm{CY}_3}$ is lifted; extending the discussion to the general case in which multiple points are lifted is straightforward. For the arguments in this section, it is convenient to label gauge groups in the $2d$ gauge theory using a pair of indices $(i,m)$, where $i$ runs over nodes in the $4d$ quiver and $m$ runs over quiver blocks.
    
Let us first consider the case in which a point is lifted using a single perfect matching $p$. From $p$, we can construct a brick matching $q_0$, given by 
    \begin{align}
         q_0=  \cup_{i=1}^{n} q^{(m)}\, ,
    \end{align} 
where $q^{(m)}$ denotes the chiral field content within the quiver block $m$. For a $(+)$ block it is given by
\beq
       q^{(m)} = \{X_{(i,m+1)(j,m)}|\mathcal{X}_{ij} \in p\} \, ,
       \label{qm+}
\eeq
    while for a $(-)$ block it is given by 
\beq
        q^{(m)} = \{X_{(i,m)(j,m+1)}|\mathcal{X}_{ij}\in p\} \, .
        \label{qm-}
\eeq
It is straightforward to verify that $q_0$ is indeed a brick matching. It covers every $E$-term of Fermi fields descending from $4d $ chiral fields in $p$ and every $J$-term from the other Fermi fields. Notice that this type of brick matching does not contain any vertical chiral $X_{(i,m)(i,m\pm 1)}$, where the signs are those of the quiver blocks. According to the prescription discussed above for determining the positions of brick matchings in the toric diagram, this fact implies that $q_0$ is not lifted, i.e. it remains on the original $T_{\rm{CY}_3}$.
    
From the previous discussion, it becomes clear that in order to construct brick matchings for the lifted points it is necessary to include vertical chiral fields. Let us denote $B_{(+)}$ and $B_{(-)}$ the sets of $(+)$ and $(-)$ blocks for the perfect matching $p$, respectively. The brick matching that descend from $p$ and have maximum and minimum vertical coordinates in $T_{\rm{CY}_4}$ are given by 
   \beq
   \begin{array}{ccccc}
    q_{max} & = & (\cup_{m \in B_{(+)} }\{X_{(i,m)(i,m+1)}\}) & \cup & (\cup_{n \in B_{(-)} } q^{(n)}) \\[.15cm]
    q_{min} & = & (\cup_{n \in B_{(-)} } \{X_{(i,m+1)(i,m)}\}) & \cup & (\cup_{n \in B_{(+)}} q^{(n)}) 
    \end{array}
    \label{qmax_qmin}
    \eeq
with the $q^{(n)}$ defined as in \eref{qm+} and \eref{qm-}. Similarly, generalizing \eref{qmax_qmin} to include vertical chiral fields in $n_+$ of the $(+)$ blocks and $n_-$ of the $(-)$ blocks, we obtain brick matchings with $n_+-n_-$ vertical displacement. In fact $q_0$ is the $n_+=n_-=0$ case of this construction. These arguments extend to the case in which multiple points in the original toric diagram are lifted. 
    
In general, the brick matchings we have just described are not all the brick matchings of the resulting theory, but they contain all the ones corresponding to the corners of $T_{\rm{CY}_4}$.

 \paragraph{Inconsistent $3d$ Printing.}

So far, we have restricted to a single perfect matching per lifted point. Let us consider what happens if we use multiple perfect matchings associated to the same point of the original toric diagram. If two or more perfect matchings from the same point are used on quiver blocks {\it with the same sign}, it is easy to check that naively applying \eref{qmax_qmin} we do not obtain brick matchings. This is a first indication of a pathology. In this case the points are not actually lifted by the expected amount, we end up with multiple brick matchings on some corners of $T_{\rm{CY}_4}$ and the resulting brane brick model is reducible. Reducibility or, equivalently, inconsistency of brane brick models and how to fix it will be the subject of \sref{section_consistency_and_reduction}. There is no problem, however, with using two different perfect matchings corresponding to the same point if the corresponding quiver blocks have different signs. Of course, we can use more perfect matchings and reduce the brane brick model at the final stage.

\section{Examples}

\label{section_examples}

In this section we present various explicit examples of $3d$ printing. The primary objective is by no means to expand the catalogue of known examples, since plenty of them are already available in the literature (see e.g. \cite{Franco:2015tna,Franco:2015tya,Franco:2016nwv,Franco:2016qxh,Franco:2016fxm,Franco:2017cjj}). Our goal is to present examples that illustrate special features of $3d$ printing and the new types of geometries it can handle. We focus on models constructed using $\mathbb{C}^3$ and conifold quiver blocks. In all cases, after determining the gauge theory we verify that it indeed corresponds to the desired CY$_4$ using fast forward algorithm introduced in \cite{Franco:2015tya}.

\subsection{Models from $\mathbbm{C}^{3}$ Quiver Blocks}
    
Let us first consider models constructed from $\mathbb{C}^3$. The corresponding $4d$ gauge theory is $\mathcal{N}=4$ super Yang-Mills (SYM) and the periodic quiver and brane tiling for it were first introduced in \cite{Hanany:2005ve,Franco:2005rj}. We will focus on examples in which more than one perfect matching are used for the lift, i.e. models that are explicitly beyond the scope of orbifold reduction.

\subsubsection{Two Perfect Matchings}

\begin{figure}[ht]
	\centering
	\includegraphics[width=12cm]{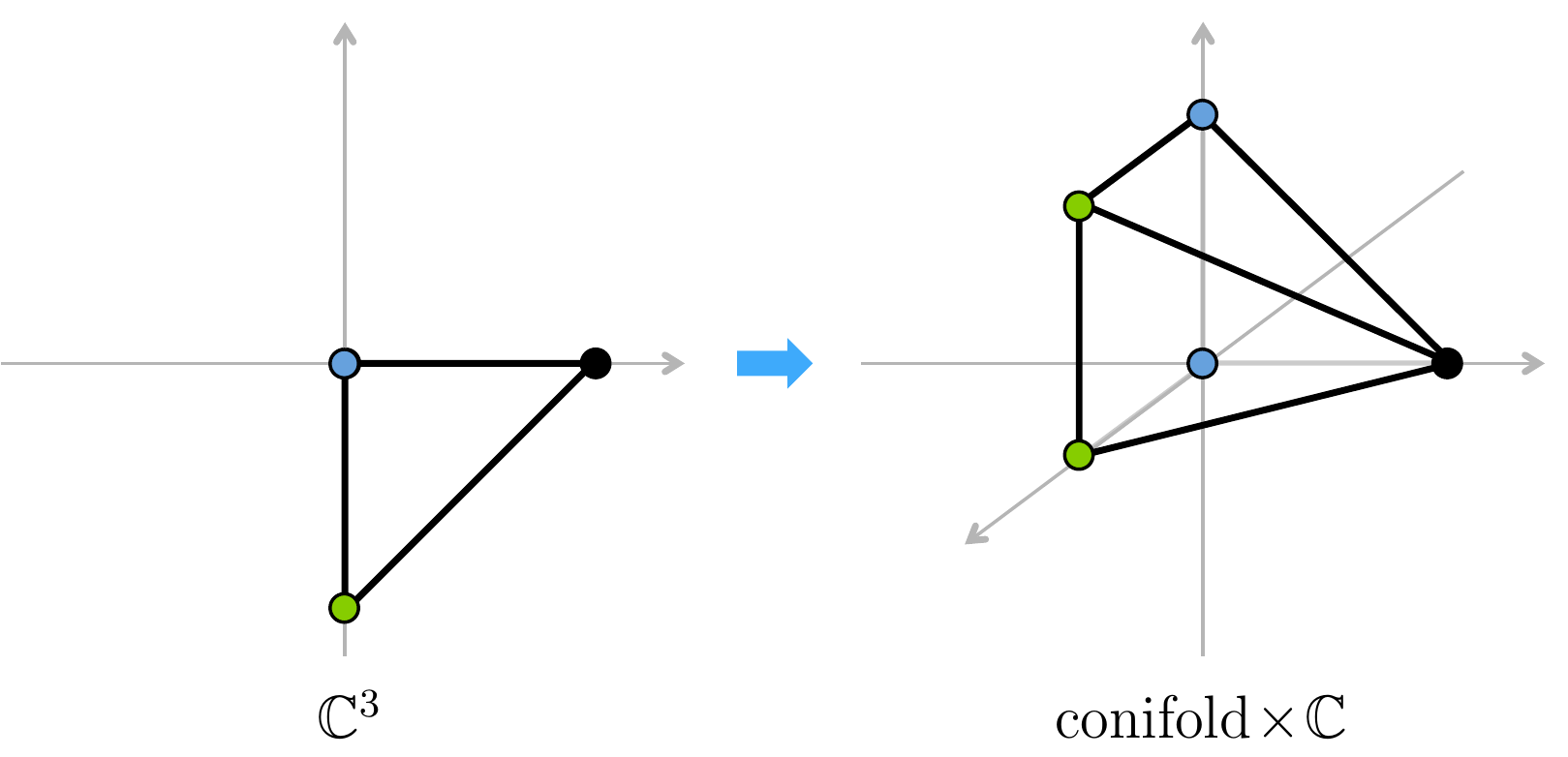}
\caption{The toric diagram for $\rm{conifold}\times \mathbb{C}$, obtained by lifting two perfect matchings of $\mathbb{C}^3$.}
	\label{C3_to_conifoldxC}
\end{figure}

Let us start from $\mathbb{C}^3$ and lift two corners of the toric diagram in the same direction, as shown in \fref{C3_to_conifoldxC}. We obtain the toric diagram for the $\rm{conifold}\times \mathbb{C}$.

\begin{figure}
                 \begin{center}
                    \tdplotsetmaincoords{80}{120} 
                      \tdplotsetrotatedcoords{0}{0}{145} 
                      \begin{tikzpicture}[scale=2.5,tdplot_rotated_coords] 
                        \tikzstyle{every node}=[circle,very thick,fill=yellow2,draw,inner sep=2pt,font=\scriptsize]
                       \draw (0.0,0.0,0.0) node(a1){$\underline{1}$};
                        \draw (0.0,1.0,0.0) node(a2){$\underline{1}$};
                        \draw (1.0,0.0,0.0) node(a3){$\underline{1}$};
                        \draw (1.0,1.0,0.0) node(a4){$\underline{1}$};
                        \draw (0.0,0.0,1.0) node(a5){$\overline{1}$};
                        \draw (0.0,1.0,1.0) node(a6){$\overline{1}$};
                        \draw (1.0,0.0,1.0) node(a7){$\overline{1}$};
                        \draw (1.0,1.0,1.0) node(a8){$\overline{1}$};
                        \draw (0.0,0.0,1.5) node(a9){$\underline{2}$};
                        \draw (0.0,1.0,1.5) node(a10){$\underline{2}$};
                        \draw (1.0,0.0,1.5) node(a11){$\underline{2}$};
                        \draw (1.0,1.0,1.5) node(a12){$\underline{2}$};
                        \draw (0.0,0.0,2.5) node(a13){$\overline{2}$};
                        \draw (0.0,1.0,2.5) node(a14){$\overline{2}$};
                        \draw (1.0,0.0,2.5) node(a15){$\overline{2}$};
                        \draw (1.0,1.0,2.5) node(a16){$\overline{2}$};
                        \draw (2.5,0.0,0.25) node(a17){$1$};
                        \draw (2.5,0.0,2.25) node(a18){$1$};
                        \draw (2.5,1.0,0.25) node(a19){$1$};
                        \draw (2.5,1.0,2.25) node(a20){$1$};
                        \draw (3.5,0.0,0.25) node(a21){$1$};
                        \draw (3.5,0.0,2.25) node(a22){$1$};
                        \draw (3.5,1.0,0.25) node(a23){$1$};
                        \draw (3.5,1.0,2.25) node(a24){$1$};
                        \draw (2.5,0.0,1.25) node(a25){$2$};
                        \draw (2.5,1.0,1.25) node(a26){$2$};
                        \draw (3.5,0.0,1.25) node(a27){$2$};
                        \draw (3.5,1.0,1.25) node(a28){$2$};
                        \draw (5.0,0.0,0.25) node(a29){$1$};
                        \draw (5.0,0.0,2.25) node(a30){$1$};
                        \draw (5.0,1.0,0.25) node(a31){$1$};
                        \draw (5.0,1.0,2.25) node(a32){$1$};
                        \draw (6.0,0.0,0.25) node(a33){$1$};
                        \draw (6.0,0.0,2.25) node(a34){$1$};
                        \draw (6.0,1.0,0.25) node(a35){$1$};
                        \draw (6.0,1.0,2.25) node(a36){$1$};
                        \draw (5.0,0.0,1.25) node(a37){$2$};
                        \draw (5.0,1.0,1.25) node(a38){$2$};
                        \draw (6.0,0.0,1.25) node(a39){$2$};
                        \draw (6.0,1.0,1.25) node(a40){$2$};
                        \draw[very thick,-latex](a40)--(a36);
                        \draw[very thick,-latex](a35)--(a40);
                        \draw[very thick,red](a38)--(a36);
                        \draw[very thick,-latex](a38)--(a35);
                        \draw[very thick,-latex](a38)--(a32);
                        \draw[very thick,-latex](a31)--(a38);
                        \draw[very thick,-latex](a28)--(a24);
                        \draw[very thick,-latex](a23)--(a28);
                        \draw[very thick,-latex,bend right = 10](a20) to (a24);
                        \draw[very thick,red](a26)--(a24);
                        \draw[very thick,-latex, bend left = 10](a26) to (a28);
                        \draw[very thick,bend right = 10,red](a26) to (a28);
                        \draw[very thick,-latex](a26)--(a23);
                        \draw[very thick,red , bend left = 10](a19) to (a23);
                        \draw[very thick,-latex](a26)--(a20);
                        \draw[very thick,-latex](a19)--(a26);
                        \draw[very thick,-latex](a12)--(a16);
                        \draw[very thick,-latex](a4)--(a8);
                        \draw[very thick,-latex](a14)--(a16);
                        \draw[very thick,red](a10)--(a16);
                        \draw[very thick,-latex](a10)--(a12);
                        \draw[very thick,red](a6)--(a8);
                        \draw[very thick,-latex](a6)--(a4);
                        \draw[very thick,red](a2)--(a4);
                        \draw[very thick,-latex](a10)--(a14);
                        \draw[very thick,-latex](a2)--(a6);
                        \draw[very thick,-latex](a36)--(a34);
                        \draw[very thick,red](a40)--(a34);
                        \draw[very thick,-latex](a40)--(a39);
                        \draw[very thick,red](a35)--(a39);
                        \draw[very thick,-latex](a35)--(a33);
                        \draw[very thick,-latex](a34)--(a38);
                        \draw[very thick,red](a33)--(a38);
                        \draw[very thick,-latex](a32)--(a30);
                        \draw[very thick,red](a38)--(a30);
                        \draw[very thick,-latex](a38)--(a37);
                        \draw[very thick,red](a31)--(a37);
                        \draw[very thick,-latex](a31)--(a29);
                        \draw[very thick,-latex](a24)--(a22);
                        \draw[very thick,red](a28)--(a22);
                        \draw[very thick,-latex](a28)--(a27);
                        \draw[very thick,red](a23)--(a27);
                        \draw[very thick,-latex](a23)--(a21);
                        \draw[very thick,red,bend left = 10](a22) to (a20);
                        \draw[very thick,-latex,bend right = 8](a22)--(a26);
                        \draw[very thick,-latex, bend left = 8](a27) to (a26);
                        \draw[very thick,red , bend right = 8](a27) to (a26);
                        \draw[very thick,red](a21)--(a26);
                        \draw[very thick,-latex , bend right = 10](a21) to (a19);
                        \draw[very thick,-latex](a20)--(a18);
                        \draw[very thick,red](a26)--(a18);
                        \draw[very thick,-latex](a26)--(a25);
                        \draw[very thick,red](a19)--(a25);
                        \draw[very thick,-latex](a19)--(a17);
                        \draw[very thick,-latex](a16)--(a15);
                        \draw[very thick,red](a12)--(a15);
                        \draw[very thick,-latex](a12)--(a11);
                        \draw[very thick,-latex](a8)--(a7);
                        \draw[very thick,red](a4)--(a7);
                        \draw[very thick,-latex](a4)--(a3);
                        \draw[very thick,red](a15)--(a14);
                        \draw[very thick,-latex](a15)--(a10);
                        \draw[very thick,red](a11)--(a10);
                        \draw[very thick,-latex](a7)--(a6);
                        \draw[very thick,red](a3)--(a6);
                        \draw[very thick,-latex](a3)--(a2);
                        \draw[very thick,-latex](a14)--(a13);
                        \draw[very thick,red](a10)--(a13);
                        \draw[very thick,-latex](a10)--(a9);
                        \draw[very thick,-latex](a6)--(a5);
                        \draw[very thick,red](a2)--(a5);
                        \draw[very thick,-latex](a2)--(a1);
                        \draw[very thick,-latex](a39)--(a34);
                        \draw[very thick,-latex](a33)--(a39);
                        \draw[very thick,red](a37)--(a34);
                        \draw[very thick,-latex](a37)--(a33);
                        \draw[very thick,-latex](a37)--(a30);
                        \draw[very thick,-latex](a29)--(a37);
                        \draw[very thick,-latex](a27)--(a22);
                        \draw[very thick,-latex](a21)--(a27);
                        \draw[very thick,-latex,bend right = 10](a18) to (a22);
                        \draw[very thick,red](a25)--(a22);
                        \draw[very thick,-latex,bend left = 10](a25) to (a27);
                        \draw[very thick,bend right = 10, red](a25) to (a27);
                        \draw[very thick,-latex](a25)--(a21);
                        \draw[very thick,bend left = 10,red](a17) to (a21);
                        \draw[very thick,-latex](a25)--(a18);
                        \draw[very thick,-latex](a17)--(a25);
                        \draw[very thick,-latex](a11)--(a15);
                        \draw[very thick,-latex](a3)--(a7);
                        \draw[very thick,-latex](a13)--(a15);
                        \draw[very thick,red](a9)--(a15);
                        \draw[very thick,-latex](a9)--(a11);
                        \draw[very thick,red](a5)--(a7);
                        \draw[very thick,-latex](a5)--(a3);
                        \draw[very thick,red](a1)--(a3);
                        \draw[very thick,-latex](a9)--(a13);
                        \draw[very thick,-latex](a1)--(a5);
                        \draw[very thick, blue,-latex] (1.5 , 0.5 , 1.2) -- (2 , 0.5 , 1.2);
                        \draw[very thick, blue,-latex] (4 , 0.5 , 1.2) -- (4.5 , 0.5 , 1.2);
                      \end{tikzpicture} 
                     \caption{$3d$ printing of the periodic quiver for the $\rm{conifold}\times \mathbb{C}$ using two $\mathbb{C}^3$ quiver blocks corresponding to different perfect matchings. Four massive chiral-Fermi pairs are integrated out in the last step.}
                     \label{C3_to_conifoldxC_quiver}
                 \end{center}
             \end{figure}
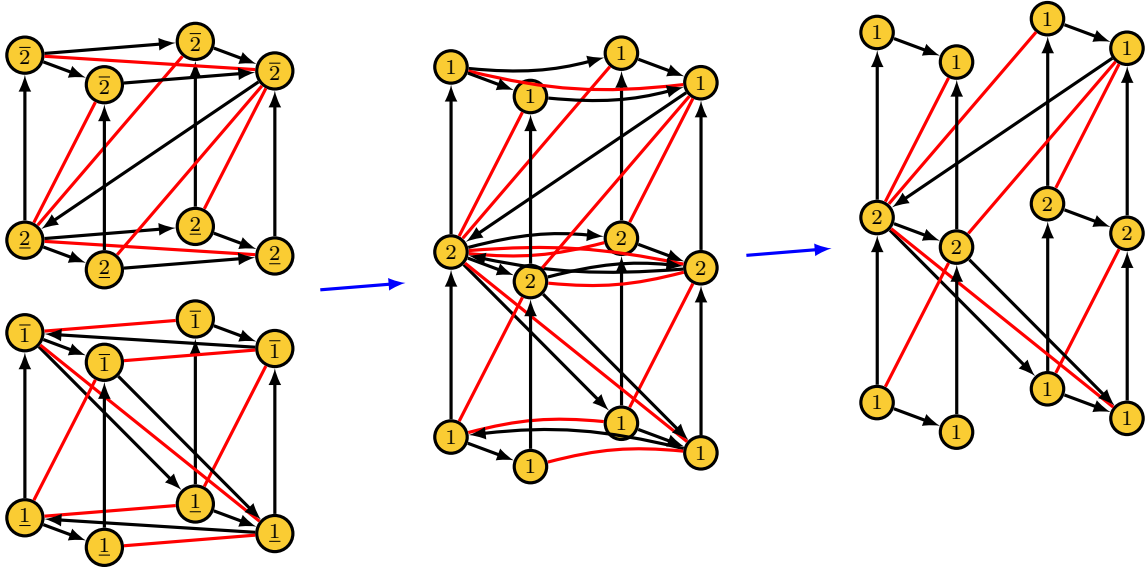

 \fref{C3_to_conifoldxC_quiver} shows the two quiver blocks and how they are glued. The perfect matching used for each of the quiver blocks can be identified from the Fermis on the boundaries. We glue the quiver blocks identifying $\underline{1}$ and $\overline{2}$ (we call the resulting node $1$) and $\overline{1}$ and $\underline{2}$ (which we call $2$).
 
 The $J$- and $E$-terms of the theory can be determined using the machinery introduced in the previous section. They are:
            \begin{align}
                 \renewcommand{\arraystretch}{1.2}
                 \begin{array}{rrclcrcl} & & J & &\phantom{abcde}& & E & \\\Lambda_{11}^{(1)}: & Y_{11}D_{11} & - & D_{11}Y_{11}  &&  Z_{12}X_{21} & - & X_{11}\\\Lambda_{11}^{(2)}: & X_{11}Y_{11} & - & Y_{11}X_{11}  &&  D_{11} & - & D_{12}Z_{21}\\\Lambda_{12}^{(1)}: & X_{21}Y_{11} & - & Y_{22}X_{21}  &&  Z_{12}D_{22} & - & D_{11}Z_{12}\\\Lambda_{12}^{(2)}: & D_{22}X_{21} & - & X_{21}D_{11}  &&  Z_{12}Y_{22} & - & Y_{11}Z_{12}\\\Lambda_{21}^{(1)}: & Y_{11}D_{12} & - & D_{12}Y_{22}  &&  Z_{21}X_{11} & - & X_{22}Z_{21}\\\Lambda_{21}^{(2)}: & D_{12}X_{22} & - & X_{11}D_{12}  &&  Z_{21}Y_{11} & - & Y_{22}Z_{21}\\\Lambda_{22}^{(1)}: & Y_{22}D_{22} & - & D_{22}Y_{22}  &&  X_{22} & - & X_{21}Z_{12}\\\Lambda_{22}^{(2)}: & X_{22}Y_{22} & - & Y_{22}X_{22}  &&  Z_{21}D_{12} & - & D_{22}\end{array}
\label{J_E_C3_to_conifoldxC_1}                
             \end{align} 

From the periodic quiver at the center of \fref{C3_to_conifoldxC_quiver} and, equivalently, the linear $E$-terms in \eref{J_E_C3_to_conifoldxC_1}, we see that this theory has four chiral-Fermi massive pairs. Integrating them out, we obtain the periodic quiver on the right of \fref{C3_to_conifoldxC_quiver}, and the following $J$- and $E$-terms:
             \begin{align}
                \renewcommand{\arraystretch}{1.2}
                 \begin{array}{rrclcrcl} & & J & &\phantom{abcde}& & E & \\\Lambda_{12}^{(1)}: & X_{21}Y_{11} & - & Y_{22}X_{21}  &&  Z_{12}Z_{21}D_{12} & - & D_{12}Z_{21}Z_{12}\\\Lambda_{12}^{(2)}: & Z_{21}D_{12}X_{21} & - & X_{21}D_{12}Z_{21}  &&  Z_{12}Y_{22} & - & Y_{11}Z_{12}\\\Lambda_{21}^{(1)}: & Y_{11}D_{12} & - & D_{12}Y_{22}  &&  Z_{21}Z_{12}X_{21} & - & X_{21}Z_{12}Z_{21}\\\Lambda_{21}^{(2)}: & D_{12}X_{21}Z_{12} & - & Z_{12}X_{21}D_{12}  &&  Z_{21}Y_{11} & - & Y_{22}Z_{21}\end{array}
             \end{align}

The theory we obtained is, as expected, the dimensional reduction of the conifold gauge theory. This is a simple example of the kind of situation illustrated in \fref{C3_and_conifold_to_D3}, in which a given $2d$ theory, or more generally  $2d$ theories for the same CY$_4$, can be reached in multiple ways. It is quite remarkable that the same gauge theory can be generated using different methods and starting from two substantially different $4d$ parent theories: by dimensional reduction of the conifold gauge theory (a minimally SUSY, chiral theory) or by $3d$ printing from $\mathcal{N}=4$ SYM (a maximally SUSY, non-chiral theory).

\subsubsection{Three Perfect Matchings}

\label{section_D3_from_C^4}

\begin{figure}[ht]
	\centering
	\includegraphics[width=12cm]{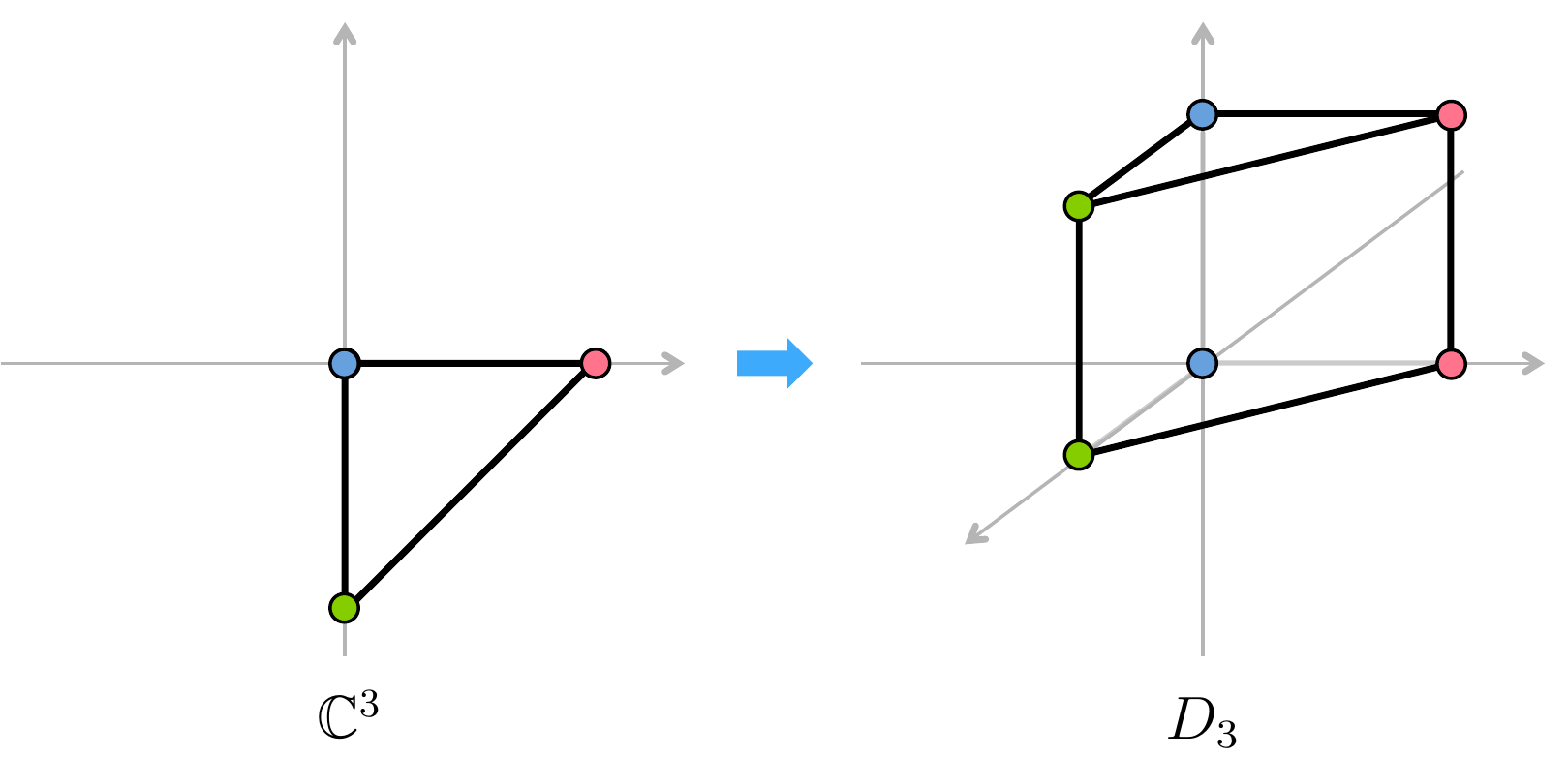}
\caption{The toric diagram for $D_3$, obtained by lifting three perfect matchings of $\mathbb{C}^3$.}
	\label{C3_to_D3}
\end{figure}

Next, let us start from $\mathbb{C}^3$ and lift the three corners of the toric diagram in the same direction, as in \fref{C3_to_D3}. We obtain the toric diagram for a geometry that is often referred to as $D_3$ \cite{Franco:2015tna}.

\begin{figure}
                \begin{center}
                  \tdplotsetmaincoords{80}{120} 
                  \tdplotsetrotatedcoords{0}{0}{150} 
                  \begin{tikzpicture}[scale=2.5,tdplot_rotated_coords] 
                    \tikzstyle{every node}=[circle,very thick,fill=yellow2,draw,inner sep=2pt,font=\scriptsize]
                    \draw (0.0,0.0,0.0) node(a1){$\underline{1}$};
                    \draw (0.0,1.0,0.0) node(a2){$\underline{1}$};
                    \draw (1.0,0.0,0.0) node(a3){$\underline{1}$};
                    \draw (1.0,1.0,0.0) node(a4){$\underline{1}$};
                    \draw (0.0,0.0,1.0) node(a5){$\overline{1}$};
                    \draw (0.0,1.0,1.0) node(a6){$\overline{1}$};
                    \draw (1.0,0.0,1.0) node(a7){$\overline{1}$};
                    \draw (1.0,1.0,1.0) node(a8){$\overline{1}$};
                    \draw (0.0,0.0,1.75) node(a9){$\underline{2}$};
                    \draw (0.0,1.0,1.75) node(a10){$\underline{2}$};
                    \draw (1.0,0.0,1.75) node(a11){$\underline{2}$};
                    \draw (1.0,1.0,1.75) node(a12){$\underline{2}$};
                    \draw (0.0,0.0,2.75) node(a13){$\overline{2}$};
                    \draw (0.0,1.0,2.75) node(a14){$\overline{2}$};
                    \draw (1.0,0.0,2.75) node(a15){$\overline{2}$};
                    \draw (1.0,1.0,2.75) node(a16){$\overline{2}$};
                    \draw (0.0,0.0,3.5) node(a17){$\underline{3}$};
                    \draw (0.0,1.0,3.5) node(a18){$\underline{3}$};
                    \draw (1.0,0.0,3.5) node(a19){$\underline{3}$};
                    \draw (1.0,1.0,3.5) node(a20){$\underline{3}$};
                    \draw (0.0,0.0,4.5) node(a21){$\overline{3}$};
                    \draw (0.0,1.0,4.5) node(a22){$\overline{3}$};
                    \draw (1.0,0.0,4.5) node(a23){$\overline{3}$};
                    \draw (1.0,1.0,4.5) node(a24){$\overline{3}$};
                    \draw (3.0,0.0,0.75) node(a25){$1$};
                    \draw (3.0,0.0,3.75) node(a26){$1$};
                    \draw (3.0,1.0,0.75) node(a27){$1$};
                    \draw (3.0,1.0,3.75) node(a28){$1$};
                    \draw (4.0,0.0,0.75) node(a29){$1$};
                    \draw (4.0,0.0,3.75) node(a30){$1$};
                    \draw (4.0,1.0,0.75) node(a31){$1$};
                    \draw (4.0,1.0,3.75) node(a32){$1$};
                    \draw (3.0,0.0,1.75) node(a33){$2$};
                    \draw (3.0,1.0,1.75) node(a34){$2$};
                    \draw (4.0,0.0,1.75) node(a35){$2$};
                    \draw (4.0,1.0,1.75) node(a36){$2$};
                    \draw (3.0,0.0,2.75) node(a37){$3$};
                    \draw (3.0,1.0,2.75) node(a38){$3$};
                    \draw (4.0,0.0,2.75) node(a39){$3$};
                    \draw (4.0,1.0,2.75) node(a40){$3$};
                    \draw[very thick,-latex](a40)--(a32);
                    \draw[very thick,-latex](a36)--(a40);
                    \draw[very thick,-latex](a31)--(a36);
                    \draw[very thick,-latex](a28)--(a32);
                    \draw[very thick,red](a38)--(a32);
                    \draw[very thick,-latex](a38)--(a36);
                    \draw[very thick,red](a27)--(a36);
                    \draw[very thick,-latex](a27)--(a31);
                    \draw[very thick,-latex](a38)--(a28);
                    \draw[very thick,-latex](a34)--(a38);
                    \draw[very thick,-latex](a27)--(a34);
                    \draw[very thick,-latex](a20)--(a24);
                    \draw[very thick,-latex](a12)--(a16);
                    \draw[very thick,-latex](a4)--(a8);
                    \draw[very thick,-latex](a22)--(a24);
                    \draw[very thick,red](a18)--(a24);
                    \draw[very thick,-latex](a18)--(a20);
                    \draw[very thick,red](a14)--(a16);
                    \draw[very thick,-latex](a14)--(a12);
                    \draw[very thick,red](a10)--(a12);
                    \draw[very thick,-latex](a6)--(a8);
                    \draw[very thick,red](a2)--(a8);
                    \draw[very thick,-latex](a2)--(a4);
                    \draw[very thick,-latex](a18)--(a22);
                    \draw[very thick,-latex](a10)--(a14);
                    \draw[very thick,-latex](a2)--(a6);
                    \draw[very thick,red](a40)--(a30);
                    \draw[very thick,-latex](a40)--(a39);
                    \draw[very thick,red](a36)--(a39);
                    \draw[very thick,-latex](a36)--(a29);
                    \draw[very thick,-latex](a30)--(a38);
                    \draw[very thick,red](a35)--(a38);
                    \draw[very thick,-latex](a35)--(a34);
                    \draw[very thick,red](a29)--(a34);
                    \draw[very thick,red](a38)--(a26);
                    \draw[very thick,-latex](a38)--(a37);
                    \draw[very thick,red](a34)--(a37);
                    \draw[very thick,-latex](a34)--(a25);
                    \draw[very thick,-latex](a24)--(a23);
                    \draw[very thick,red](a20)--(a23);
                    \draw[very thick,-latex](a20)--(a19);
                    \draw[very thick,-latex](a16)--(a15);
                    \draw[very thick,red](a12)--(a15);
                    \draw[very thick,-latex](a12)--(a11);
                    \draw[very thick,red](a8)--(a7);
                    \draw[very thick,-latex](a8)--(a3);
                    \draw[very thick,red](a4)--(a3);
                    \draw[very thick,red](a23)--(a22);
                    \draw[very thick,-latex](a23)--(a18);
                    \draw[very thick,red](a19)--(a18);
                    \draw[very thick,-latex](a15)--(a14);
                    \draw[very thick,red](a11)--(a14);
                    \draw[very thick,-latex](a11)--(a10);
                    \draw[very thick,-latex](a7)--(a6);
                    \draw[very thick,red](a3)--(a6);
                    \draw[very thick,-latex](a3)--(a2);
                    \draw[very thick,-latex](a22)--(a21);
                    \draw[very thick,red](a18)--(a21);
                    \draw[very thick,-latex](a18)--(a17);
                    \draw[very thick,-latex](a14)--(a13);
                    \draw[very thick,red](a10)--(a13);
                    \draw[very thick,-latex](a10)--(a9);
                    \draw[very thick,red](a6)--(a5);
                    \draw[very thick,-latex](a6)--(a1);
                    \draw[very thick,red](a2)--(a1);
                    \draw[very thick,-latex](a39)--(a30);
                    \draw[very thick,-latex](a35)--(a39);
                    \draw[very thick,-latex](a29)--(a35);
                    \draw[very thick,-latex](a26)--(a30);
                    \draw[very thick,red](a37)--(a30);
                    \draw[very thick,-latex](a37)--(a35);
                    \draw[very thick,red](a25)--(a35);
                    \draw[very thick,-latex](a25)--(a29);
                    \draw[very thick,-latex](a37)--(a26);
                    \draw[very thick,-latex](a33)--(a37);
                    \draw[very thick,-latex](a25)--(a33);
                    \draw[very thick,-latex](a19)--(a23);
                    \draw[very thick,-latex](a11)--(a15);
                    \draw[very thick,-latex](a3)--(a7);
                    \draw[very thick,-latex](a21)--(a23);
                    \draw[very thick,red](a17)--(a23);
                    \draw[very thick,-latex](a17)--(a19);
                    \draw[very thick,red](a13)--(a15);
                    \draw[very thick,-latex](a13)--(a11);
                    \draw[very thick,red](a9)--(a11);
                    \draw[very thick,-latex](a5)--(a7);
                    \draw[very thick,red](a1)--(a7);
                    \draw[very thick,-latex](a1)--(a3);
                    \draw[very thick,-latex](a17)--(a21);
                    \draw[very thick,-latex](a9)--(a13);
                    \draw[very thick,-latex](a1)--(a5);
                    \draw[very thick, blue , -latex] (1.6,0.5,2.25) -- (2.4,0.5,2.25);
                  \end{tikzpicture} 
                    \caption{$3d$ printing of the periodic quiver for $D_3$ using three $\mathbb{C}^3$ quiver blocks corresponding to different perfect matchings. Massive chiral-Fermi pairs are integrated out.}
                    \label{C3_to_D3_quiver}
                \end{center}
            \end{figure}
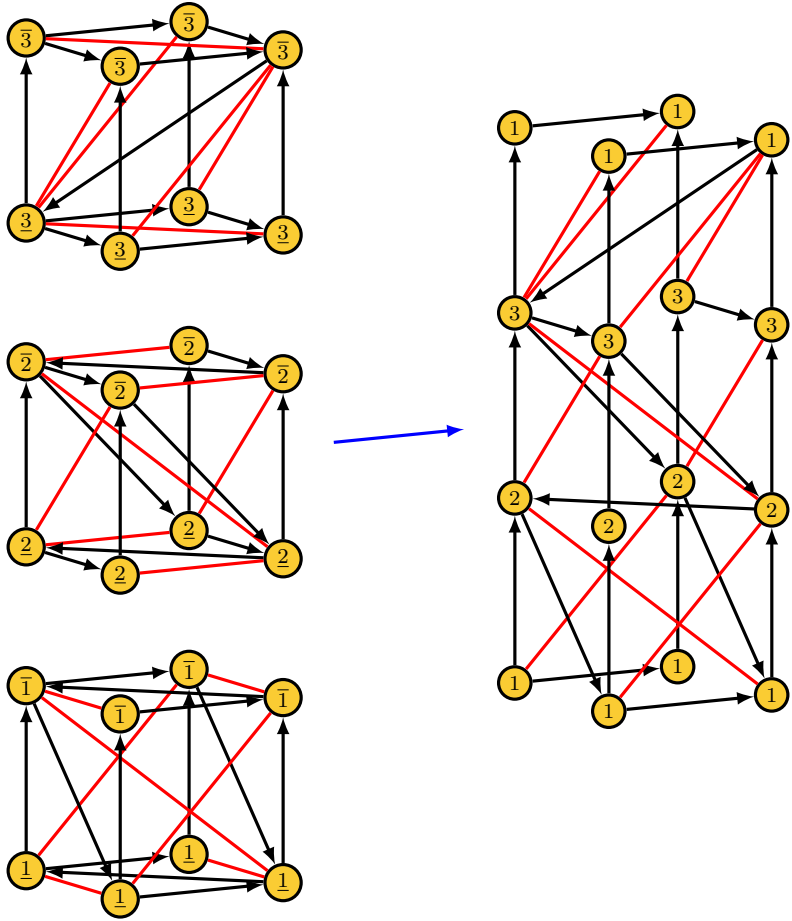

The corresponding $3d$ printing of the periodic quiver is shown in \fref{C3_to_D3_quiver}.

The $J$- and $E$-terms are:
           \begin{align}
                            \renewcommand{\arraystretch}{1.2}
                \begin{array}{rrclcrcl} & & J & &\phantom{abcde}& & E & \\\Lambda_{12}^{(1)}: & X_{21}Y_{11} & - & Z_{23}Y_{32}X_{21}  &&  Z_{12}D_{22} & - & D_{13}Z_{31}Z_{12}\\\Lambda_{12}^{(2)}: & D_{22}X_{21} & - & X_{21}D_{13}Z_{31}  &&  Z_{12}Z_{23}Y_{32} & - & Y_{11}Z_{12}\\\Lambda_{23}^{(1)}: & X_{33}Y_{32} & - & Y_{32}X_{21}Z_{12}  &&  Z_{23}Z_{31}D_{13} & - & D_{22}Z_{23}\\\Lambda_{23}^{(2)}: & Y_{32}D_{22} & - & Z_{31}D_{13}Y_{32}  &&  Z_{23}X_{33} & - & X_{21}Z_{12}Z_{23}\\\Lambda_{31}^{(1)}: & Y_{11}D_{13} & - & D_{13}Y_{32}Z_{23}  &&  Z_{31}Z_{12}X_{21} & - & X_{33}Z_{31}\\\Lambda_{31}^{(2)}: & D_{13}X_{33} & - & Z_{12}X_{21}D_{13}  &&  Z_{31}Y_{11} & - & Y_{32}Z_{23}Z_{31}\end{array} \label{d3je}      
            \end{align}     
This is precisely the gauge theory for $D_3$ originally found in \cite{Franco:2015tna} by partial resolution of the $\mathbb{C}^4/(\mathbb{Z}_2 \times \mathbb{Z}_2 \times \mathbb{Z}_2)$ orbifold. We note that this example illustrates the versatility of $3d$ printing, since this theory {\it cannot} be obtained by dimensional reduction, orbifolding or orbifold reduction.

\subsection{Models from Conifold Quiver Blocks} 

We will now consider models constructed out of conifold quiver blocks. The $4d$ $\mathcal{N}=1$ gauge theory for the conifold was introduced in \cite{Klebanov:1998hh} and its periodic quiver and brane tiling first appeared in \cite{Hanany:2005ve,Franco:2005rj}. We will consider an example in which two points in the toric diagram are lifted in the same direction, i.e. with two $(+)$ quiver blocks, and another one in which the same points are lifted in opposite directions, namely with the same quiver blocks but with $(+)$ and $(-)$ orientations.

\subsubsection{Two Perfect Matchings}

\begin{figure}[ht]
	\centering
	\includegraphics[width=12cm]{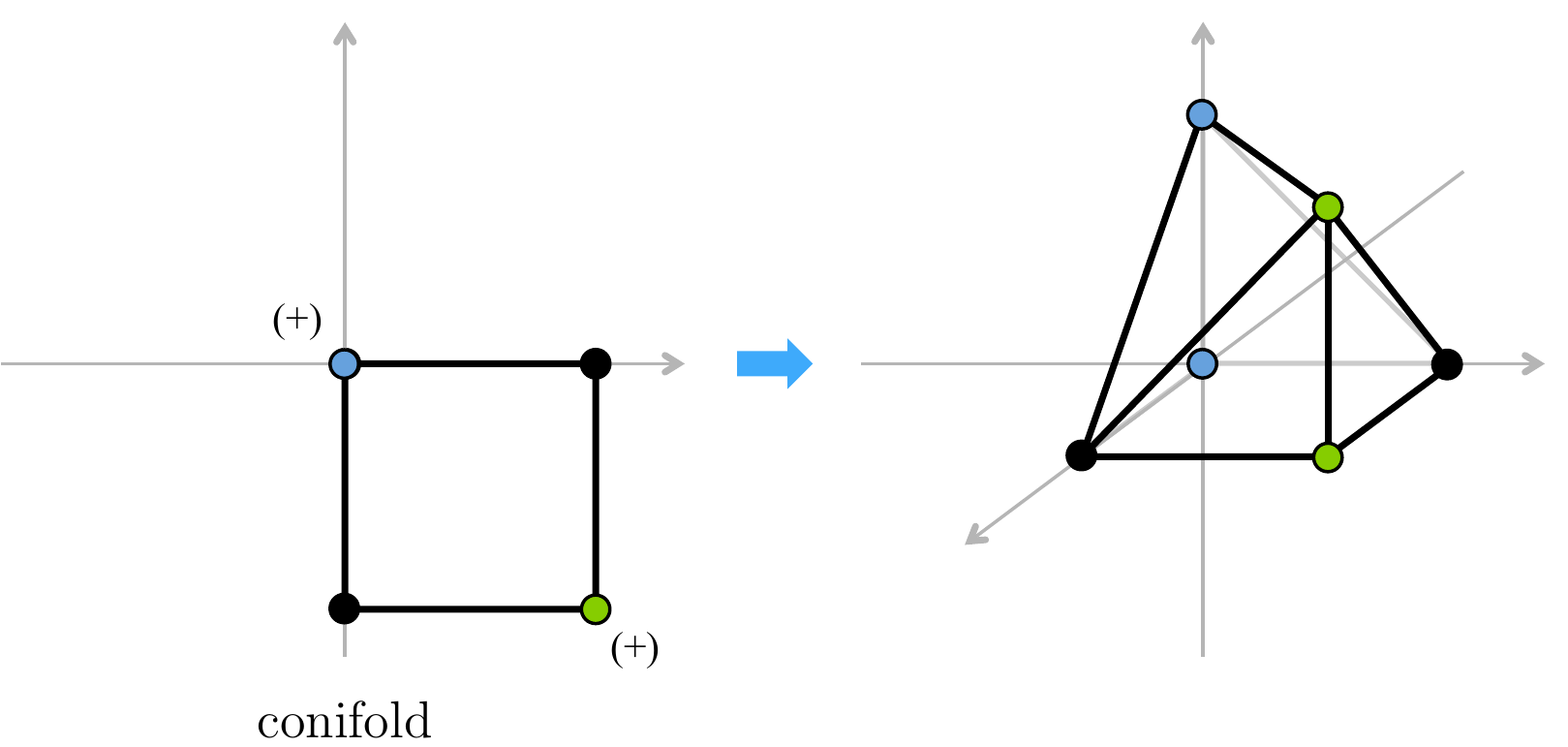}
\caption{The toric diagram of $H_4$, obtained by lifting two perfect matchings of the conifold.}
	\label{conifold_to_new_geometry}
\end{figure}

Let us lift two opposite corners of the conifold toric diagram in the same direction, as shown in \fref{conifold_to_new_geometry}. The resulting geometry was first considered in \cite{Franco:2017cjj}, where it was called $H_4$.
           
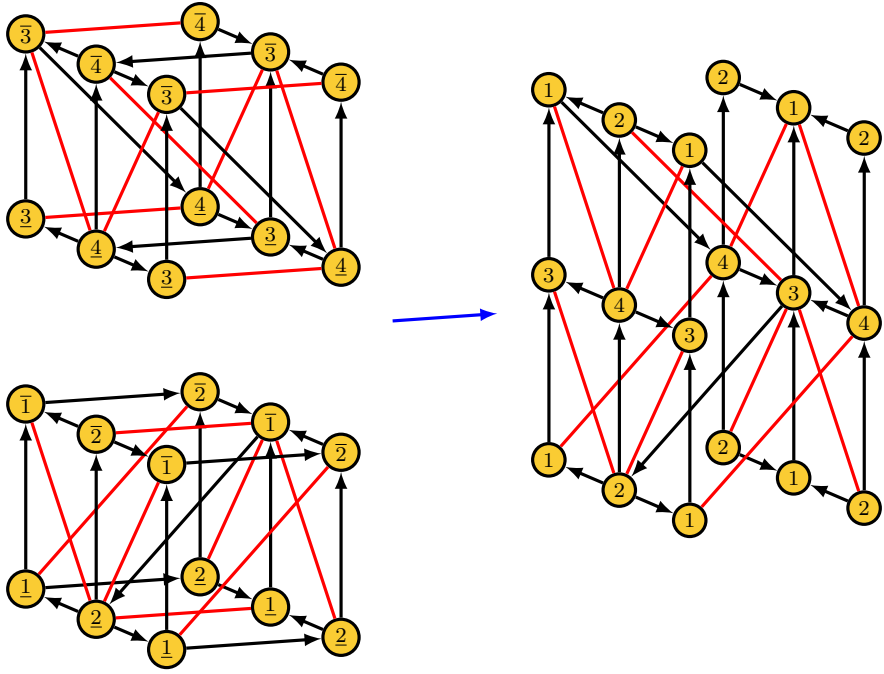
\begin{figure}[ht]
                \begin{center}
                        \tdplotsetmaincoords{80}{118} 
                        \tdplotsetrotatedcoords{0}{0}{140} 
                        \begin{tikzpicture}[scale=2.5,tdplot_rotated_coords] 
                            \tikzstyle{every node}=[circle,very thick,fill=yellow2,draw,inner sep=2pt,font=\scriptsize]
                            \draw (0.0,0.0,0.0) node(a1){$\underline{1}$};
                            \draw (0.0,2.0,0.0) node(a2){$\underline{1}$};
                            \draw (1.0,1.0,0.0) node(a3){$\underline{1}$};
                            \draw (0.0,1.0,0.0) node(a4){$\underline{2}$};
                            \draw (1.0,2.0,0.0) node(a5){$\underline{2}$};
                            \draw (1.0,0.0,0.0) node(a6){$\underline{2}$};
                            \draw (0.0,0.0,1.0) node(a7){$\overline{1}$};
                            \draw (0.0,2.0,1.0) node(a8){$\overline{1}$};
                            \draw (1.0,1.0,1.0) node(a9){$\overline{1}$};
                            \draw (0.0,1.0,1.0) node(a10){$\overline{2}$};
                            \draw (1.0,2.0,1.0) node(a11){$\overline{2}$};
                            \draw (1.0,0.0,1.0) node(a12){$\overline{2}$};
                            \draw (0.0,0.0,2.0) node(a13){$\underline{3}$};
                            \draw (0.0,2.0,2.0) node(a14){$\underline{3}$};
                            \draw (1.0,1.0,2.0) node(a15){$\underline{3}$};
                            \draw (0.0,1.0,2.0) node(a16){$\underline{4}$};
                            \draw (1.0,2.0,2.0) node(a17){$\underline{4}$};
                            \draw (1.0,0.0,2.0) node(a18){$\underline{4}$};
                            \draw (0.0,0.0,3.0) node(a19){$\overline{3}$};
                            \draw (0.0,2.0,3.0) node(a20){$\overline{3}$};
                            \draw (1.0,1.0,3.0) node(a21){$\overline{3}$};
                            \draw (0.0,1.0,3.0) node(a22){$\overline{4}$};
                            \draw (1.0,2.0,3.0) node(a23){$\overline{4}$};
                            \draw (1.0,0.0,3.0) node(a24){$\overline{4}$};
                            \draw (3.0,0.0,0.5) node(a25){$1$};
                            \draw (3.0,0.0,2.5) node(a26){$1$};
                            \draw (3.0,2.0,0.5) node(a27){$1$};
                            \draw (3.0,2.0,2.5) node(a28){$1$};
                            \draw (4.0,1.0,0.5) node(a29){$1$};
                            \draw (4.0,1.0,2.5) node(a30){$1$};
                            \draw (3.0,1.0,0.5) node(a31){$2$};
                            \draw (3.0,1.0,2.5) node(a32){$2$};
                            \draw (4.0,2.0,0.5) node(a33){$2$};
                            \draw (4.0,2.0,2.5) node(a34){$2$};
                            \draw (4.0,0.0,0.5) node(a35){$2$};
                            \draw (4.0,0.0,2.5) node(a36){$2$};
                            \draw (3.0,0.0,1.5) node(a37){$3$};
                            \draw (3.0,2.0,1.5) node(a38){$3$};
                            \draw (4.0,1.0,1.5) node(a39){$3$};
                            \draw (3.0,1.0,1.5) node(a40){$4$};
                            \draw (4.0,2.0,1.5) node(a41){$4$};
                            \draw (4.0,0.0,1.5) node(a42){$4$};
                            \draw[very thick,-latex](a41)--(a34);
                            \draw[very thick,-latex](a33)--(a41);
                            \draw[very thick,-latex](a28)--(a41);
                            \draw[very thick,red](a27)--(a41);
                            \draw[very thick,-latex](a38)--(a28);
                            \draw[very thick,-latex](a27)--(a38);
                            \draw[very thick,-latex](a17)--(a23);
                            \draw[very thick,-latex](a5)--(a11);
                            \draw[very thick,red](a20)--(a23);
                            \draw[very thick,-latex](a20)--(a17);
                            \draw[very thick,red](a14)--(a17);
                            \draw[very thick,-latex](a8)--(a11);
                            \draw[very thick,red](a2)--(a11);
                            \draw[very thick,-latex](a2)--(a5);
                            \draw[very thick,-latex](a14)--(a20);
                            \draw[very thick,-latex](a2)--(a8);
                            \draw[very thick,-latex](a34)--(a30);
                            \draw[very thick,red](a41)--(a30);
                            \draw[very thick,-latex](a41)--(a39);
                            \draw[very thick,red](a33)--(a39);
                            \draw[very thick,-latex](a33)--(a29);
                            \draw[very thick,-latex](a32)--(a28);
                            \draw[very thick,red](a40)--(a28);
                            \draw[very thick,-latex](a40)--(a38);
                            \draw[very thick,red](a31)--(a38);
                            \draw[very thick,-latex](a31)--(a27);
                            \draw[very thick,-latex](a23)--(a21);
                            \draw[very thick,red](a17)--(a21);
                            \draw[very thick,-latex](a17)--(a15);
                            \draw[very thick,-latex](a11)--(a9);
                            \draw[very thick,red](a5)--(a9);
                            \draw[very thick,-latex](a5)--(a3);
                            \draw[very thick,-latex](a22)--(a20);
                            \draw[very thick,red](a16)--(a20);
                            \draw[very thick,-latex](a16)--(a14);
                            \draw[very thick,-latex](a10)--(a8);
                            \draw[very thick,red](a4)--(a8);
                            \draw[very thick,-latex](a4)--(a2);
                            \draw[very thick,-latex](a39)--(a30);
                            \draw[very thick,-latex](a29)--(a39);
                            \draw[very thick,red](a39)--(a32);
                            \draw[very thick,-latex](a39)--(a31);
                            \draw[very thick,-latex](a40)--(a32);
                            \draw[very thick,-latex](a31)--(a40);
                            \draw[very thick,-latex](a15)--(a21);
                            \draw[very thick,-latex](a3)--(a9);
                            \draw[very thick,-latex](a21)--(a22);
                            \draw[very thick,red](a15)--(a22);
                            \draw[very thick,-latex](a15)--(a16);
                            \draw[very thick,red](a9)--(a10);
                            \draw[very thick,-latex](a9)--(a4);
                            \draw[very thick,red](a3)--(a4);
                            \draw[very thick,-latex](a16)--(a22);
                            \draw[very thick,-latex](a4)--(a10);
                            \draw[very thick,-latex](a36)--(a30);
                            \draw[very thick,red](a42)--(a30);
                            \draw[very thick,-latex](a42)--(a39);
                            \draw[very thick,red](a35)--(a39);
                            \draw[very thick,-latex](a35)--(a29);
                            \draw[very thick,-latex](a32)--(a26);
                            \draw[very thick,red](a40)--(a26);
                            \draw[very thick,-latex](a40)--(a37);
                            \draw[very thick,red](a31)--(a37);
                            \draw[very thick,-latex](a31)--(a25);
                            \draw[very thick,-latex](a24)--(a21);
                            \draw[very thick,red](a18)--(a21);
                            \draw[very thick,-latex](a18)--(a15);
                            \draw[very thick,-latex](a12)--(a9);
                            \draw[very thick,red](a6)--(a9);
                            \draw[very thick,-latex](a6)--(a3);
                            \draw[very thick,-latex](a22)--(a19);
                            \draw[very thick,red](a16)--(a19);
                            \draw[very thick,-latex](a16)--(a13);
                            \draw[very thick,-latex](a10)--(a7);
                            \draw[very thick,red](a4)--(a7);
                            \draw[very thick,-latex](a4)--(a1);
                            \draw[very thick,-latex](a42)--(a36);
                            \draw[very thick,-latex](a35)--(a42);
                            \draw[very thick,-latex](a26)--(a42);
                            \draw[very thick,red](a25)--(a42);
                            \draw[very thick,-latex](a37)--(a26);
                            \draw[very thick,-latex](a25)--(a37);
                            \draw[very thick,-latex](a18)--(a24);
                            \draw[very thick,-latex](a6)--(a12);
                            \draw[very thick,red](a19)--(a24);
                            \draw[very thick,-latex](a19)--(a18);
                            \draw[very thick,red](a13)--(a18);
                            \draw[very thick,-latex](a7)--(a12);
                            \draw[very thick,red](a1)--(a12);
                            \draw[very thick,-latex](a1)--(a6);
                            \draw[very thick,-latex](a13)--(a19);
                            \draw[very thick,-latex](a1)--(a7);
                            \draw[very thick , blue , -latex](1.7,1,1.5) -- (2.3,1,1.5);
                        \end{tikzpicture} 
                    \caption{$3d$ printing of a periodic quiver for $H_4$ using two conifold quiver blocks.}
                    \label{conifold_to_new_geometry_quiver}
                \end{center}
            \end{figure}

We construct the periodic quiver using two conifold quiver blocks for the appropriate perfect matchings. This process is shown in \fref{conifold_to_new_geometry_quiver}. The $J$- and $E$-terms are:       
\begin{align}
                \renewcommand{\arraystretch}{1.2}
                \begin{array}{rrclcrcl} & & J & &\phantom{abcde}& & E & \\
                \Lambda_{14}: & Y_{43}X_{32}X_{21} & - & X_{43}X_{32}Y_{21}  &&  Z_{13}Z_{31}Y_{14} & - & Y_{14}Z_{42}Z_{24}\\
                \Lambda_{41}^{(1)}: & Y_{14}Y_{43}X_{32}Z_{24} & - & Z_{13}X_{32}Y_{21}Y_{14}  &&  Z_{42}X_{21} & - & X_{43}Z_{31}\\
                \Lambda_{41}^{(2)}: & Z_{13}X_{32}X_{21}Y_{14} & - & Y_{14}X_{43}X_{32}Z_{24}  &&  Z_{42}Y_{21} & - & Y_{43}Z_{31} \\
                \Lambda_{32}: & X_{21}Y_{14}Y_{43} & - & Y_{21}Y_{14}X_{43}  &&  Z_{31}Z_{13}X_{32} & - & X_{32}Z_{24}Z_{42} \\
                \Lambda_{23}^{(1)}: & Z_{31}Y_{14}Y_{43}X_{32} & - & X_{32}Y_{21}Y_{14}Z_{42}  &&  Z_{24}X_{43} & - & X_{21}Z_{13}\\                
                \Lambda_{23}^{(2)}: & X_{32}X_{21}Y_{14}Z_{42} & - & Z_{31}Y_{14}X_{43}X_{32}  &&  Z_{24}Y_{43} & - & Y_{21}Z_{13}
                \end{array}\label{uuje}
\end{align}
 
There are three Fermis between nodes 1 and 4, and three between 2 and 3. Given the $\Lambda \leftrightarrow \bar{\Lambda}$ symmetry of $2d$ $(0,2)$ theories, it is possible to orient all Fermis connecting each pair of nodes in the same direction. Instead of doing this, we choose to make the orientation of the parent $4d$ chiral fields manifest. Moreover, given the structure of the $J$- and $E$-terms in \eref{uuje}, there is a clear pairing of some of these Fermis, which our notation emphasizes. Our results coincide with one of the two phases presented for $H_4$ in \cite{Franco:2017cjj}. It is worth emphasizing that, as comparison between this example and \cite{Franco:2017cjj} illustrates, periodic quivers obtained from quiver blocks tend to be better organized than the ones that arise from partial resolution, even though they fully agree.

\subsubsection{$Q^{1,1,1}$}
        
\begin{figure}[ht]
	\centering
	\includegraphics[width=12cm]{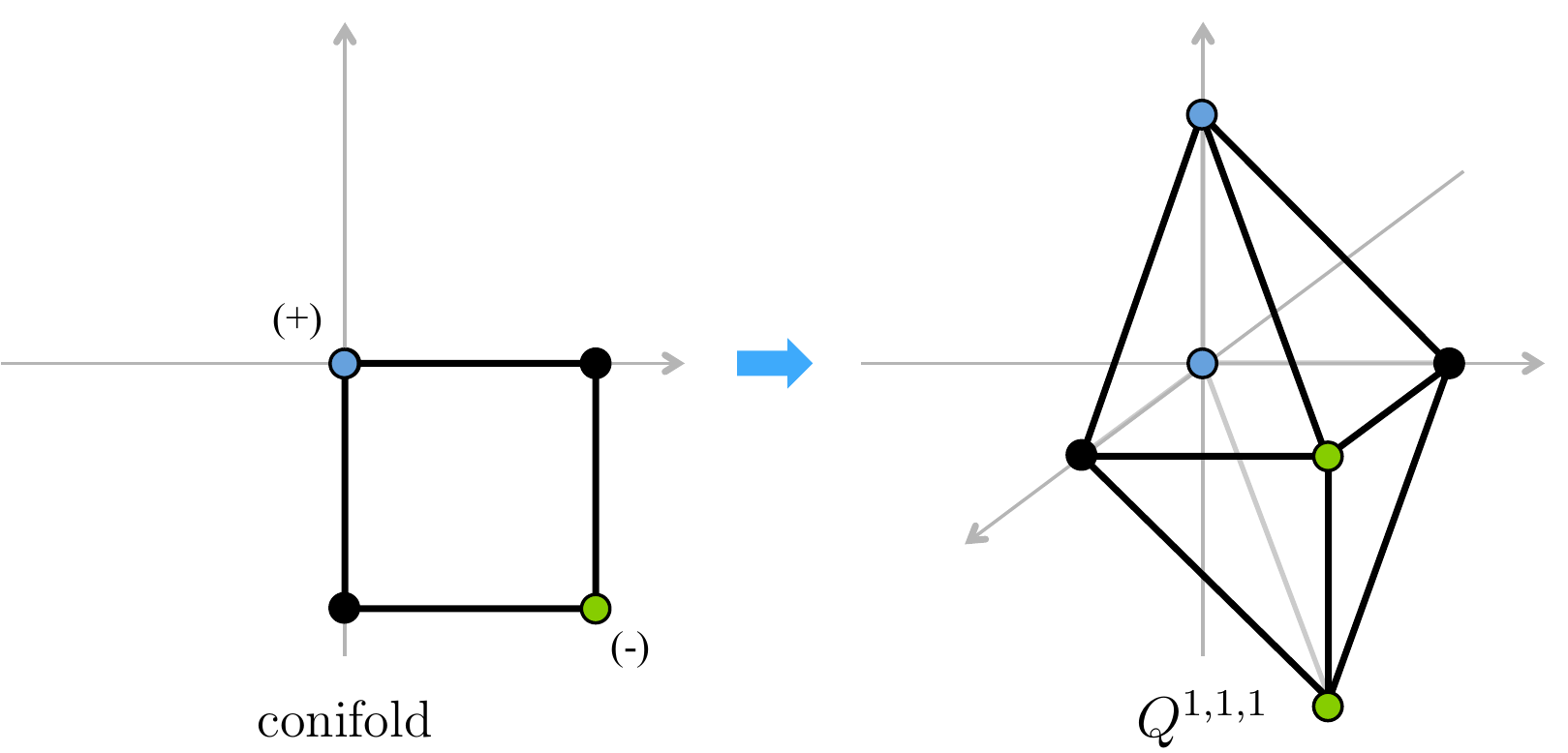}
\caption{The toric diagram for $Q^{1,1,1}$, obtained by lifting two perfect matchings of the conifold in opposite directions.}
	\label{conifold_to_Q111}
\end{figure}

 Let us now consider the the real cone over the $7d$ Sasaki-Einstein manifold $Q^{1,1,1}$, which is the homogeneous coset space
 \beq
 {SU(2) \times SU(2) \times SU(2) \over U(1) \times U(1)} 
 \eeq
 and has a $U(1)_R \times SU(2)^3$ isometry \cite{DAuria:1983sda}. For brevity, we will refer to the full cone as $Q^{1,1,1}$. Its toric diagram can also be constructed by lifting two opposite corners of the conifold toric diagram, but doing so in opposite directions as shown in \fref{conifold_to_Q111}.       
        
\begin{figure}[ht]
                \begin{center}
                        \tdplotsetmaincoords{80}{118} 
                        \tdplotsetrotatedcoords{0}{0}{140} 
                        \begin{tikzpicture}[scale=2.5,tdplot_rotated_coords] 
                            \tikzstyle{every node}=[circle,very thick,fill=yellow2,draw,inner sep=2pt,font=\scriptsize]
                            \draw (0.0,0.0,0.0) node(a1){$\underline{1}$};
                            \draw (0.0,2.0,0.0) node(a2){$\underline{1}$};
                            \draw (1.0,1.0,0.0) node(a3){$\underline{1}$};
                            \draw (0.0,1.0,0.0) node(a4){$\underline{2}$};
                            \draw (1.0,2.0,0.0) node(a5){$\underline{2}$};
                            \draw (1.0,0.0,0.0) node(a6){$\underline{2}$};
                            \draw (0.0,0.0,1.0) node(a7){$\overline{1}$};
                            \draw (0.0,2.0,1.0) node(a8){$\overline{1}$};
                            \draw (1.0,1.0,1.0) node(a9){$\overline{1}$};
                            \draw (0.0,1.0,1.0) node(a10){$\overline{2}$};
                            \draw (1.0,2.0,1.0) node(a11){$\overline{2}$};
                            \draw (1.0,0.0,1.0) node(a12){$\overline{2}$};
                            \draw (0.0,0.0,3.0) node(a13){$\underline{3}$};
                            \draw (0.0,2.0,3.0) node(a14){$\underline{3}$};
                            \draw (1.0,1.0,3.0) node(a15){$\underline{3}$};
                            \draw (0.0,1.0,3.0) node(a16){$\underline{4}$};
                            \draw (1.0,2.0,3.0) node(a17){$\underline{4}$};
                            \draw (1.0,0.0,3.0) node(a18){$\underline{4}$};
                            \draw (0.0,0.0,2.0) node(a19){$\overline{3}$};
                            \draw (0.0,2.0,2.0) node(a20){$\overline{3}$};
                            \draw (1.0,1.0,2.0) node(a21){$\overline{3}$};
                            \draw (0.0,1.0,2.0) node(a22){$\overline{4}$};
                            \draw (1.0,2.0,2.0) node(a23){$\overline{4}$};
                            \draw (1.0,0.0,2.0) node(a24){$\overline{4}$};
                            \draw (3.0,0.0,0.5) node(a25){$1$};
                            \draw (3.0,0.0,2.5) node(a26){$1$};
                            \draw (3.0,2.0,0.5) node(a27){$1$};
                            \draw (3.0,2.0,2.5) node(a28){$1$};
                            \draw (4.0,1.0,0.5) node(a29){$1$};
                            \draw (4.0,1.0,2.5) node(a30){$1$};
                            \draw (3.0,1.0,0.5) node(a31){$2$};
                            \draw (3.0,1.0,2.5) node(a32){$2$};
                            \draw (4.0,2.0,0.5) node(a33){$2$};
                            \draw (4.0,2.0,2.5) node(a34){$2$};
                            \draw (4.0,0.0,0.5) node(a35){$2$};
                            \draw (4.0,0.0,2.5) node(a36){$2$};
                            \draw (3.0,0.0,1.5) node(a37){$3$};
                            \draw (3.0,2.0,1.5) node(a38){$3$};
                            \draw (4.0,1.0,1.5) node(a39){$3$};
                            \draw (3.0,1.0,1.5) node(a40){$4$};
                            \draw (4.0,2.0,1.5) node(a41){$4$};
                            \draw (4.0,0.0,1.5) node(a42){$4$};
                            \draw[very thick,-latex](a34)--(a41);
                            \draw[very thick,-latex](a33)--(a41);
                            \draw[very thick,-latex](a38)--(a34);
                            \draw[very thick,red](a27)--(a41);
                            \draw[very thick,-latex](a28)--(a38);
                            \draw[very thick,-latex](a27)--(a38);
                            \draw[very thick,-latex](a17)--(a23);
                            \draw[very thick,-latex](a5)--(a11);
                            \draw[very thick,red](a14)--(a17);
                            \draw[very thick,-latex](a20)--(a17);
                            \draw[very thick,red](a20)--(a23);
                            \draw[very thick,-latex](a8)--(a11);
                            \draw[very thick,red](a2)--(a11);
                            \draw[very thick,-latex](a2)--(a5);
                            \draw[very thick,-latex](a14)--(a20);
                            \draw[very thick,-latex](a2)--(a8);
                            \draw[very thick,-latex](a34)--(a30);
                            \draw[very thick,red](a34)--(a39);
                            \draw[very thick,-latex](a41)--(a39);
                            \draw[very thick,red](a33)--(a39);
                            \draw[very thick,-latex](a33)--(a29);
                            \draw[very thick,-latex](a32)--(a28);
                            \draw[very thick,red](a32)--(a38);
                            \draw[very thick,-latex](a40)--(a38);
                            \draw[very thick,red](a31)--(a38);
                            \draw[very thick,-latex](a31)--(a27);
                            \draw[very thick,-latex](a17)--(a15);
                            \draw[very thick,red](a17)--(a21);
                            \draw[very thick,-latex](a23)--(a21);
                            \draw[very thick,-latex](a11)--(a9);
                            \draw[very thick,red](a5)--(a9);
                            \draw[very thick,-latex](a5)--(a3);
                            \draw[very thick,-latex](a16)--(a14);
                            \draw[very thick,red](a16)--(a20);
                            \draw[very thick,-latex](a22)--(a20);
                            \draw[very thick,-latex](a10)--(a8);
                            \draw[very thick,red](a4)--(a8);
                            \draw[very thick,-latex](a4)--(a2);
                            \draw[very thick,-latex](a30)--(a39);
                            \draw[very thick,-latex](a29)--(a39);
                            \draw[very thick,red](a30)--(a40);
                            \draw[very thick,-latex](a39)--(a31);
                            \draw[very thick,-latex](a32)--(a40);
                            \draw[very thick,-latex](a31)--(a40);
                            \draw[very thick,-latex](a15)--(a21);
                            \draw[very thick,-latex](a3)--(a9);
                            \draw[very thick,-latex](a15)--(a16);
                            \draw[very thick,red](a15)--(a22);
                            \draw[very thick,-latex](a21)--(a22);
                            \draw[very thick,red](a9)--(a10);
                            \draw[very thick,-latex](a9)--(a4);
                            \draw[very thick,red](a3)--(a4);
                            \draw[very thick,-latex](a16)--(a22);
                            \draw[very thick,-latex](a4)--(a10);
                            \draw[very thick,-latex](a36)--(a30);
                            \draw[very thick,red](a36)--(a39);
                            \draw[very thick,-latex](a42)--(a39);
                            \draw[very thick,red](a35)--(a39);
                            \draw[very thick,-latex](a35)--(a29);
                            \draw[very thick,-latex](a32)--(a26);
                            \draw[very thick,red](a32)--(a37);
                            \draw[very thick,-latex](a40)--(a37);
                            \draw[very thick,red](a31)--(a37);
                            \draw[very thick,-latex](a31)--(a25);
                            \draw[very thick,-latex](a18)--(a15);
                            \draw[very thick,red](a18)--(a21);
                            \draw[very thick,-latex](a24)--(a21);
                            \draw[very thick,-latex](a12)--(a9);
                            \draw[very thick,red](a6)--(a9);
                            \draw[very thick,-latex](a6)--(a3);
                            \draw[very thick,-latex](a16)--(a13);
                            \draw[very thick,red](a16)--(a19);
                            \draw[very thick,-latex](a22)--(a19);
                            \draw[very thick,-latex](a10)--(a7);
                            \draw[very thick,red](a4)--(a7);
                            \draw[very thick,-latex](a4)--(a1);
                            \draw[very thick,-latex](a36)--(a42);
                            \draw[very thick,-latex](a35)--(a42);
                            \draw[very thick,-latex](a37)--(a36);
                            \draw[very thick,red](a25)--(a42);
                            \draw[very thick,-latex](a26)--(a37);
                            \draw[very thick,-latex](a25)--(a37);
                            \draw[very thick,-latex](a18)--(a24);
                            \draw[very thick,-latex](a6)--(a12);
                            \draw[very thick,red](a13)--(a18);
                            \draw[very thick,-latex](a19)--(a18);
                            \draw[very thick,red](a19)--(a24);
                            \draw[very thick,-latex](a7)--(a12);
                            \draw[very thick,red](a1)--(a12);
                            \draw[very thick,-latex](a1)--(a6);
                            \draw[very thick,-latex](a13)--(a19);
                            \draw[very thick,-latex](a1)--(a7);
                            \draw[very thick , blue , -latex](1.7,1,1.5) -- (2.3,1,1.5);
                        \end{tikzpicture} 
                    \caption{$3d$ printing of the periodic quiver for $Q^{1,1,1}$ using two conifold quiver blocks.}
                    \label{coni2pud}
                \end{center}
            \end{figure}
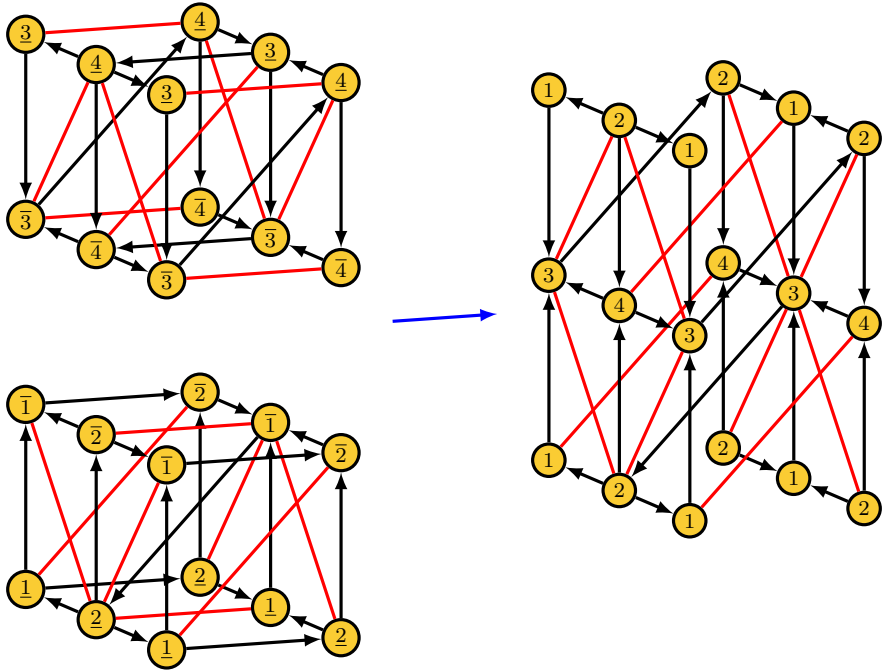

The periodic quiver is built using the same blocks employed for the theory in the previous section, but combining them with $(+)$ and $(-)$ orientations in order to achieve the desired lift of the toric diagram. This is shown in \fref{coni2pud}.

The $J$- and $E$-terms are:        
\begin{align}
                \renewcommand{\arraystretch}{1.2}
               \begin{array}{rrclcrcl} & & J & &\phantom{abcde}& & E & \\\Lambda_{14}^{(1)}: & Y_{43}X_{32}X_{21} & - & X_{43}X_{32}Y_{21}  &&  Z_{13}^{-}Y_{32}Z_{24}^{+} & - & Z_{13}^{+}Y_{32}Z_{24}^{-}\\\Lambda_{14}^{(2)}: & X_{43}Y_{32}Y_{21} & - & Y_{43}Y_{32}X_{21}  &&  Z_{13}^{-}X_{32}Z_{24}^{+} & - & Z_{13}^{+}X_{32}Z_{24}^{-}
               \end{array} \nonumber
\end{align}
               
 \begin{align}
                \renewcommand{\arraystretch}{1.2}
               \begin{array}{rrclcrcl}               
               \Lambda_{23}^{(1)}: & Y_{32}Z_{24}^{-}Y_{43}X_{32} & - & X_{32}Y_{21}Z_{13}^{-}Y_{32}  &&  X_{21}Z_{13}^{+} & - & Z_{24}^{+}X_{43}\\\Lambda_{23}^{(2)}: & X_{32}X_{21}Z_{13}^{-}Y_{32} & - & Y_{32}Z_{24}^{-}X_{43}X_{32}  &&  Y_{21}Z_{13}^{+} & - & Z_{24}^{+}Y_{43}\\\Lambda_{23}^{(3)}: & Y_{32}Y_{21}Z_{13}^{+}X_{32} & - & X_{32}Z_{24}^{+}Y_{43}Y_{32}  &&  Z_{24}^{-}X_{43} & - & X_{21}Z_{13}^{-}\\\Lambda_{23}^{(4)}: & X_{32}Z_{24}^{+}X_{43}Y_{32} & - & Y_{32}X_{21}Z_{13}^{+}X_{32}  &&  Z_{24}^{-}Y_{43} & - & Y_{21}Z_{13}^{-}\end{array}
\end{align}
Our results are in perfect agreement with the previous determination of this theory \cite{Franco:2015tna,Franco:2015tya}. As shown in \cite{Franco:2015tna}, the full $U(1)_R \times SU(2)^3$ global symmetry associated to the isometry is not manifest at the level of the Lagrangian but emerges on the moduli space. We note that until now, like for $D_3$ and $H_4$, the only tool available for finding this gauge theory was partial resolution.

\section{Toric Phases of $Q^{1,1,1}/\mathbb{Z}_2$}

\label{section_phases_Q111/Z2}

A common feature of all models considered in \sref{section_examples} is that there is a single perfect matching for every point in the toric diagrams of the parent CY$_3$ geometries, $\mathbb{C}^3$ and the conifold. This implies that there is a unique quiver block, up to orientation, for lifting any point. In this section we consider a more general example, in which a point in the toric diagram can be lifted using quiver blocks for different perfect matchings.

\begin{figure}[ht]
	\centering
	\includegraphics[width=11cm]{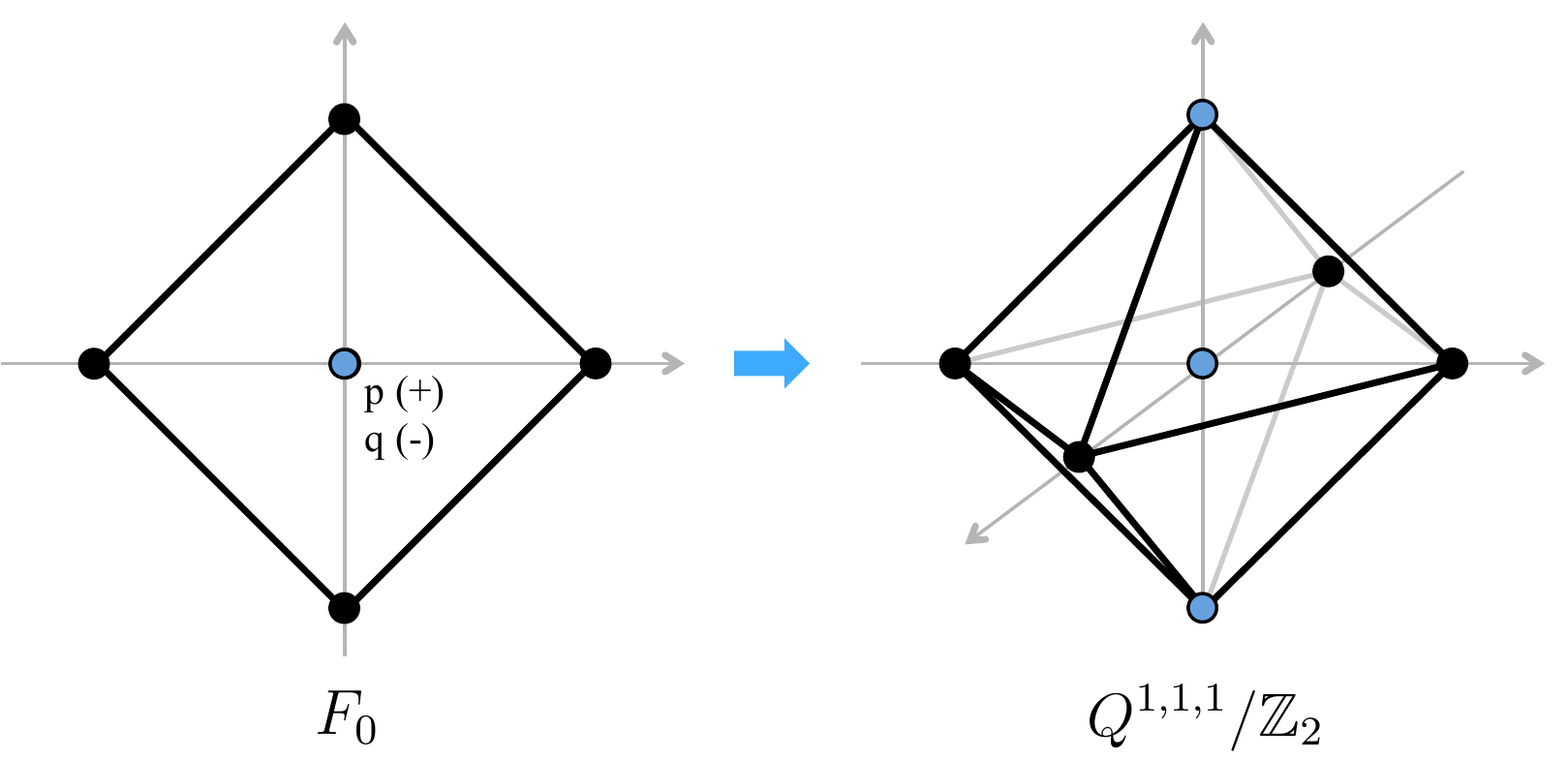}
\caption{The toric diagram for $Q^{1,1,1}/\mathbb{Z}_2$, obtained by lifting central perfect matchings of $F_0$ in opposite directions.}
	\label{F0_to_Q111Z2}
\end{figure}

Let us consider $F_0$, whose toric diagram is shown in \fref{F0_to_Q111Z2}. This geometry is associated to two toric $4d$ $\mathcal{N}=1$ gauge theories, which are usually designated phases 1 and 2 \cite{Feng:2001xr,Feng:2002zw}.\footnote{The quivers for both phases of $F_0$ have 4 nodes. We refer to phase 2 as the one with 12 chiral fields and phase 1 as the one with 8 chiral fields. Some references in the literature swap the names of these two phases.} The central point in the toric diagram corresponds to 4 and 5 perfect matchings for phases 1 and 2, respectively. We will lift this point in the positive and negative directions to produce the toric diagram of $Q^{1,1,1}/\mathbb{Z}_2$, as in \fref{F0_to_Q111Z2}. 

Let us start from phase 2 of $F_0$ and construct a periodic quiver for $Q^{1,1,1}/\mathbb{Z}_2$ using quiver blocks for two different perfect matchings, as shown in \fref{3d_printing_phase_D}. The choice of perfect matchings is easily determined by looking at the Fermi fields on the boundaries of each quiver block. We take one orientation to be $(+)$ and the other one to be $(-)$ to achieve the desired lift. The $J$- and $E$-terms for this theory are:

\begin{figure} 
                        \tdplotsetmaincoords{80}{118} 
                        \tdplotsetrotatedcoords{0}{0}{140} 
                                                \begin{tikzpicture}[scale=2.4,tdplot_rotated_coords] 
                            \tikzstyle{every node}=[circle,very thick,fill=yellow2,draw,inner sep=2pt,font=\scriptsize]
        \draw (0.0,0.0,0.0) node(a1){$\underline{1}$};
        \draw (0.0,2.0,0.0) node(a2){$\underline{1}$};
        \draw (2.0,0.0,0.0) node(a3){$\underline{1}$};
        \draw (2.0,2.0,0.0) node(a4){$\underline{1}$};
        \draw (1.0,0.0,0.0) node(a5){$\underline{2}$};
        \draw (1.0,2.0,0.0) node(a6){$\underline{2}$};
        \draw (0.0,1.0,0.0) node(a7){$\underline{3}$};
        \draw (2.0,1.0,0.0) node(a8){$\underline{3}$};
        \draw (1.0,1.0,0.0) node(a9){$\underline{4}$};
        \draw (0.0,0.0,1.0) node(a10){$\overline{1}$};
        \draw (0.0,2.0,1.0) node(a11){$\overline{1}$};
        \draw (2.0,0.0,1.0) node(a12){$\overline{1}$};
        \draw (2.0,2.0,1.0) node(a13){$\overline{1}$};
        \draw (1.0,0.0,1.0) node(a14){$\overline{2}$};
        \draw (1.0,2.0,1.0) node(a15){$\overline{2}$};
        \draw (0.0,1.0,1.0) node(a16){$\overline{3}$};
        \draw (2.0,1.0,1.0) node(a17){$\overline{3}$};
        \draw (1.0,1.0,1.0) node(a18){$\overline{4}$};
        \draw (0.0,0.0,3.0) node(a19){$\underline{5}$};
        \draw (0.0,2.0,3.0) node(a20){$\underline{5}$};
        \draw (2.0,0.0,3.0) node(a21){$\underline{5}$};
        \draw (2.0,2.0,3.0) node(a22){$\underline{5}$};
        \draw (1.0,0.0,3.0) node(a23){$\underline{6}$};
        \draw (1.0,2.0,3.0) node(a24){$\underline{6}$};
        \draw (0.0,1.0,3.0) node(a25){$\underline{7}$};
        \draw (2.0,1.0,3.0) node(a26){$\underline{7}$};
        \draw (1.0,1.0,3.0) node(a27){$\underline{8}$};
        \draw (0.0,0.0,2.0) node(a28){$\overline{5}$};
        \draw (0.0,2.0,2.0) node(a29){$\overline{5}$};
        \draw (2.0,0.0,2.0) node(a30){$\overline{5}$};
        \draw (2.0,2.0,2.0) node(a31){$\overline{5}$};
        \draw (1.0,0.0,2.0) node(a32){$\overline{6}$};
        \draw (1.0,2.0,2.0) node(a33){$\overline{6}$};
        \draw (0.0,1.0,2.0) node(a34){$\overline{7}$};
        \draw (2.0,1.0,2.0) node(a35){$\overline{7}$};
        \draw (1.0,1.0,2.0) node(a36){$\overline{8}$};
        \draw (4.0,0.0,0.5) node(a37){$1$};
        \draw (4.0,0.0,2.5) node(a38){$1$};
        \draw (4.0,2.0,0.5) node(a39){$1$};
        \draw (4.0,2.0,2.5) node(a40){$1$};
        \draw (6.0,0.0,0.5) node(a41){$1$};
        \draw (6.0,0.0,2.5) node(a42){$1$};
        \draw (6.0,2.0,0.5) node(a43){$1$};
        \draw (6.0,2.0,2.5) node(a44){$1$};
        \draw (5.0,0.0,0.5) node(a45){$2$};
        \draw (5.0,0.0,2.5) node(a46){$2$};
        \draw (5.0,2.0,0.5) node(a47){$2$};
        \draw (5.0,2.0,2.5) node(a48){$2$};
        \draw (4.0,1.0,0.5) node(a49){$3$};
        \draw (4.0,1.0,2.5) node(a50){$3$};
        \draw (6.0,1.0,0.5) node(a51){$3$};
        \draw (6.0,1.0,2.5) node(a52){$3$};
        \draw (5.0,1.0,0.5) node(a53){$4$};
        \draw (5.0,1.0,2.5) node(a54){$4$};
        \draw (4.0,0.0,1.5) node(a55){$5$};
        \draw (4.0,2.0,1.5) node(a56){$5$};
        \draw (6.0,0.0,1.5) node(a57){$5$};
        \draw (6.0,2.0,1.5) node(a58){$5$};
        \draw (5.0,0.0,1.5) node(a59){$6$};
        \draw (5.0,2.0,1.5) node(a60){$6$};
        \draw (4.0,1.0,1.5) node(a61){$7$};
        \draw (6.0,1.0,1.5) node(a62){$7$};
        \draw (5.0,1.0,1.5) node(a63){$8$};
        \draw[very thick,-latex](a44)--(a58);
        \draw[very thick,-latex](a58)--(a52);
        \draw[very thick,-latex](a43)--(a58);
        \draw[very thick,red](a43)--(a62);
        \draw[very thick,red](a44)--(a60);
        \draw[very thick,-latex](a58)--(a47);
        \draw[very thick,-latex](a54)--(a44);
        \draw[very thick,red](a54)--(a58);
        \draw[very thick,-latex](a63)--(a58);
        \draw[very thick,red](a53)--(a58);
        \draw[very thick,-latex](a53)--(a43);
        \draw[very thick,-latex](a48)--(a60);
        \draw[very thick,-latex](a60)--(a54);
        \draw[very thick,-latex](a47)--(a60);
        \draw[very thick,red](a47)--(a63);
        \draw[very thick,-latex](a54)--(a40);
        \draw[very thick,red](a54)--(a56);
        \draw[very thick,-latex](a63)--(a56);
        \draw[very thick,red](a53)--(a56);
        \draw[very thick,-latex](a53)--(a39);
        \draw[very thick,red](a40)--(a60);
        \draw[very thick,-latex](a56)--(a47);
        \draw[very thick,-latex](a40)--(a56);
        \draw[very thick,-latex](a56)--(a50);
        \draw[very thick,-latex](a39)--(a56);
        \draw[very thick,red](a39)--(a61);
        \draw[very thick,red](a22)--(a26);
        \draw[very thick,-latex](a22)--(a31);
        \draw[very thick,-latex](a31)--(a26);
        \draw[very thick,red](a31)--(a35);
        \draw[very thick,-latex](a13)--(a17);
        \draw[very thick,-latex](a4)--(a13);
        \draw[very thick,red](a4)--(a17);
        \draw[very thick,-latex](a4)--(a8);
        \draw[very thick,-latex](a22)--(a24);
        \draw[very thick,red](a22)--(a33);
        \draw[very thick,-latex](a31)--(a33);
        \draw[very thick,red](a13)--(a15);
        \draw[very thick,-latex](a13)--(a6);
        \draw[very thick,red](a4)--(a6);
        \draw[very thick,-latex](a27)--(a22);
        \draw[very thick,red](a27)--(a31);
        \draw[very thick,-latex](a36)--(a31);
        \draw[very thick,-latex](a18)--(a13);
        \draw[very thick,red](a9)--(a13);
        \draw[very thick,-latex](a9)--(a4);
        \draw[very thick,red](a24)--(a27);
        \draw[very thick,-latex](a24)--(a33);
        \draw[very thick,-latex](a33)--(a27);
        \draw[very thick,red](a33)--(a36);
        \draw[very thick,-latex](a15)--(a18);
        \draw[very thick,-latex](a6)--(a15);
        \draw[very thick,red](a6)--(a18);
        \draw[very thick,-latex](a6)--(a9);
        \draw[very thick,-latex](a27)--(a20);
        \draw[very thick,red](a27)--(a29);
        \draw[very thick,-latex](a36)--(a29);
        \draw[very thick,-latex](a18)--(a11);
        \draw[very thick,red](a9)--(a11);
        \draw[very thick,-latex](a9)--(a2);
        \draw[very thick,-latex](a20)--(a24);
        \draw[very thick,red](a20)--(a33);
        \draw[very thick,-latex](a29)--(a33);
        \draw[very thick,red](a11)--(a15);
        \draw[very thick,-latex](a11)--(a6);
        \draw[very thick,red](a2)--(a6);
        \draw[very thick,red](a20)--(a25);
        \draw[very thick,-latex](a20)--(a29);
        \draw[very thick,-latex](a29)--(a25);
        \draw[very thick,red](a29)--(a34);
        \draw[very thick,-latex](a11)--(a16);
        \draw[very thick,-latex](a2)--(a11);
        \draw[very thick,red](a2)--(a16);
        \draw[very thick,-latex](a2)--(a7);
        \draw[very thick,-latex](a52)--(a62);
        \draw[very thick,-latex](a57)--(a52);
        \draw[very thick,-latex](a51)--(a62);
        \draw[very thick,red](a41)--(a62);
        \draw[very thick,red](a52)--(a63);
        \draw[very thick,-latex](a62)--(a53);
        \draw[very thick,-latex](a54)--(a42);
        \draw[very thick,red](a54)--(a57);
        \draw[very thick,-latex](a63)--(a57);
        \draw[very thick,red](a53)--(a57);
        \draw[very thick,-latex](a53)--(a41);
        \draw[very thick,-latex](a54)--(a63);
        \draw[very thick,-latex](a59)--(a54);
        \draw[very thick,-latex](a53)--(a63);
        \draw[very thick,red](a45)--(a63);
        \draw[very thick,-latex](a54)--(a38);
        \draw[very thick,red](a54)--(a55);
        \draw[very thick,-latex](a63)--(a55);
        \draw[very thick,red](a53)--(a55);
        \draw[very thick,-latex](a53)--(a37);
        \draw[very thick,red](a50)--(a63);
        \draw[very thick,-latex](a61)--(a53);
        \draw[very thick,-latex](a50)--(a61);
        \draw[very thick,-latex](a55)--(a50);
        \draw[very thick,-latex](a49)--(a61);
        \draw[very thick,red](a37)--(a61);
        \draw[very thick,red](a21)--(a26);
        \draw[very thick,-latex](a26)--(a35);
        \draw[very thick,-latex](a30)--(a26);
        \draw[very thick,red](a30)--(a35);
        \draw[very thick,-latex](a12)--(a17);
        \draw[very thick,-latex](a8)--(a17);
        \draw[very thick,red](a3)--(a17);
        \draw[very thick,-latex](a3)--(a8);
        \draw[very thick,-latex](a26)--(a27);
        \draw[very thick,red](a26)--(a36);
        \draw[very thick,-latex](a35)--(a36);
        \draw[very thick,red](a17)--(a18);
        \draw[very thick,-latex](a17)--(a9);
        \draw[very thick,red](a8)--(a9);
        \draw[very thick,-latex](a27)--(a21);
        \draw[very thick,red](a27)--(a30);
        \draw[very thick,-latex](a36)--(a30);
        \draw[very thick,-latex](a18)--(a12);
        \draw[very thick,red](a9)--(a12);
        \draw[very thick,-latex](a9)--(a3);
        \draw[very thick,red](a23)--(a27);
        \draw[very thick,-latex](a27)--(a36);
        \draw[very thick,-latex](a32)--(a27);
        \draw[very thick,red](a32)--(a36);
        \draw[very thick,-latex](a14)--(a18);
        \draw[very thick,-latex](a9)--(a18);
        \draw[very thick,red](a5)--(a18);
        \draw[very thick,-latex](a5)--(a9);
        \draw[very thick,-latex](a27)--(a19);
        \draw[very thick,red](a27)--(a28);
        \draw[very thick,-latex](a36)--(a28);
        \draw[very thick,-latex](a18)--(a10);
        \draw[very thick,red](a9)--(a10);
        \draw[very thick,-latex](a9)--(a1);
        \draw[very thick,-latex](a25)--(a27);
        \draw[very thick,red](a25)--(a36);
        \draw[very thick,-latex](a34)--(a36);
        \draw[very thick,red](a16)--(a18);
        \draw[very thick,-latex](a16)--(a9);
        \draw[very thick,red](a7)--(a9);
        \draw[very thick,red](a19)--(a25);
        \draw[very thick,-latex](a25)--(a34);
        \draw[very thick,-latex](a28)--(a25);
        \draw[very thick,red](a28)--(a34);
        \draw[very thick,-latex](a10)--(a16);
        \draw[very thick,-latex](a7)--(a16);
        \draw[very thick,red](a1)--(a16);
        \draw[very thick,-latex](a1)--(a7);
        \draw[very thick,-latex](a42)--(a57);
        \draw[very thick,-latex](a41)--(a57);
        \draw[very thick,red](a42)--(a59);
        \draw[very thick,-latex](a57)--(a45);
        \draw[very thick,-latex](a46)--(a59);
        \draw[very thick,-latex](a45)--(a59);
        \draw[very thick,red](a38)--(a59);
        \draw[very thick,-latex](a55)--(a45);
        \draw[very thick,-latex](a38)--(a55);
        \draw[very thick,-latex](a37)--(a55);
        \draw[very thick,-latex](a21)--(a30);
        \draw[very thick,-latex](a3)--(a12);
        \draw[very thick,-latex](a21)--(a23);
        \draw[very thick,red](a21)--(a32);
        \draw[very thick,-latex](a30)--(a32);
        \draw[very thick,red](a12)--(a14);
        \draw[very thick,-latex](a12)--(a5);
        \draw[very thick,red](a3)--(a5);
        \draw[very thick,-latex](a23)--(a32);
        \draw[very thick,-latex](a5)--(a14);
        \draw[very thick,-latex](a19)--(a23);
        \draw[very thick,red](a19)--(a32);
        \draw[very thick,-latex](a28)--(a32);
        \draw[very thick,red](a10)--(a14);
        \draw[very thick,-latex](a10)--(a5);
        \draw[very thick,red](a1)--(a5);
        \draw[very thick,-latex](a19)--(a28);
        \draw[very thick,-latex](a1)--(a10);
        \draw (1.0,0.0,1.0) node(a15){$\overline{2}$};
        \draw (5.0,0.0,1.5) node(a59){$6$};
        \draw[very thick , blue , -latex](2.7,1,1.5) -- (3.3,1,1.5);
    \end{tikzpicture}
    \caption{$3d$ printing for phase $D$ of $Q^{1,1,1} /\mathbbm{Z}_{2}$.}
    \label{3d_printing_phase_D}
\end{figure}
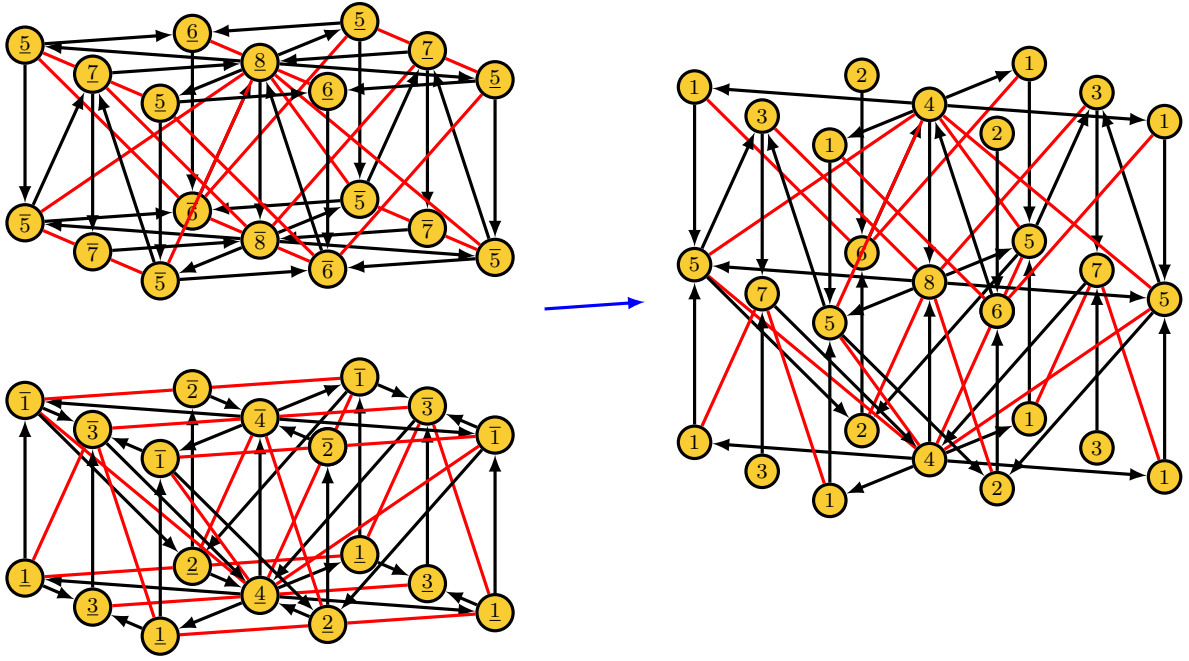

\vspace{-.5cm}
\begin{align}
    \begin{array}{rrclcrcl} & & J & &\phantom{abcde}& & E & \\ 
        \Lambda_{16}^{+}: & X_{64}^{+}X_{41}^{--} & - & X_{64}^{-}X_{41}^{-+}  &&  X_{15}^{+}X_{52}^{+}X_{26}^{-} & - & X_{15}^{-}X_{52}^{+}X_{26}^{+}\\ 
        \Lambda_{16}^{-}: & X_{64}^{-}X_{41}^{++} & - & X_{64}^{+}X_{41}^{+-}  &&  X_{15}^{+}X_{52}^{-}X_{26}^{-} & - & X_{15}^{-}X_{52}^{-}X_{26}^{+}\\ 
        \Lambda_{17}^{+}: & X_{74}^{-}X_{41}^{+-} & - & X_{74}^{+}X_{41}^{--}  &&  X_{15}^{+}X_{53}^{+}X_{37}^{-} & - & X_{15}^{-}X_{53}^{+}X_{37}^{+}\\ 
        \Lambda_{17}^{-}: & X_{74}^{+}X_{41}^{-+} & - & X_{74}^{-}X_{41}^{++}  &&  X_{15}^{+}X_{53}^{-}X_{37}^{-} & - & X_{15}^{-}X_{53}^{-}X_{37}^{+}\\ 
        \Lambda_{28}^{+}: & X_{85}^{--}X_{52}^{+} & - & X_{85}^{+-}X_{52}^{-}  &&  X_{26}^{+}X_{64}^{+}X_{48}^{-} & - & X_{26}^{-}X_{64}^{+}X_{48}^{+}\\ 
        \Lambda_{28}^{-}: & X_{85}^{++}X_{52}^{-} & - & X_{85}^{-+}X_{52}^{+}  &&  X_{26}^{+}X_{64}^{-}X_{48}^{-} & - & X_{26}^{-}X_{64}^{-}X_{48}^{+}\\ 
        \Lambda_{38}^{+}: & X_{85}^{-+}X_{53}^{-} & - & X_{85}^{--}X_{53}^{+}  &&  X_{37}^{+}X_{74}^{+}X_{48}^{-} & - & X_{37}^{-}X_{74}^{+}X_{48}^{+}\\ 
        \Lambda_{38}^{-}: & X_{85}^{+-}X_{53}^{+} & - & X_{85}^{++}X_{53}^{-}  &&  X_{37}^{+}X_{74}^{-}X_{48}^{-} & - & X_{37}^{-}X_{74}^{-}X_{48}^{+}\\    
        \Lambda_{45}^{+++}: & X_{52}^{-}X_{26}^{-}X_{64}^{-} & - & X_{53}^{-}X_{37}^{-}X_{74}^{-}  &&  X_{48}^{+}X_{85}^{++} & - & X_{41}^{++}X_{15}^{+}\\ 
        \Lambda_{45}^{++-}: & X_{52}^{-}X_{26}^{+}X_{64}^{-} & - & X_{53}^{-}X_{37}^{+}X_{74}^{-}  &&  X_{41}^{++}X_{15}^{-} & - & X_{48}^{-}X_{85}^{++}\\ 
        \Lambda_{45}^{+-+}: & X_{53}^{+}X_{37}^{-}X_{74}^{-} & - & X_{52}^{-}X_{26}^{-}X_{64}^{+}  &&  X_{48}^{+}X_{85}^{+-} & - & X_{41}^{+-}X_{15}^{+}\\ 
        \Lambda_{45}^{+--}: & X_{53}^{+}X_{37}^{+}X_{74}^{-} & - & X_{52}^{-}X_{26}^{+}X_{64}^{+}  &&  X_{41}^{+-}X_{15}^{-} & - & X_{48}^{-}X_{85}^{+-}\\ 
        \Lambda_{45}^{-++}: & X_{53}^{-}X_{37}^{-}X_{74}^{+} & - & X_{52}^{+}X_{26}^{-}X_{64}^{-}  &&  X_{48}^{+}X_{85}^{-+} & - & X_{41}^{-+}X_{15}^{+}\\ 
        \Lambda_{45}^{-+-}: & X_{53}^{-}X_{37}^{+}X_{74}^{+} & - & X_{52}^{+}X_{26}^{+}X_{64}^{-}  &&  X_{41}^{-+}X_{15}^{-} & - & X_{48}^{-}X_{85}^{-+}\\ 
        \Lambda_{45}^{--+}: & X_{52}^{+}X_{26}^{-}X_{64}^{+} & - & X_{53}^{+}X_{37}^{-}X_{74}^{+}  &&  X_{48}^{+}X_{85}^{--} & - & X_{41}^{--}X_{15}^{+}\\ 
        \Lambda_{45}^{---}: & X_{52}^{+}X_{26}^{+}X_{64}^{+} & - & X_{53}^{+}X_{37}^{+}X_{74}^{+}  &&  X_{41}^{--}X_{15}^{-} & - & X_{48}^{-}X_{85}^{--}
    \end{array}
\end{align}

Until now, 5 toric phases for $Q^{1,1,1}/\mathbb{Z}_2$ had been identified in the literature \cite{Franco:2016nwv}. As expected from having the same underlying CY$_4$, all of them have been shown to be related by triality. However, the theory we have just constructed, which we denote phase $D$, does not correspond to any of the known phases.\footnote{In \cite{Franco:2016nwv}, a different theory was called phase $D$. Motivated by the general classification of $Q^{1,1,1}/\mathbb{Z}_2$ phases that will be presented in the next section, we have decided to change the nomenclature.} It is then natural to ask whether this theory is related to the other phases by triality. In addition, it would be interesting to map the space of toric triality phases for this geometry. Below we address both questions.

\subsection{The Triality Web}

\label{section_triality_web}

\begin{figure} 
\begin{center}
    \tdplotsetmaincoords{80}{118} 
    \tdplotsetrotatedcoords{0}{0}{140} 
    \begin{tikzpicture}[scale=2.5,tdplot_rotated_coords] 
        \tikzstyle{every node}=[circle,very thick,fill=yellow2,draw,inner sep=2pt,font=\scriptsize]
        \draw (0.0,0.0,0.0) node(a1){$1$};
        \draw (0.0,0.0,2.0) node(a2){$1$};
        \draw (0.0,2.0,0.0) node(a3){$1$};
        \draw (0.0,2.0,2.0) node(a4){$1$};
        \draw (2.0,0.0,0.0) node(a5){$1$};
        \draw (2.0,0.0,2.0) node(a6){$1$};
        \draw (2.0,2.0,0.0) node(a7){$1$};
        \draw (2.0,2.0,2.0) node(a8){$1$};
        \draw (1.0,0.0,0.0) node(a9){$2$};
        \draw (1.0,0.0,2.0) node(a10){$2$};
        \draw (1.0,2.0,0.0) node(a11){$2$};
        \draw (1.0,2.0,2.0) node(a12){$2$};
        \draw (0.0,1.0,0.0) node(a13){$3$};
        \draw (0.0,1.0,2.0) node(a14){$3$};
        \draw (2.0,1.0,0.0) node(a15){$3$};
        \draw (2.0,1.0,2.0) node(a16){$3$};
        \draw (1.0,1.0,0.0) node(a17){$4$};
        \draw (1.0,1.0,2.0) node(a18){$4$};
        \draw (0.0,0.0,1.0) node(a19){$5$};
        \draw (0.0,2.0,1.0) node(a20){$5$};
        \draw (2.0,0.0,1.0) node(a21){$5$};
        \draw (2.0,2.0,1.0) node(a22){$5$};
        \draw (1.0,0.0,1.0) node(a23){$6$};
        \draw (1.0,2.0,1.0) node(a24){$6$};
        \draw (0.0,1.0,1.0) node(a25){$7$};
        \draw (2.0,1.0,1.0) node(a26){$7$};
        \draw (1.0,1.0,1.0) node(a27){$8$};
        \draw[very thick,-latex](a16)--(a8);
        \draw[very thick,red](a8)--(a22);
        \draw[very thick,-latex](a22)--(a26);
        \draw[very thick,red](a7)--(a22);
        \draw[very thick,-latex](a15)--(a7);
        \draw[very thick,-latex](a8)--(a12);
        \draw[very thick,-latex](a7)--(a11);
        \draw[very thick,red](a12)--(a16);
        \draw[very thick,-latex](a24)--(a22);
        \draw[very thick,red](a24)--(a26);
        \draw[very thick,red](a11)--(a15);
        \draw[very thick,-latex](a18)--(a12);
        \draw[very thick,-latex](a12)--(a24);
        \draw[very thick,-latex](a24)--(a27);
        \draw[very thick,-latex](a11)--(a24);
        \draw[very thick,-latex](a17)--(a11);
        \draw[very thick,red](a12)--(a14);
        \draw[very thick,-latex](a24)--(a20);
        \draw[very thick,red](a24)--(a25);
        \draw[very thick,red](a11)--(a13);
        \draw[very thick,-latex](a4)--(a12);
        \draw[very thick,-latex](a3)--(a11);
        \draw[very thick,-latex](a14)--(a4);
        \draw[very thick,red](a4)--(a20);
        \draw[very thick,-latex](a20)--(a25);
        \draw[very thick,red](a3)--(a20);
        \draw[very thick,-latex](a13)--(a3);
        \draw[very thick,-latex](a16)--(a6);
        \draw[very thick,-latex](a26)--(a16);
        \draw[very thick,-latex](a21)--(a26);
        \draw[very thick,-latex](a26)--(a15);
        \draw[very thick,-latex](a15)--(a5);
        \draw[very thick,-latex](a16)--(a18);
        \draw[very thick,-latex](a15)--(a17);
        \draw[very thick,red](a10)--(a16);
        \draw[very thick,-latex](a27)--(a26);
        \draw[very thick,red](a23)--(a26);
        \draw[very thick,red](a9)--(a15);
        \draw[very thick,-latex](a18)--(a10);
        \draw[very thick,red](a18)--(a27);
        \draw[very thick,-latex](a23)--(a27);
        \draw[very thick,red](a17)--(a27);
        \draw[very thick,-latex](a17)--(a9);
        \draw[very thick,red](a10)--(a14);
        \draw[very thick,-latex](a27)--(a25);
        \draw[very thick,red](a23)--(a25);
        \draw[very thick,red](a9)--(a13);
        \draw[very thick,-latex](a14)--(a18);
        \draw[very thick,-latex](a13)--(a17);
        \draw[very thick,-latex](a14)--(a2);
        \draw[very thick,-latex](a25)--(a14);
        \draw[very thick,-latex](a19)--(a25);
        \draw[very thick,-latex](a25)--(a13);
        \draw[very thick,-latex](a13)--(a1);
        \draw[very thick,red](a6)--(a21);
        \draw[very thick,red](a5)--(a21);
        \draw[very thick,-latex](a6)--(a10);
        \draw[very thick,-latex](a5)--(a9);
        \draw[very thick,-latex](a23)--(a21);
        \draw[very thick,-latex](a10)--(a23);
        \draw[very thick,-latex](a9)--(a23);
        \draw[very thick,-latex](a23)--(a19);
        \draw[very thick,-latex](a2)--(a10);
        \draw[very thick,-latex](a1)--(a9);
        \draw[very thick,red](a2)--(a19);
        \draw[very thick,red](a1)--(a19);
    \end{tikzpicture} 
    \end{center}
    \caption{Periodic quiver for phase $A$ of $Q^{1,1,1}/\mathbbm{Z}_{2}$.}
        \label{phase_A}
\end{figure}
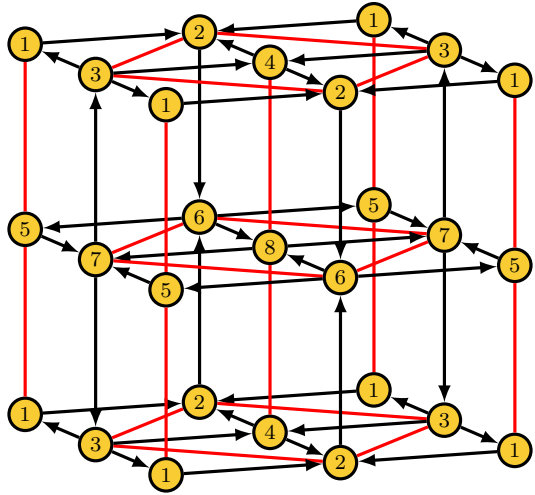

Let us first consider phase $A$ \cite{Franco:2016nwv}, whose periodic quiver is shown in \fref{phase_A}. The corresponding $J$- and $E$-terms are:
\begin{align}
\renewcommand{\arraystretch}{1.2}
    \begin{array}{rrclcrcl} & & J & &\phantom{abcde}& & E & \\ 
    \Lambda_{15}^{+}: & X_{57}^{+}X_{73}^{-}X_{31}^{-} & - & X_{57}^{-}X_{73}^{-}X_{31}^{+}  &&  X_{12}^{+}X_{26}^{+}X_{65}^{-} & - & X_{12}^{-}X_{26}^{+}X_{65}^{+}\\ 
    \Lambda_{15}^{-}: & X_{57}^{+}X_{73}^{+}X_{31}^{-} & - & X_{57}^{-}X_{73}^{+}X_{31}^{+}  &&  X_{12}^{-}X_{26}^{-}X_{65}^{+} & - & X_{12}^{+}X_{26}^{-}X_{65}^{-}\\
    \Lambda_{48}^{+}: & X_{87}^{+}X_{73}^{-}X_{34}^{-} & - & X_{87}^{-}X_{73}^{-}X_{34}^{+}  &&  X_{42}^{-}X_{26}^{+}X_{68}^{+} & - & X_{42}^{+}X_{26}^{+}X_{68}^{-}\\
    \Lambda_{48}^{-}: & X_{87}^{+}X_{73}^{+}X_{34}^{-} & - & X_{87}^{-}X_{73}^{+}X_{34}^{+}  &&  X_{42}^{+}X_{26}^{-}X_{68}^{-} & - & X_{42}^{-}X_{26}^{-}X_{68}^{+} \\
    \Lambda_{32}^{++}: & X_{26}^{+}X_{68}^{-}X_{87}^{-}X_{73}^{-} & - & X_{26}^{-}X_{68}^{-}X_{87}^{-}X_{73}^{+}  &&  X_{31}^{+}X_{12}^{+} & - & X_{34}^{+}X_{42}^{+}\\ 
    \Lambda_{32}^{+-}: & X_{26}^{+}X_{68}^{+}X_{87}^{-}X_{73}^{-} & - & X_{26}^{-}X_{68}^{+}X_{87}^{-}X_{73}^{+}  &&  X_{34}^{+}X_{42}^{-} & - & X_{31}^{-}X_{12}^{+}\\ 
    \Lambda_{32}^{-+}: & X_{26}^{+}X_{68}^{-}X_{87}^{+}X_{73}^{-} & - & X_{26}^{-}X_{68}^{-}X_{87}^{+}X_{73}^{+}  &&  X_{34}^{-}X_{42}^{+} & - & X_{31}^{+}X_{12}^{-}\\ 
    \Lambda_{32}^{--}: & X_{26}^{+}X_{68}^{+}X_{87}^{+}X_{73}^{-} & - & X_{26}^{-}X_{68}^{+}X_{87}^{+}X_{73}^{+}  &&  X_{31}^{-}X_{12}^{-} & - & X_{34}^{-}X_{42}^{-}\\ 
    \Lambda_{67}^{++}: & X_{73}^{+}X_{31}^{-}X_{12}^{-}X_{26}^{-} & - & X_{73}^{-}X_{31}^{-}X_{12}^{-}X_{26}^{+}  &&  X_{68}^{+}X_{87}^{+} & - & X_{65}^{+}X_{57}^{+}\\ 
    \Lambda_{67}^{+-}: & X_{73}^{+}X_{31}^{+}X_{12}^{-}X_{26}^{-} & - & X_{73}^{-}X_{31}^{+}X_{12}^{-}X_{26}^{+}  &&  X_{65}^{+}X_{57}^{-} & - & X_{68}^{-}X_{87}^{+}\\ 
    \Lambda_{67}^{-+}: & X_{73}^{+}X_{31}^{-}X_{12}^{+}X_{26}^{-} & - & X_{73}^{-}X_{31}^{-}X_{12}^{+}X_{26}^{+}  &&  X_{65}^{-}X_{57}^{+} & - & X_{68}^{+}X_{87}^{-}\\ 
    \Lambda_{67}^{--}: & X_{73}^{+}X_{31}^{+}X_{12}^{+}X_{26}^{-} & - & X_{73}^{-}X_{31}^{+}X_{12}^{+}X_{26}^{+}  &&  X_{68}^{-}X_{87}^{-} & - & X_{65}^{-}X_{57}^{-}
    \end{array}
\end{align}

We will map the web of toric phases of $Q^{1,1,1}/\mathbb{Z}_2$ as follows. Starting from phase $A$, we will perform triality or inverse triality on every node that results in a toric theory, namely in a theory that is described by a periodic quiver. Those nodes are characterized by having two incoming chiral arrows, in the case of triality, or two outgoing chiral arrows, in the case of inverse triality. We will iterate this process on the resulting phases after exhausting all possibilities. It is reasonable to assume that this procedure generates all toric phases, i.e. that, up to relabeling of nodes, all toric phases are connected in this way. It is in principle possible that some different toric phases can only be connected by a sequence of triality transformations that passes through non-toric phases. All existing classifications of toric phases in the similar context of CY 3-folds (see e.g. \cite{Cachazo:2001sg,Franco:2003ja,Franco:2004jz}), for which the connection is through Seiberg duality, suggest that this never occurs.

Proceeding as explained, we find 14 toric phases for this geometry, namely we discover 9 new phases that did not appear in previous studies of $Q^{1,1,1}/\mathbb{Z}_2$ \cite{Franco:2016nwv}. We distinguish phases modulo relabeling of nodes. The periodic quivers for all these theories are presented in Appendix \sref{section_toric_phases_Q111/Z2}. In Table \ref{phases_Q111_Z2} we collect some basic information characterizing these theories that facilitates their comparison. The phases have been ordered according to the total number of Fermi fields, which we denote $F$. For each phase, we provide a sequence of triality transformations connecting it to phase $A$ in the form shown in \fref{phase_A}.\footnote{This sequence is of course not unique.} In the ``Fermi Multiplicities" column we give the multiplicity of Fermi fields for the 8 nodes in the quiver. For example, $4\times \mathbf{2} + 4\times \mathbf{4}$ indicates that the corresponding theory has 4 nodes with 2 Fermis and 4 nodes with 4 Fermis. Finally, $m_{\mbox{(0,0,0)}}$ is the number of brick matchings associated to the central point in the $Q^{1,1,1}/\mathbb{Z}_2$ toric diagram (see \fref{F0_to_Q111Z2}). 

\begin{table}[H]
\centering
\begin{tabular}{|c|c|c|c|c|} \hline Name & Triality Path & F & Fermi Multiplicities & $m_{(0,0,0)}$ \\ \hline 
       \ \ \, A   \ $\star$ &  & 12 & $4\times \mathbf{2}   + 4\times \mathbf{4}$   & 10 \\ 
       \ \ \,  B   \ $\star$  & 1 & 12 & $4\times \mathbf{2}   + 4\times \mathbf{4}$   & 11 \\ 
        C   & 15 & 16 & $4\times \mathbf{2}   + 2\times \mathbf{4}   + 2\times \mathbf{8}$   & 12 \\
        D   & 147 & 16 & $4\times \mathbf{2}   + 2\times \mathbf{4}   + 2\times \mathbf{8}$   & 13 \\
       \ \ \,  E   \ $\star$ & 14 & 16 & $2\times \mathbf{2}   + 5\times \mathbf{4}   + 1\times \mathbf{8}$   & 15 \\ 
       \ \ \,  F   \ $\star$ & 12 & 16 & $2\times \mathbf{2}   + 4\times \mathbf{4}   + 2\times \mathbf{6}$   & 14 \\ 
        G   & 157 & 20 & $3\times \mathbf{2}   + 1\times \mathbf{4}   + 2\times \mathbf{6}   + 1\times \mathbf{8}   + 1\times \mathbf{10}$   & 13\\ 
        H   & 3 & 20 & $2\times \mathbf{2}   + 4\times \mathbf{4}   + 1\times \mathbf{8}   + 1\times \mathbf{12}$   & 14 \\ 
        I   & 13 & 20 & $2\times \mathbf{2}   + 3\times \mathbf{4}   + 1\times \mathbf{6}   + 1\times \mathbf{8}   + 1\times \mathbf{10}$    & 14 \\ 
        J   & 132 & 24 & $2\times \mathbf{2}   + 1\times \mathbf{4}   + 2\times \mathbf{6}   + 2\times \mathbf{8}   + 1\times \mathbf{12}$   \ \, & 19 \\ 
        K   & 1572 & 24 & $2\times \mathbf{2}   + 4\times \mathbf{6}   + 2\times \mathbf{10}$   & 14 \\ 
        L   & 142 & 24 & $6\times \mathbf{4}   + 2\times \mathbf{12}$    & 19\\ 
       \ \ \,  M   \ $\star$ & 36 & 28 & $6\times \mathbf{4}   + 2\times \mathbf{16}$   & 16 \\ 
        N   & 136 & 28 & $4\times \mathbf{4}   + 4\times \mathbf{10}$   & 16\\ \hline
\end{tabular}
\caption{Basic information regarding the 14 toric phases of $\mathbb{Q}^{1,1,1}/\mathbb{Z}_2$.}
\label{phases_Q111_Z2}
\end{table}
 
We indicate the phases that have previously appeared in \cite{Franco:2016nwv} with a star. As anticipated, we have labeled theories differently. Phases $A$, $B$, $E$, $F$ and $M$ here correspond to phases $A$, $B$, $D$, $C$ and $S$ of \cite{Franco:2016nwv}, respectively.
    
Finally, Table \ref{connection_phases_Q111/Z2} summarizes how the 14 phases are interconnected by triality.  In this table, for each of the phases we consider the labeling of nodes given in Appendix \sref{section_toric_phases_Q111/Z2}, i.e. the one obtained by acting on \fref{phase_A} with the sequences of trialities in Table \ref{phases_Q111_Z2}. In each column, we indicate the phases obtained by acting with triality or inverse triality on the corresponding node. The underline indicates phases obtained by inverse triality while the blanks correspond to the nodes for which triality does not give a toric phase. Some entries contain a single theory, because in those cases only triality or inverse triality, but not both, result in a toric phase.    

\begin{table}[H]
\centering
\begin{tabular}{|C{.7cm}|C{1.1cm}|C{1.1cm}|C{1.1cm}|C{1.1cm}|C{1.1cm}|C{1.1cm}|C{1.1cm}|C{1.1cm}|} \hline &1&2&3&4&5&6&7&8 \\ \hline
A & B , \underline{B} & \underline{H} & H & B , \underline{B} & B , \underline{B} & H & \underline{H} & B , \underline{B} \\ 
B & B , \underline{A} & F , \underline{E} & \underline{I} & E , \underline{F} & C & I & \underline{C} & A , \underline{B} \\
C & \underline{B} & G , \underline{D} &  & H , \underline{I} & \underline{B} &  & G , \underline{D} & H , \underline{I} \\ 
D & C , \underline{G} & E &  & C , \underline{G} & G , \underline{C} &  & \underline{E} & G , \underline{C} \\ 
E & F , \underline{B} & \underline{\mbox{L}} &  & F , \underline{B} & H & J & \underline{D} & H \\ 
F & I & E , \underline{B} & \underline{G} &  & G &  & \underline{I} & B , \underline{E} \\ 
G & \underline{F} & K , \underline{G} &  & I , \underline{J} &  &  & D , \underline{C} &  \\ 
H & \underline{E} &  & \underline{A} & \underline{E} & I , \underline{C} & M &  & I , \underline{C} \\ 
I & \underline{F} & J , \underline{G} & \underline{B} &  &  & \underline{N} &  & C , \underline{H} \\ 
J &  & G , \underline{I} & \underline{E} &  &  &  &  & G , \underline{I} \\ 
K &  & G , \underline{G} &  &  &  &  & G , \underline{G} &  \\ 
\mbox{L} & E & \underline{E} &  & E & \underline{E} &  & E & \underline{E} \\
M & \underline{H} &  & \underline{H} & \underline{H} & \underline{H} & \underline{H} &  & \underline{H} \\
N & I &  & I &  &  & \underline{I} &  & \underline{I} \\ \hline
\end{tabular}
\caption{Triality connections between the 14 toric phases of $\mathbb{Q}^{1,1,1}/\mathbb{Z}_2$.}
\label{connection_phases_Q111/Z2}
\end{table}

\section{Consistency and Reduction}

\label{section_consistency_and_reduction}

An important question when constructing brane brick models is whether they are {\it consistent} or, equivalently, {\it irreducible}. The analogue problem for brane tilings and, more generally, bipartite graphs on Riemann surfaces has been extensively studied (see e.g. \cite{MR2908565,Hanany:2005ss,Gulotta:2008ef,MR2809653,MR2805177,MR2911380,Franco:2012mm,Hanany:2015tgh} and references therein).  In this section we take the first steps on this issue for brane brick models, proposing natural generalizations of the brane tiling case.

\subsection{Diagnosing Reducibility}

There are various equivalent criteria for identifying inconsistent, i.e. reducible, brane tilings. Arguably one of the simplest to implement is given by the mismatch between the number of gauge groups in the quiver and the normalized area of the corresponding toric diagram \cite{Franco:2017jeo}. Several explicit examples of inconsistent brane tilings can be found in \cite{Davey:2009bp}. This condition generalizes straightforwardly to brane brick models: we claim that a brane brick model is inconsistent whenever its number of gauge groups is larger than the normalized volume of the toric diagram computed from it.

\subsection{Reducing Brane Brick Models}

Given an inconsistent brane brick model, it can be turned into a consistent one by {\it reduction}. Reduction is defined as a process that reduces the number of gauge groups while preserving the toric diagram. Below we discuss two reduction mechanisms, which generalize similar operations for brane tilings.

\subsubsection*{Higgsing}

The first method for reduction is by higgsings that preserve the toric diagram.\footnote{Here and in what follows, we have in mind classical higgsing.} The number of bifundamental chiral fields to be turned on in order to completely reduce the brane brick model equals the difference between the number of gauge groups and the volume of the toric diagram. In general, there might be multiple sets of bifundamentals that achieve the desired reduction. Searching for such higgsings can be systematized exploiting the correspondence between brick matchings, which corresponds to GLSM fields, and fields in the quiver. The map between these two sets of objects is often encoded in the so-called $P$-matrix \cite{Franco:2015tya}. It is important to emphasize that the necessary higgsings might only become available after performing some triality transformation(s). In \sref{section_D3_from_conifold} we will present an example illustrating this phenomenon.

\subsubsection*{Generalized Bubble Reduction}

An alternative mechanism for reduction can be identified by considering the familiar case of brane tilings. Inconsistent brane tilings can also be reduced using {\it bubble reduction} \cite{Postnikov:2006kva,ArkaniHamed:2012nw}. In terms of the dual quiver, a bubble corresponds to a node with one incoming and one outgoing bifundamental arrows. When all ranks are equal, this corresponds to an $N_f=N_c$ gauge group. Bubble reduction is equivalent to formally applying Seiberg duality to such a node. The dualized node disappears, there are no magnetic flavors and we are only left with the corresponding mesons. It is important to remark that an inconsistent brane tiling might not exhibit explicit bubbles. In general, it is necessary to apply Seiberg duality transformations in order to make bubbles appear.

\begin{figure}[ht]
	\centering
	\includegraphics[width=12cm]{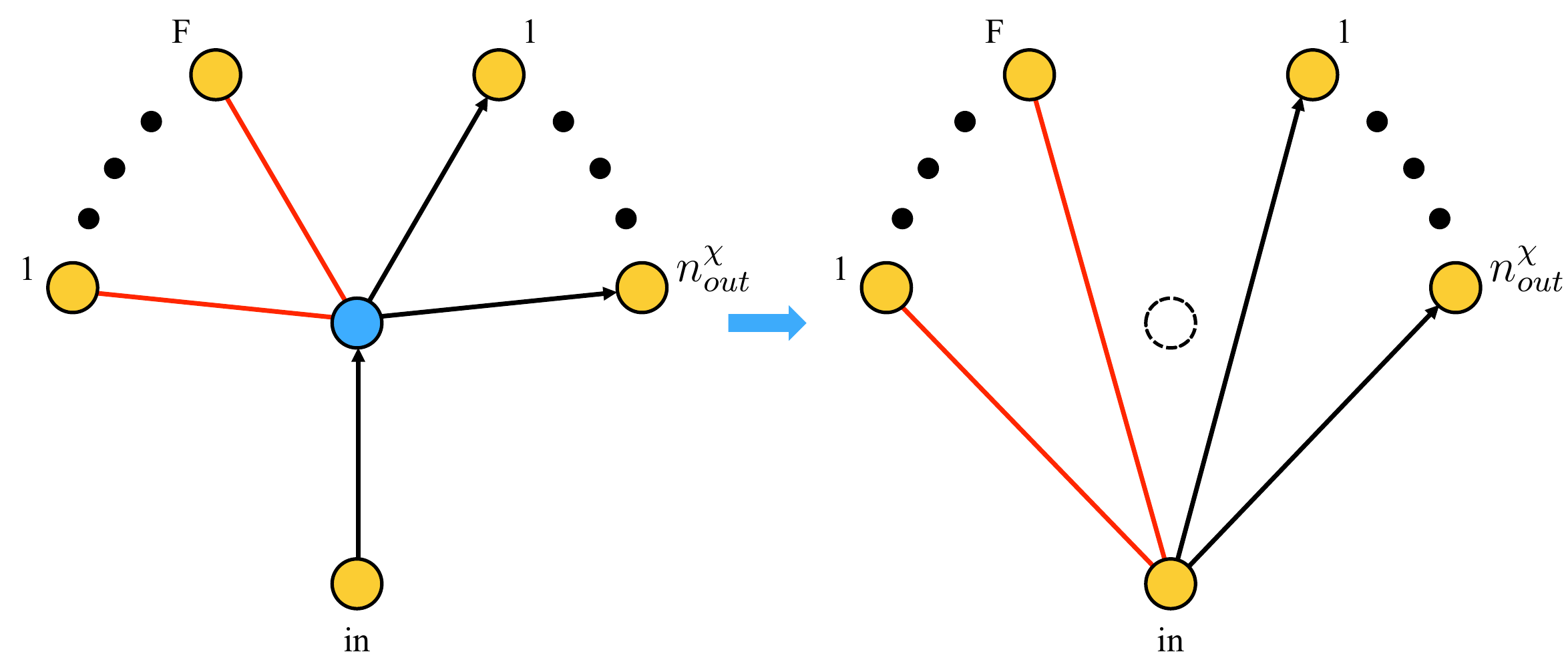}
\caption{A node with a single incoming chiral field corresponds to a bubble. Due to anomaly cancellation, $n^\chi_{out}=F+1$. The bubble can be removed by formal triality. We can alternatively think of this process as giving a non-zero vev to the incoming chiral.}
	\label{2d_bubble_reduction}
\end{figure}

The previous discussion suggests how to generalize the concept of bubble to brane brick models. In this context, a bubble corresponds to a node in the quiver that would disappear when formally applying triality or inverse triality. We refer the reader to \cite{Gadde:2013lxa} for the triality rules. Specifically, a bubble that disappears by acting with triality corresponds to a node with a single incoming chiral arrow and $n^\chi_{out}=F+1$, where $n^\chi_{out}$ and $F$ are the numbers of outgoing chiral arrows and Fermi lines, respectively. Similarly, a bubble that disappears by acting with inverse triality corresponds to a node with a single outgoing chiral arrow and $n^\chi_{in}=F+1$. \fref{2d_bubble_reduction} shows the removal of a bubble using triality. While the node associated to the bubble disappears, triality generates mesons connecting the other nodes. It is interesting to note that exactly the same theory is obtained by higgsing the theory with a vev for the incoming chiral, which results in the merging of the bubble node and the $in$ node.\footnote{The reduction by higgsing discussed in the previous section is, however, a more general operation, not restricted to this particular case.} The elimination of a bubble with inverse triality is completely analogous. 

The example we will present in \sref{section_reduction_3d_printing} suggests that there can be inconsistent brane brick models for which it is impossible to make bubbles explicit by sequences of trialities. Better understanding the conditions under which this is possible is certainly desirable and we leave it for future work.

\subsection{Reduction and $3d$ Printing}

\label{section_reduction_3d_printing}

The previous discussion of consistency applies to general brane brick models. Let us now focus on brane brick models constructed via $3d$ printing. It is clear that $3d$ printing can easily give rise to theories where the number of gauge groups is larger than the volume of the corresponding toric diagram and are hence inconsistent. This generically happens when lifting multiple points in the toric diagram or when, as mentioned in \sref{section_3d_printing_and_geometry}, different perfect matchings are used to lift the a point of the toric diagram in the same direction.

Even in these cases, when combined with reduction, $3d$ printing provides a systematic approach for constructing gauge theories associated to desired toric CY 4-folds. The procedure works as follows. First, an inconsistent brane brick model with the desired toric diagram is generated using $3d$ printing. Finally, this brane tiling is reduced while preserving the toric diagram until producing a consistent theory, i.e. one with the appropriate number of gauge groups.

\subsubsection{An Example}

\label{section_D3_from_conifold}

\begin{figure}[ht]
	\centering
	\includegraphics[width=12cm]{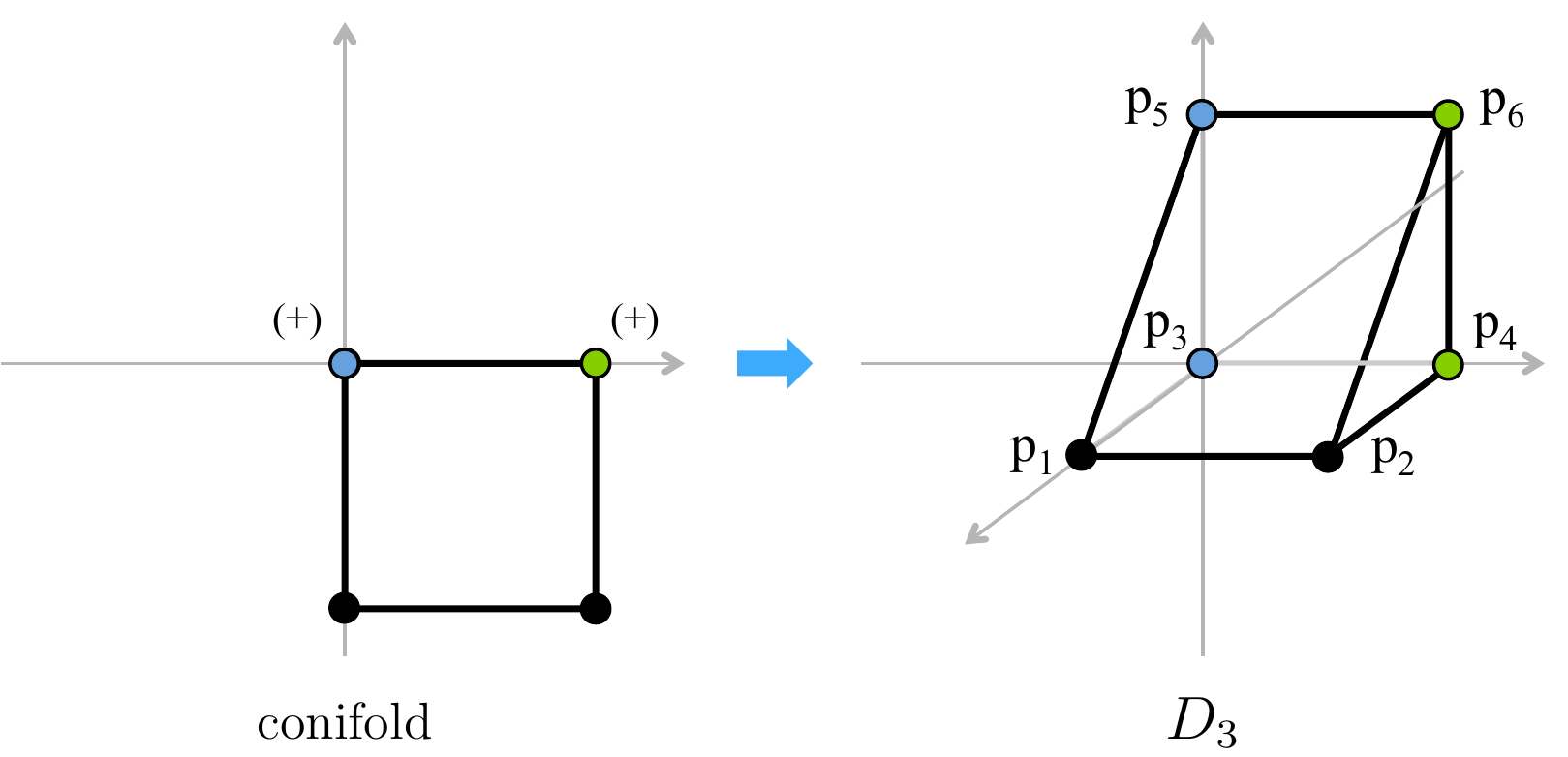}
\caption{The toric diagram for $D_3$, obtained by lifting two perfect matchings of the conifold.}
	\label{conifold_to_D3}
\end{figure}

Let us generate the toric diagram of $D_3$ by lifting two corners of the conifold, as shown in \fref{conifold_to_D3}. This example was previously mentioned in \fref{C3_and_conifold_to_D3}, to illustrate how a given CY$_4$ can be obtained by lifting different CY$_3$'s.

\begin{figure}
        \begin{center}
                \tdplotsetmaincoords{80}{130} 
                \tdplotsetrotatedcoords{0}{0}{111} 
                \begin{tikzpicture}[scale=2.5,tdplot_rotated_coords] 
                    \tikzstyle{every node}=[circle,very thick,fill=yellow2,draw,inner sep=2pt,font=\scriptsize]
                    \draw (0.0,0.0,0.0) node(a1){$\underline{1}$};
                    \draw (0.0,2.0,0.0) node(a2){$\underline{1}$};
                    \draw (1.0,0.0,0.0) node(a3){$\underline{1}$};
                    \draw (1.0,2.0,0.0) node(a4){$\underline{1}$};
                    \draw (0.0,1.0,0.0) node(a5){$\underline{2}$};
                    \draw (1.0,1.0,0.0) node(a6){$\underline{2}$};
                    \draw (0.0,0.0,1.0) node(a7){$\overline{1}$};
                    \draw (0.0,2.0,1.0) node(a8){$\overline{1}$};
                    \draw (1.0,0.0,1.0) node(a9){$\overline{1}$};
                    \draw (1.0,2.0,1.0) node(a10){$\overline{1}$};
                    \draw (0.0,1.0,1.0) node(a11){$\overline{2}$};
                    \draw (1.0,1.0,1.0) node(a12){$\overline{2}$};
                    \draw (0.0,0.0,2.0) node(a13){$\underline{3}$};
                    \draw (0.0,2.0,2.0) node(a14){$\underline{3}$};
                    \draw (1.0,0.0,2.0) node(a15){$\underline{3}$};
                    \draw (1.0,2.0,2.0) node(a16){$\underline{3}$};
                    \draw (0.0,1.0,2.0) node(a17){$\underline{4}$};
                    \draw (1.0,1.0,2.0) node(a18){$\underline{4}$};
                    \draw (0.0,0.0,3.0) node(a19){$\overline{3}$};
                    \draw (0.0,2.0,3.0) node(a20){$\overline{3}$};
                    \draw (1.0,0.0,3.0) node(a21){$\overline{3}$};
                    \draw (1.0,2.0,3.0) node(a22){$\overline{3}$};
                    \draw (0.0,1.0,3.0) node(a23){$\overline{4}$};
                    \draw (1.0,1.0,3.0) node(a24){$\overline{4}$};
                    \draw (3.0,0.0,0.5) node(a25){$1$};
                    \draw (3.0,0.0,2.5) node(a26){$1$};
                    \draw (3.0,2.0,0.5) node(a27){$1$};
                    \draw (3.0,2.0,2.5) node(a28){$1$};
                    \draw (4.0,0.0,0.5) node(a29){$1$};
                    \draw (4.0,0.0,2.5) node(a30){$1$};
                    \draw (4.0,2.0,0.5) node(a31){$1$};
                    \draw (4.0,2.0,2.5) node(a32){$1$};
                    \draw (3.0,1.0,0.5) node(a33){$2$};
                    \draw (3.0,1.0,2.5) node(a34){$2$};
                    \draw (4.0,1.0,0.5) node(a35){$2$};
                    \draw (4.0,1.0,2.5) node(a36){$2$};
                    \draw (3.0,0.0,1.5) node(a37){$3$};
                    \draw (3.0,2.0,1.5) node(a38){$3$};
                    \draw (4.0,0.0,1.5) node(a39){$3$};
                    \draw (4.0,2.0,1.5) node(a40){$3$};
                    \draw (3.0,1.0,1.5) node(a41){$4$};
                    \draw (4.0,1.0,1.5) node(a42){$4$};
                    \draw[very thick,-latex](a40)--(a32);
                    \draw[very thick,-latex](a31)--(a40);
                    \draw[very thick,-latex](a38)--(a28);
                    \draw[very thick,-latex](a27)--(a38);
                    \draw[very thick,-latex](a16)--(a22);
                    \draw[very thick,-latex](a4)--(a10);
                    \draw[very thick,-latex](a14)--(a20);
                    \draw[very thick,-latex](a2)--(a8);
                    \draw[very thick,-latex](a36)--(a32);
                    \draw[very thick,red](a42)--(a32);
                    \draw[very thick,-latex](a42)--(a40);
                    \draw[very thick,red](a35)--(a40);
                    \draw[very thick,-latex](a35)--(a31);
                    \draw[very thick,red](a38)--(a36);
                    \draw[very thick,-latex](a38)--(a35);
                    \draw[very thick,-latex](a34)--(a28);
                    \draw[very thick,red](a41)--(a28);
                    \draw[very thick,-latex](a41)--(a38);
                    \draw[very thick,red](a33)--(a38);
                    \draw[very thick,-latex](a33)--(a27);
                    \draw[very thick,-latex](a24)--(a22);
                    \draw[very thick,red](a18)--(a22);
                    \draw[very thick,-latex](a18)--(a16);
                    \draw[very thick,-latex](a12)--(a10);
                    \draw[very thick,red](a6)--(a10);
                    \draw[very thick,-latex](a6)--(a4);
                    \draw[very thick,-latex](a20)--(a24);
                    \draw[very thick,red](a14)--(a24);
                    \draw[very thick,-latex](a14)--(a18);
                    \draw[very thick,red](a8)--(a12);
                    \draw[very thick,-latex](a8)--(a6);
                    \draw[very thick,red](a2)--(a6);
                    \draw[very thick,-latex](a23)--(a20);
                    \draw[very thick,red](a17)--(a20);
                    \draw[very thick,-latex](a17)--(a14);
                    \draw[very thick,-latex](a11)--(a8);
                    \draw[very thick,red](a5)--(a8);
                    \draw[very thick,-latex](a5)--(a2);
                    \draw[very thick,-latex](a42)--(a36);
                    \draw[very thick,-latex](a35)--(a42);
                    \draw[very thick,-latex](a41)--(a34);
                    \draw[very thick,-latex](a33)--(a41);
                    \draw[very thick,-latex](a18)--(a24);
                    \draw[very thick,-latex](a6)--(a12);
                    \draw[very thick,-latex](a17)--(a23);
                    \draw[very thick,-latex](a5)--(a11);
                    \draw[very thick,-latex](a30)--(a36);
                    \draw[very thick,red](a39)--(a36);
                    \draw[very thick,-latex](a39)--(a42);
                    \draw[very thick,red](a29)--(a42);
                    \draw[very thick,-latex](a29)--(a35);
                    \draw[very thick,-latex](a36)--(a37);
                    \draw[very thick,red](a35)--(a37);
                    \draw[very thick,-latex](a26)--(a34);
                    \draw[very thick,red](a37)--(a34);
                    \draw[very thick,-latex](a37)--(a41);
                    \draw[very thick,red](a25)--(a41);
                    \draw[very thick,-latex](a25)--(a33);
                    \draw[very thick,-latex](a21)--(a24);
                    \draw[very thick,red](a15)--(a24);
                    \draw[very thick,-latex](a15)--(a18);
                    \draw[very thick,-latex](a9)--(a12);
                    \draw[very thick,red](a3)--(a12);
                    \draw[very thick,-latex](a3)--(a6);
                    \draw[very thick,red](a24)--(a19);
                    \draw[very thick,-latex](a24)--(a13);
                    \draw[very thick,red](a18)--(a13);
                    \draw[very thick,-latex](a12)--(a7);
                    \draw[very thick,red](a6)--(a7);
                    \draw[very thick,-latex](a6)--(a1);
                    \draw[very thick,-latex](a19)--(a23);
                    \draw[very thick,red](a13)--(a23);
                    \draw[very thick,-latex](a13)--(a17);
                    \draw[very thick,-latex](a7)--(a11);
                    \draw[very thick,red](a1)--(a11);
                    \draw[very thick,-latex](a1)--(a5);
                    \draw[very thick,-latex](a39)--(a30);
                    \draw[very thick,-latex](a29)--(a39);
                    \draw[very thick,-latex](a37)--(a26);
                    \draw[very thick,-latex](a25)--(a37);
                    \draw[very thick,-latex](a15)--(a21);
                    \draw[very thick,-latex](a3)--(a9);
                    \draw[very thick,-latex](a13)--(a19);
                    \draw[very thick,-latex](a1)--(a7);
                    \draw[very thick , blue , -latex](1.7,1,1.5) -- (2.3,1,1.5);
                \end{tikzpicture} 
            \caption{$3d$ printing of an inconsistent periodic quiver for $D_3$.}
            \label{uucd}
        \end{center}
    \end{figure}
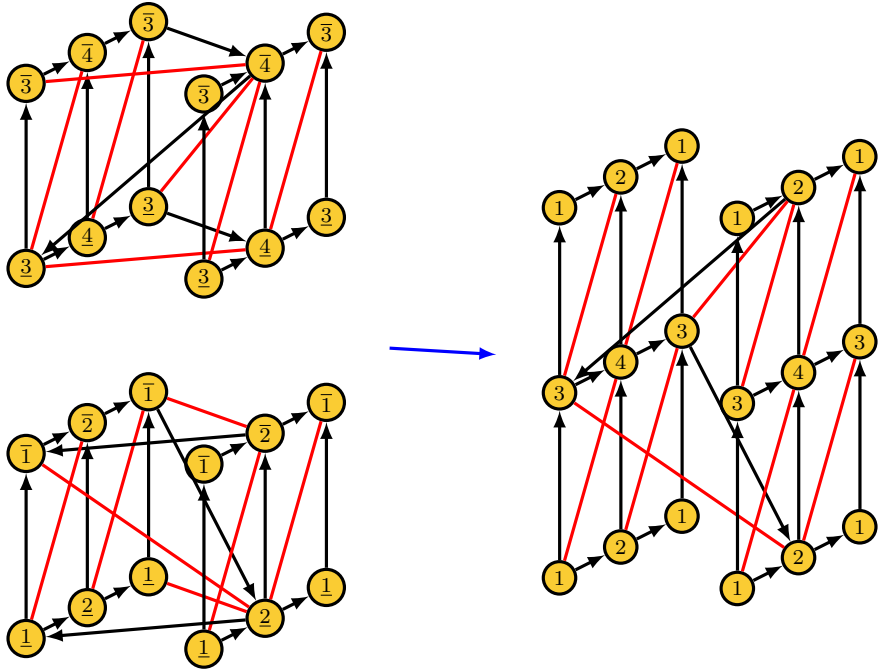

\fref{uucd} shows the periodic quiver obtained via the corresponding $3d$ printing. The $J$- and $E$- terms are:
     \begin{align}
        \renewcommand{\arraystretch}{1.2}
        \begin{array}{rrclcrcl} & & J & &\phantom{abcde}& & E & \\\Lambda_{14}: & Y_{43}X_{32}X_{23}Z_{31} & - & Z_{42}X_{23}X_{32}Y_{21}  &&  Z_{13}Y_{34} & - & Y_{12}Z_{24}\\\Lambda_{23}^{(1)}: & Y_{34}Y_{43}X_{32} & - & X_{32}Y_{21}Y_{12}  &&  Z_{24}Z_{42}X_{23} & - & X_{23}Z_{31}Z_{13}\\\Lambda_{23}^{(2)}: & X_{32}X_{23}Z_{31}Y_{12} & - & Y_{34}Z_{42}X_{23}X_{32}  &&  Z_{24}Y_{43} & - & Y_{21}Z_{13}\\\Lambda_{32}^{(1)}: & X_{23}Y_{34}Y_{43} & - & Y_{21}Y_{12}X_{23}  &&  Z_{31}Z_{13}X_{32} & - & X_{32}Z_{24}Z_{42}\\\Lambda_{32}^{(2)}: & Y_{21}Z_{13}X_{32}X_{23} & - & X_{23}X_{32}Z_{24}Y_{43}  &&  Z_{31}Y_{12} & - & Y_{34}Z_{42}\\\Lambda_{41}: & Z_{13}X_{32}X_{23}Y_{34} & - & Y_{12}X_{23}X_{32}Z_{24}  &&  Z_{42}Y_{21} & - & Y_{43}Z_{31}\end{array} \label{uucdje}
    \end{align}
 
 This theory is clearly inconsistent and needs to be reduced, since it has 4 gauge groups, while the normalized volume of the $D_{3}$ toric diagram is 3.
 
A quick inspection of the quiver reveals that it does not have any explicit bubble, namely a node with only one incoming or one outgoing chiral arrow. Let us then determine whether it is possible to reduce this theory by higgsing. For this, it is useful to consider the $P$-matrix, which summarizes the map between points in chiral fields in the quiver and brick matchings, which in turn correspond to points in the toric diagram. It is given by
\begin{align}
        \begin{array}{c|cccccc}&p_{1}&p_{2}&p_{3}&p_{4}&p_{5}&p_{6}\\ \hline X_{23}&0&0&0&1&0&0\\X_{32}&0&0&1&0&0&0\\Z_{31}&0&0&0&0&0&1\\Z_{42}&0&0&0&0&0&1\\Y_{12}&1&0&0&0&0&0\\Y_{34}&1&0&0&0&0&0\\Y_{21}&0&1&0&0&0&0\\Y_{43}&0&1&0&0&0&0\\Z_{13}&0&0&0&0&1&0\\Z_{24}&0&0&0&0&1&0\\\end{array}
\label{d3cdpm}
\end{align}
It is clear that it is impossible to turn a vev for any chiral field while preserving the toric diagram. Any vev would remove a brick matching. We conclude that it is impossible to reduce this phase by higgsing. It is worth noting that it always possible to reduce inconsistent brane tilings by higgsing, so this appears to be a novel feature of brane brick models.

\subsubsection*{Triality and Reduction}
    
\begin{figure}
            \begin{center}
                  \tdplotsetmaincoords{81}{110} 
                    \tdplotsetrotatedcoords{76.5}{0}{110} 
                   \begin{tikzpicture}[scale=2.5,xscale = 1.2 ,tdplot_rotated_coords] 
                        \tikzstyle{every node}=[circle,very thick,fill=yellow2,draw,inner sep=2pt,font=\footnotesize]
        \draw (0.0,0.0,0.0) node(a1){$1$};
        \draw (0.0,0.0,2.0) node(a2){$1$};
        \draw (0.0,2.0,0.0) node(a3){$1$};
        \draw (0.0,2.0,2.0) node(a4){$1$};
        \draw (2.0,0.0,0.0) node(a5){$1$};
        \draw (2.0,0.0,2.0) node(a6){$1$};
        \draw (2.0,2.0,0.0) node(a7){$1$};
        \draw (2.0,2.0,2.0) node(a8){$1$};
        \draw (1.0,0.0,0.0) node(a9){$2$};
        \draw (1.0,0.0,2.0) node(a10){$2$};
        \draw (1.0,2.0,0.0) node(a11){$2$};
        \draw (1.0,2.0,2.0) node(a12){$2$};
        \draw (0.0,0.5,1.0) node(a13){$3$};
        \draw (2.0,0.5,1.0) node(a14){$3$};
        \draw (1.0,1.5,1.0) node(a15){$4$};
        \draw[very thick,-latex](a14)--(a8);
        \draw[very thick,-latex](a7)--(a15);
        \draw[very thick,red](a14)--(a4);
        \draw[very thick,red](a15)--(a12);
        \draw[very thick,-latex](a15)--(a11);
        \draw[very thick,-latex](a11)--(a12);
        \draw[very thick,-latex](a12)--(a4);
        \draw[very thick,red](a11)--(a4);
        \draw[very thick,-latex](a11)--(a3);
        \draw[very thick,-latex](a4)--(a15);
        \draw[very thick,-latex](a13)--(a4);
        \draw[very thick,-latex](a15)--(a14);
        \draw[very thick,red](a13)--(a15);
        \draw[very thick,-latex](a14)--(a13);
        \draw[very thick,-latex](a10)--(a14);
        \draw[very thick,red](a9)--(a14);
        \draw[very thick,red](a13)--(a10);
        \draw[very thick,-latex](a13)--(a9);
        \draw[very thick,-latex](a9)--(a10);
        \draw[very thick,-latex](a10)--(a2);
        \draw[very thick,red](a9)--(a2);
        \draw[very thick,-latex](a9)--(a1);
    \end{tikzpicture} 
                \caption{Periodic quiver obtained by performing a triality transformation on node 4 of \fref{uucd}.}
                \label{uucd4t}
            \end{center}
        \end{figure}
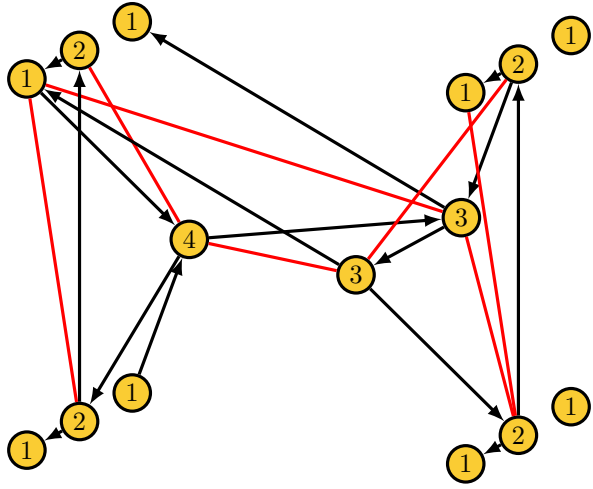

Performing a triality transformation on node $4$ of the theory in \fref{uucd} we obtain another toric phase, which is described by the periodic quiver shown in \fref{uucd4t}. Its $J$- and $E$- terms are:
        \begin{align}  
            \renewcommand{\arraystretch}{1.2}
            \begin{array}{rccccrcl} & & J & &\phantom{abcde}& & E & \\\Lambda_{12}: & X_{21}^{}X_{14}^{+}X_{43}^{}X_{31}^{} & - & X_{22}^{}X_{21}^{}  &&  X_{14}^{-}X_{42}^{} & - & X_{14}^{+}X_{42}^{}X_{23}^{}X_{32}^{}\\\Lambda_{13}: & X_{31}^{}X_{14}^{+}X_{42}^{}X_{21}^{} & - & X_{33}^{}X_{31}^{}  &&  X_{14}^{-}X_{43}^{} & - & X_{14}^{+}X_{43}^{}X_{32}^{}X_{23}^{}\\\Lambda_{23}^{(1)}: & X_{32}^{}X_{21}^{}X_{14}^{+}X_{42}^{} & - & X_{33}^{}X_{32}^{}  &&  X_{22}^{}X_{23}^{} & - & X_{23}^{}X_{31}^{}X_{14}^{+}X_{43}^{}\\\Lambda_{23}^{(2)}: & X_{31}^{}X_{14}^{+}X_{43}^{}X_{32}^{} & - & X_{32}^{}X_{22}^{}  &&  X_{21}^{}X_{14}^{+}X_{42}^{}X_{23}^{} & - & X_{23}^{}X_{33}^{}\\\Lambda_{24}: & X_{43}^{}X_{31}^{}X_{14}^{+}X_{42}^{} & - & X_{42}^{}X_{22}^{}  &&  X_{23}^{}X_{32}^{}X_{21}^{}X_{14}^{+} & - & X_{21}^{}X_{14}^{-}\\\Lambda_{34}: & X_{42}^{}X_{21}^{}X_{14}^{+}X_{43}^{} & - & X_{43}^{}X_{33}^{}  &&  X_{32}^{}X_{23}^{}X_{31}^{}X_{14}^{+} & - & X_{31}^{}X_{14}^{-}\end{array}
        \end{align}
    
The $P$-matrix is
\begin{align}
        \begin{array}{c|cccccc}&p_{1}&p_{2}&p_{3}&p_{4}&p_{5}&p_{6}\\ \hline X_{21}^{}&0&0&0&0&1&0\\X_{33}^{}&0&0&0&0&1&1\\X_{42}^{}&0&0&0&0&0&1\\X_{22}^{}&1&1&0&0&0&0\\X_{31}^{}&0&1&0&0&0&0\\X_{43}^{}&1&0&0&0&0&0\\X_{14}^{-}&0&0&1&1&0&0\\X_{23}^{}&0&0&0&1&0&0\\X_{32}^{}&0&0&1&0&0&0\\ X_{14}^{+} & 0 & 0 & 0 & 0 & 0 & 0 \end{array}
\end{align}
Remarkably $X_{14}^{+}$ is not contained in any brick matching. The reason for this is that it participates in both $J$- and $E$-terms of every Fermi field.  Giving a non-zero vev to $X_{14}^{+}$ preserves the toric diagram and hence leads to the desired reduction. This higgsing identifies nodes $1$ and $4$ and, up to a trivial relabeling of nodes, precisely gives rise to the $D_3$ theory presented in \sref{section_D3_from_C^4}.  We have thus provided two alternative ways of systematically constructing the $D_3$ theory: $3d$ printing from $\mathbb{C}^3$, and $3d$ printing from the conifold plus reduction.

\subsection{Reduction for Higher Dimensional Calabi-Yau's}

\label{section_reduction_general_m}

Before concluding, we would like to collect some thoughts regarding the consistency/ reducibility of quivers associated to higher dimensional CY singularities. 

It was recently shown in \cite{Franco:2017lpa}, that  singular CY $(m+2)$-folds are associated to graded quivers with potentials. The degree of the arrows in the quiver is an integer $c$, which lies in the $0\leq c \leq m$ range. We refer the reader to \cite{Franco:2017lpa} for a detailed discussion of these theories. For $m=0,1,2,3$, these quivers can be interpreted as minimally supersymmetric gauge theories in $d=6,4,2,0$. For all $m$, we expect there is also a physical interpretation in terms of the category of branes in the topological B-model on CY $(m+2)$-folds \cite{to_appear}.

When the CY $(m+2)$-folds are toric, the full theories, namely the quivers and the potentials, can be encoded in terms of periodic quivers on $\mathbb{T}^{m+1}$. This construction generalizes the known cases of $m\leq 3$. Similarly, we expect the graphs dual to the periodic quivers to be a powerful bridge between geometry and quiver theories. These objects would generalize elliptic models for $m=0$ \cite{Brunner:1997gf}, brane tilings for $m=1$ \cite{Franco:2005rj}, brane brick models for $m=2$ \cite{Franco:2015tya} and brane hyperbrick models for $m=3$ \cite{Franco:2016tcm}. 

For general $m$ it is natural to expect that, generalizing the $m=1$ and $2$ cases, inconsistency manifests as a mismatch between the number of nodes in the quivers and the normalized volumes of the corresponding toric diagrams.

We expect these theories can be render consistent by higgsing or bubble reduction. Let us elaborate on what we mean as bubble reduction for general $m$. In \cite{Franco:2017lpa}, it was noted that for any $m$, graded quivers with potentials admit order $(m+1)$ mutations that coincide with the dualities of the corresponding gauge theories for $m\leq 3$ (see also \cite{MR3590528} for related work). We envisage that bubbles correspond to nodes that would disappear by application of these mutations. These are nodes with a single incoming or outgoing degree 0 arrow. Degree 0 arrows generalize chiral fields to arbitrary $m$.

\section{Conclusions and Outlook}

\label{section_conclusions}

Recently, there has been significant progress in the connection between the geometry of toric CY 4-folds and the $2d$ $(0,2)$ gauge theories on D1-branes probing them. Milestones include the discovery of brane brick models \cite{Franco:2015tya} and their understanding in terms of mirror symmetry \cite{Franco:2016qxh}. Another important line of progress involves methods for relating CY 4-folds to CY 3-folds and their associated gauge theories. Orbifold reduction \cite{Franco:2016fxm} was the first step in this direction and the $3d$ printing algorithm introduced in this paper considerably supersedes it. Such methods are certainly useful from a practical point of view, since they are efficient tools for easily finding the gauge theories associated to rather general CY 4-folds. More importantly, they also lead to conceptual insights by connecting gauge theories in different dimensions, their associated CY's and the underlying combinatorial objects (brane tilings and brane brick models).

We presented various examples illustrating the power of $3d$ printing over earlier techniques. We were able to derive gauge theories for $D_3$, $H_4$ and $Q^{1,1,1}$ almost effortlessly. Previously, these geometries could only be dealt with using the straightforward but practically involved process of partial resolution. Similarly, we managed to generate an unknown triality phase for $Q^{1,1,1}/\mathbb{Z}_2$, which in turn motivated a full classification of the toric triality phases for this geometry. These examples exploit the two novel properties of $3d$ printing: the possibility of simultaneously lifting multiple points of $T_{\rm{CY}_3}$ and of using more than one perfect matching for lifting a given point.

We anticipate that one of the most important applications of $3d$ printing and its generalizations will be in the context of higher dimensional CY's. Singular CY $(m + 2)$-folds are associated to graded quivers (of maximum degree $m$) with potentials \cite{Franco:2017lpa}. The physical relevance of these theories is expected to be in terms of the category of branes in the topological B-model. For toric CY $(m + 2)$-folds, these theories are fully encoded by periodic quivers on $\mathbb{T}^{m+1}$ or, equivalently, the dual graphs generalizing brane tilings and brane brick models. Constructing such theories for $m>2$ is a challenging open question. In a forthcoming paper \cite{to_appear2}, we will introduce a substantial generalization of $3d$ printing that, starting from the theories associated to a toric CY$_{m+2}$ and a toric CY$_{m'+2}$, generates the quiver theory for a toric CY$_{m+m'+3}$. $3d$ printing and its natural generalization to higher dimensional CY's correspond to the simple case of $m'=0$. However, the new method for arbitrary $m$ and $m'$, constructs the quiver theories for rather general toric CY's. It is reasonable to expect that this procedure may also give rise to a useful algorithm for constructing fractional branes and exceptional collections for the corresponding geometries. We refer the reader to \cite{Wijnholt:2002qz,Herzog:2003zc,Herzog:2004qw,Aspinwall:2004vm,Aspinwall:2004bs,Herzog:2006bu,Closset:2017yte} for discussions of the CY 3-fold and 4-fold cases. 

It would also be worth studying the combinatorics of triality, determining whether the freedom in the initial $4d$ phase and lifted perfect matchings can account for all triality duals of a $3d$ theory. For example, it is immediately clear that many of the triality duals for $\mathbb{Q}^{1,1,1}/\mathbb{Z}_2$ summarized in Appendix \sref{section_toric_phases_Q111/Z2} can be obtained from the two phases of $F_0$ using different perfect matchings for $3d$ printing.

Finally, it would be interesting to explore whether, and if so how, $3d$ printing and its generalizations are related to approaches for connecting gauge theories in different dimensions through compactification, see e.g. \cite{Niarchos:2012ah, Aharony:2013dha,Aharony:2013kma,Benini:2015bwz,Hwang:2017nop,Aharony:2017adm,Closset:2017xsc}.

\acknowledgments

We would like to thank S. Lee, G. Musiker, R.-K. Seong and C. Vafa for earlier collaborations on related topics. We gratefully acknowledges support from the Simons Center for Geometry and Physics, Stony Brook University, where some of the research for this paper was performed during the 2017 Simons Summer Workshop. Our work is supported by the U.S. National Science Foundation grant PHY-1518967 and by a PSC-CUNY award. 


\appendix

\newpage

\section{The Toric Phases of $Q^{1,1,1}/\mathbb{Z}_2$}

\label{section_toric_phases_Q111/Z2}

Here we present the periodic quivers for the 14 toric phases of $Q^{1,1,1}/\mathbb{Z}_2$. The connections between these theories were discussed in \sref{section_triality_web}.

\begin{table}[H]
\centering

\caption{Periodic quivers for the 14 toric phases of $\mathbb{Q}^{1,1,1}/\mathbb{Z}_2$.}
\end{table}

\bibliographystyle{JHEP}
\bibliography{mybib}

\end{document}